\pdfoutput=1
\documentclass[11pt,twoside,a4paper,cmspaper,final,collab]{cms-tdr}
\def\svnVersion{1083207-D}\def\svnDate{2021/09/10}\def\cmsCernNoTag{CERN-EP-2021-022}\def\cmsCernDate{\today}\def\cmsMessage{"Published in Physical Review D as \href{http://dx.doi.org/10.1103/PhysRevD.104.052001}{\doi{10.1103/PhysRevD.104.052001}.}"}
\begin{document}\cmsNoteHeader{SUS-19-010}

\newcommand{\lsp}{\PSGczDo}
\newcommand{\stopq}{\PSQt}
\newcommand{\charginopm}{\PSGcpmDo}
\newcommand{\charginop}{\PSGcpDo}
\newcommand{\tstar}{\HepParticle{\PQt}{}{(\ast)}}
\newcommand{\znn}{\ensuremath{\PZ\to\PGn\PAGn}\xspace}
\newcommand{\zparennunujets}{\ensuremath{\PZ(\PGn\PAGn)+\text{jets}}\xspace}
\newcommand{\zparenll}{\ensuremath{\PZ(\Pell^+\Pell^-)}\xspace}
\newcommand{\ttZ}{\ensuremath{\ttbar\PZ}\xspace}
\newcommand{\ttH}{\ensuremath{\ttbar\PH}\xspace}
\newcommand{\WW}{\ensuremath{\PW\PW}\xspace}
\newcommand{\WZ}{\ensuremath{\PW\PZ}\xspace}
\newcommand{\ZZ}{\ensuremath{\PZ\PZ}\xspace}
\newcommand{\wjets}{\ensuremath{\PW+\text{jets}}\xspace}
\newcommand{\zjets}{\ensuremath{\PZ+\text{jets}}\xspace}
\newcommand{\gjets}{\ensuremath{\PGg+\text{jets}}\xspace}
\newcommand{\dyjets}{\ensuremath{\text{DY}+\text{jets}}\xspace}
\newcommand{\ljets}{\ensuremath{\Pell+\text{jets}}\xspace}
\newcommand{\mgluino}{\ensuremath{m_{\PSg}}\xspace}
\newcommand{\mlsp}{\ensuremath{m_{\lsp}}\xspace}
\newcommand{\mchargino}{\ensuremath{m_{\charginopm}}\xspace}
\newcommand{\mstop}{\ensuremath{m_{\stopq}}\xspace}
\newcommand{\mtop}{\ensuremath{m_{\PQt}}\xspace}
\newcommand{\mW}{\ensuremath{m_{\PW}}\xspace}
\newcommand{\dm}{\ensuremath{\Delta m}\xspace}
\newcommand{\ptb}{\ensuremath{\pt^{\PQb}}\xspace}
\newcommand{\ptISR}{\ensuremath{\pt^{\text{ISR}}}\xspace}
\newcommand{\ptsum}{\ensuremath{\pt^\text{sum}}\xspace}
\newcommand{\mll}{\ensuremath{m_{\Pell^+\Pell^-}}\xspace}
\newcommand{\jet}[1]{\ensuremath{\mathrm{j}_{#1}}\xspace}
\newcommand{\metsig}{\ensuremath{\ptmiss/\sqrt{\smash[b]{\HT}}}\xspace}
\newcommand{\miniiso}{\ensuremath{I_\text{mini}}\xspace}
\newcommand{\Nj}{\ensuremath{N_{\text{j}}}\xspace}
\newcommand{\Nb}{\ensuremath{N_{\PQb}}\xspace}
\newcommand{\Nw}{\ensuremath{N_{\PW}}\xspace}
\newcommand{\Nt}{\ensuremath{N_{\PQt}}\xspace}
\newcommand{\Nres}{\ensuremath{N_{\text{res}}}\xspace}
\newcommand{\Nsv}{\ensuremath{N_{\text{SV}}}\xspace}
\newcommand{\mTb}{\ensuremath{\mT^{\PQb}}\xspace}
\newcommand{\RZ}{\ensuremath{R_{\PZ}}\xspace}
\newcommand{\SGamma}{\ensuremath{S_{\PGg}}\xspace}

\ifthenelse{\boolean{cms@external}}{\renewcommand{\NA}{\ensuremath{\cdots}}}{}
\ifthenelse{\boolean{cms@external}}{\newcommand{\NAdescription}{An ellipsis\xspace}}{\newcommand{\NAdescription}{A dash\xspace}}
\newlength\cmsTabSkip\setlength{\cmsTabSkip}{1ex}
\newcommand{\cmsTable}[1]{\resizebox{\textwidth}{!}{#1}}
\ifthenelse{\boolean{cms@external}}{\providecommand{\cmsLeft}{upper\xspace}}{\providecommand{\cmsLeft}{left\xspace}}
\ifthenelse{\boolean{cms@external}}{\providecommand{\cmsRight}{lower\xspace}}{\providecommand{\cmsRight}{right\xspace}}

\newcommand{\TF}[1]{\ensuremath{\mathit{TF}_\text{#1}}\xspace}
\newcommand{\NMC}[1]{\ensuremath{N_\text{MC}^{#1}}\xspace}
\newcommand{\NP}[1]{\ensuremath{N_\text{pred}^{#1}}\xspace}
\newcommand{\ND}[1]{\ensuremath{N_\text{data}^{#1}}\xspace}
\newcommand{\dphi}{\ensuremath{\Delta \phi}\xspace}

\cmsNoteHeader{SUS-19-010}
\title{Search for top squark production in fully-hadronic final states in proton-proton collisions at \texorpdfstring{$\sqrt{s} = 13\TeV$}{sqrt(s) = 13 TeV}}

\date{\today}

\abstract{
A search for production of the supersymmetric partners of the top quark, top squarks, is presented.
The search is based on proton-proton collision events containing multiple jets, no leptons, and large transverse momentum imbalance.
The data were collected with the CMS detector at the CERN LHC at a center-of-mass energy of 13\TeV, and correspond to an integrated luminosity of 137\fbinv.
The targeted signal production scenarios are direct and gluino-mediated top squark production, including scenarios in which the top squark and neutralino masses are nearly degenerate.
The search utilizes novel algorithms based on deep neural networks that identify hadronically decaying top quarks and \PW bosons, which are expected in many of the targeted signal models.
No statistically significant excess of events is observed relative to the expectation from the standard model, and
limits on the top squark production cross section are obtained in the context of simplified supersymmetric models for various production and decay modes.
Exclusion limits as high as 1310\GeV are established at the 95\% confidence level on the mass of the top squark for direct top squark production models,
and as high as 2260\GeV on the mass of the gluino for gluino-mediated top squark production models.
These results represent a significant improvement over the results of previous searches for supersymmetry by CMS in the same final state.
}

\hypersetup{%
pdfauthor={CMS Collaboration},%
pdftitle={Search for top squark production in fully-hadronic final states in proton-proton collisions at sqrt(s) = 13 TeV},%
pdfsubject={CMS},%
pdfkeywords={CMS, supersymmetry, top squark, top quark tagging}}

\maketitle

\section{Introduction}
\label{sec:introduction}

The standard model (SM) of particle physics correctly predicts a wide range of phenomena.
Nonetheless, the SM has well-known shortcomings, such as an instability in the calculation of higher-order corrections to the Higgs boson mass, known as the fine-tuning (or hierarchy) problem~\cite{Barbieri:1987fn}.
There is also an abundance of experimental observations, including the existence of dark matter, that cannot be explained within the context of the SM alone~\cite{Rubin:1970zza}.
Supersymmetry (SUSY)~\cite{Wess:1973kz,Fayet:1976cr,Barbieri:1982eh,Nilles:1983ge,Haber:1984rc,Martin:1997ns} is an extension of the SM that could help explain some of these shortcomings by introducing an additional symmetry between the fermions and the bosons.
As a result, it predicts a supersymmetric partner particle (superpartner) for each SM particle.
The quantum numbers for each superpartner are the same as the quantum numbers for the corresponding SM particle with the exception of the spin, which differs by a half-integer unit.
The superpartners of quarks, gluons, and Higgs bosons are squarks \PSQ, gluinos \PSg, and higgsinos, respectively.
The neutral and charged higgsinos mix with the superpartners of the neutral and charged electroweak gauge bosons to form neutralinos \PSGcz and charginos \PSGcpm.

Divergent quantum loop corrections to the Higgs boson mass due to virtual SM particles can be canceled by corresponding contributions from virtual SUSY particles~\cite{deCarlos:1993rbr,Dimopoulos:1995mi},
which may resolve the fine-tuning problem.
The symmetry proposed by SUSY is not exact, as no SUSY particles have been observed yet and they must therefore be more massive than their SM counterparts.
The stabilizing features of SUSY can still survive if SUSY particles are not much heavier than their SM counterparts.
Superpartners of third-generation quarks play particularly important roles in this consideration, as the third-generation quarks and squarks
have large couplings to the Higgs boson, and therefore produce the largest corrections.
In so-called natural models of SUSY~\cite{Papucci:2011wy,Brust:2011tb,Delgado:2012eu},
the third-generation squarks (top squarks and bottom squarks), gluino, and higgsinos are expected to have masses no larger than a few \TeVns~\cite{Han:2016xet,Baer:2016wkz,Baer:2016bwh}.
At the same time, null results from SUSY searches at the CERN LHC so far also suggest that the first two generation squarks have much larger masses~\cite{Feng:2013pwa} and are expected to be decoupled at the LHC energy.
These considerations provide strong motivations to search for top squark production at the LHC.

In SUSY models with $R$-parity conservation~\cite{Farrar:1978xj}, SUSY particles are produced in pairs, and the lightest supersymmetric particle (LSP) is stable.
The lightest neutralino \lsp is assumed to be the LSP, which would be a good candidate for weakly-interacting massive particle dark matter.
The \lsp interacts only weakly, so it does not leave a signal in the CMS detector.
Because the \lsp is present in the decay chain of top squarks, it provides a powerful experimental probe for signal events: a large momentum imbalance in the plane perpendicular to the beam axis.

Dedicated searches for top squarks in proton-proton ($\Pp\Pp$) collision events at $\sqrt{s} = 13\TeV$ have been carried out by both the
ATLAS~\cite{Aaboud:2016lwz,Aaboud:2017ejf,Aaboud:2017nfd,Aaboud:2017ayj,Aaboud:2017opj,Aaboud:2017nmi,Aaboud:2017aeu,Aaboud:2018kya,Aaboud:2018zjf,Aaboud:2018hdl,Aaboud:2019trc,Aad:2020sgw,Aad:2020uwr,Aad:2020qwe,Aad:2020srt,Aad:2019ftg,Aad:2020aob,Aad:2021hjy} and
CMS~\cite{Sirunyan:2016jpr,Khachatryan:2017rhw,Sirunyan:2017xse,SUS-16-049,SUS-16-050,Sirunyan:2017kiw,Sirunyan:2017leh,Sirunyan:2018iwl,Sirunyan:2018rlj,Sirunyan:2019zyu,SUS-19-009,SUS-19-011} Collaborations.
In this paper, we present a search for production of top squarks in fully-hadronic final states.
This search is interpreted in $R$-parity conserving SUSY models in which the top squarks are produced in pairs or are produced via cascade decays of pair-produced gluinos.
The data were collected by the CMS detector at the LHC in 2016--2018 and correspond to an integrated luminosity of 137\fbinv
of $\Pp\Pp$ collisions at a center-of-mass energy of 13\TeV.
The search presented in this paper is an extension of the analyses presented in Refs.~\cite{SUS-16-049,SUS-16-050}, using novel top quark and \PW boson tagging algorithms, reoptimized search bins, and a data sample about four times larger.

The top quark and \PW boson tagging algorithms identify hadronically decaying top quarks and \PW bosons produced in SUSY particle decay chains.
At high momentum, the decay products of a hadronically decaying top quark or \PW boson tend to merge into a single large-radius jet.
At lower momentum, hadronic top quark decays can be resolved as three smaller-size jets.
Separate algorithms are employed to benefit from these two different classes of decay product kinematic properties.
Previous top squark searches already used hadronic top quark and \PW boson tagging algorithms, which were based on jet properties and decision trees~\cite{SUS-16-049,SUS-16-050}; however, the ones utilized in the present search benefit from the use of deep neural networks~\cite{deepCSV,JME-18-002,SUS-19-009}.
These tagging algorithms are critical for improving the sensitivity of the search to models with on-shell top quarks and \PW bosons in the final state.
Separate search bins with different top quark and \PW boson multiplicities help to maintain high sensitivities for both direct and gluino-mediated top squark production scenarios with different decay modes.

For SUSY models with compressed mass spectra, \ie, models with a small mass difference between the top squark and the LSP,
the search benefits from dedicated search bins that require a high transverse momentum (\pt) jet originating from initial-state radiation (ISR).
Signal events from models with compressed mass spectra generally leave little visible energy in the detector, making such events difficult to identify.
These events can still be detected if the top squark pair recoils against a high-\pt jet arising from ISR.
Some of the models with compressed mass spectra will yield low-\pt bottom quarks in the final states.
The search utilizes an algorithm to identify jets originating from the hadronization of \PQb quarks (\PQb jets)~\cite{deepCSV}.
To improve the sensitivity to models with compressed mass spectra, we also employ another algorithm that is specifically optimized for the identification of low-\pt \PQb quarks (soft \PQb quarks)~\cite{SUS-16-049}.

The paper is organized as follows:
Section~\ref{sec:signal_models} describes the signal models considered in this search.
Sections~\ref{sec:CMSdetector} and \ref{sec:simulation} discuss the CMS detector and the simulated data samples used in this analysis.
The event reconstruction and event selection procedures are presented in Sections~\ref{sec:reconstruction} and \ref{sec:selection}, respectively.
The background prediction methods are described in Section~\ref{sec:background}.
Results and interpretations are detailed in Section~\ref{sec:results} and a summary is presented in Section~\ref{sec:summary}.

\section{Signal models}
\label{sec:signal_models}

Top squark pairs may be produced in many different SUSY models.
In any given SUSY model, the mechanisms by which top squark pairs are produced depend on the parameters of the model.
In this search, we target production scenarios that are motivated by natural SUSY,
in which $R$-parity is conserved and the top squark is produced directly in pairs
or in cascade decays of the gluino.
The signal topologies are characterized by the so-called simplified model spectra~\cite{bib-sms-3,bib-sms-2,bib-sms-4,CMS-SMS} with a small number of parameters describing the masses of the SUSY particles.
In the following paragraphs we describe the specific models, as well as our choices for the parameters of those models, that we use for the interpretations presented in Section~\ref{sec:results}.

\begin{figure*}[th!]
\centering
\includegraphics[width=0.32\linewidth]{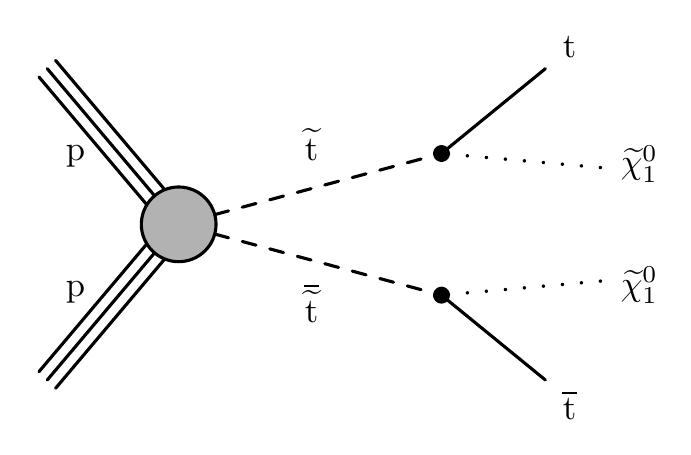}
\includegraphics[width=0.32\linewidth]{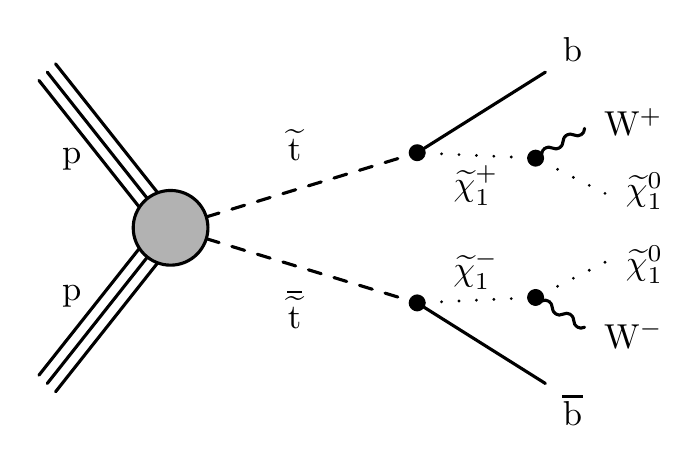}
\includegraphics[width=0.32\linewidth]{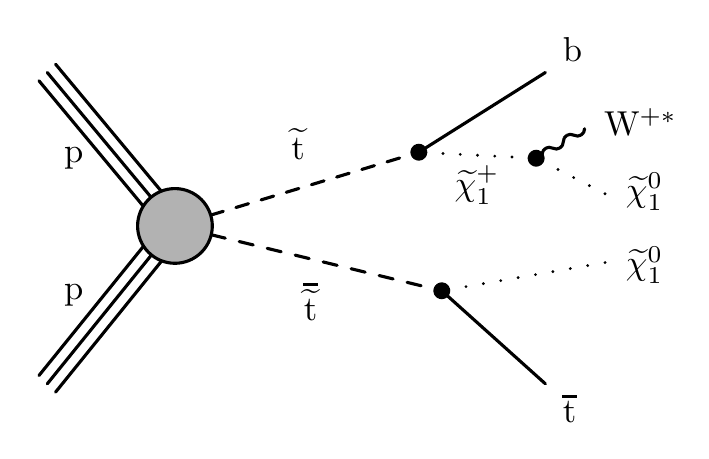}
\includegraphics[width=0.32\linewidth]{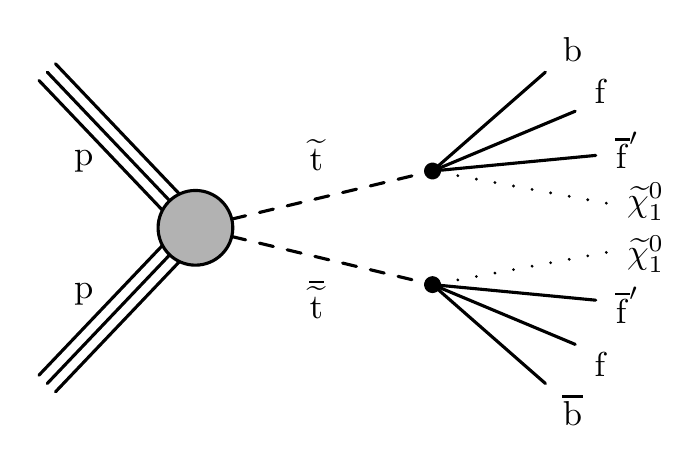}
\includegraphics[width=0.32\linewidth]{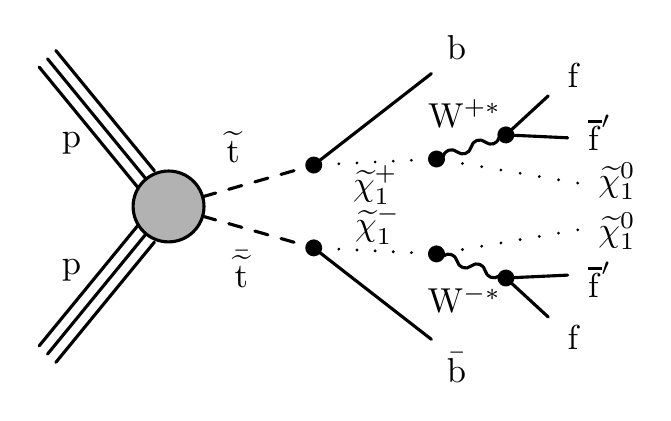}
\includegraphics[width=0.32\linewidth]{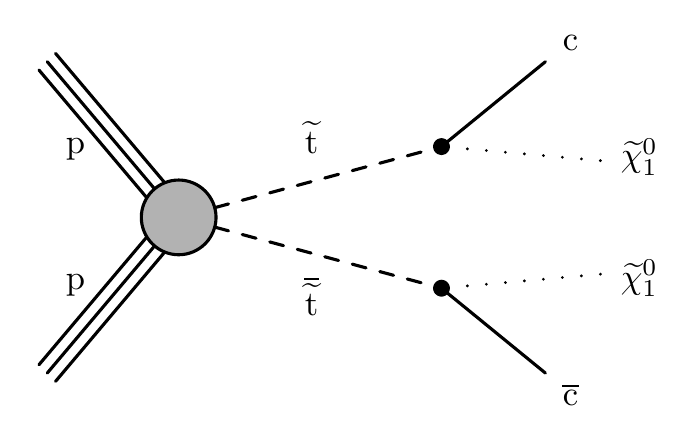}
\caption{
  Diagrams for the direct top squark production scenarios considered in this study:
  the T2tt (upper left), T2bW (upper middle), T2tb (upper right), T2ttC (lower left), T2bWC (lower middle), and T2cc (lower right)
  simplified models.
}
\label{fig:qq-event-diagrams}
\end{figure*}

For direct top squark pair production, the models shown in Fig.~\ref{fig:qq-event-diagrams} are considered.
Depending on the specific details of the SUSY model and the mass hierarchy of the SUSY particles,
the top squark decays in a variety of modes.
In particular, the mass difference \dm between the top squark \stopq and the LSP has a large impact on the decay modes of the top squark.
When ${\dm = \mstop - \mlsp}$ is larger than the \PW boson mass \mW, the two decays considered in this search are
${\stopq\to\tstar\lsp}$ with ${\tstar\to\PQb\PWp}$ and ${\stopq\to\PQb\charginop}$ with ${\charginop\to\PWp\lsp}$, as well as their charge conjugate modes.
The former decay mode corresponds to the model denoted by ``T2tt'',
and the latter to the model denoted by ``T2bW''.
In the mixed decay model ``T2tb'', the top squark decays in either of the above decay modes with a branching fraction of 50\%.
For T2tb the compressed mass spectrum of ${\mchargino-\mlsp = 5\GeV}$ is assumed,
which is a likely scenario when \charginopm and \lsp belong to the same gauge eigenstate.
For T2bW, the \charginopm mass is assumed to be the arithmetic mean of the top squark and LSP masses, as in earlier searches~\cite{SUS-16-049,SUS-19-009}.
With this assumption of moderate \charginopm mass, the \charginopm decays to an on-shell \PW boson and \lsp, and the \PW boson produces high momentum objects in the final state.
Sensitivity to the final states expected from these models is enhanced by the application of top quark and \PW boson taggers~\cite{JME-18-002}.

When \dm is smaller than \mW, the decay of the top squark to an on-shell top quark or an on-shell \PW boson is kinematically forbidden.
In such scenarios, the top squark may decay, as in the above T2tt and T2bW models, but via off-shell top quarks or \PW bosons.
The models with these decay modes are denoted by ``T2ttC'' and ``T2bWC'', respectively, where ``C'' represents the compressed mass spectrum between the top squark and the LSP.
Another possible decay mode is the loop-induced flavor changing neutral current process ${\stopq\to\PQc\lsp}$.
The model with this decay mode is referred to as ``T2cc''.
These models with small \dm are phenomenologically well motivated because of the compatibility between their prediction of the relic density of dark matter~\cite{Balazs:2004bu,Ellis:2001nx,Ellis:2014ipa} and cosmological observations;
however, signatures expected from these models
are challenging to search for experimentally because of the lack of high-\pt particles from such top squark decays.
This search gains sensitivity to these models by requiring a high-\pt jet expected to be from ISR, which gives higher \pt to particles from top squark decays, and also by using a soft \PQb quark identification algorithm.

\begin{figure*}[th!]
\centering
\includegraphics[width=0.32\linewidth]{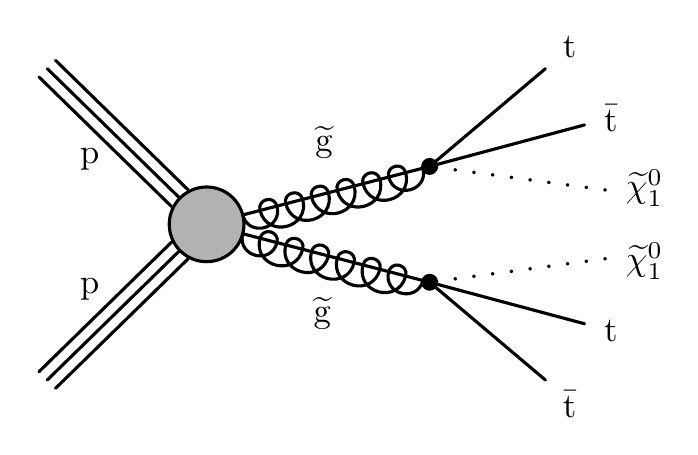}
\includegraphics[width=0.32\linewidth]{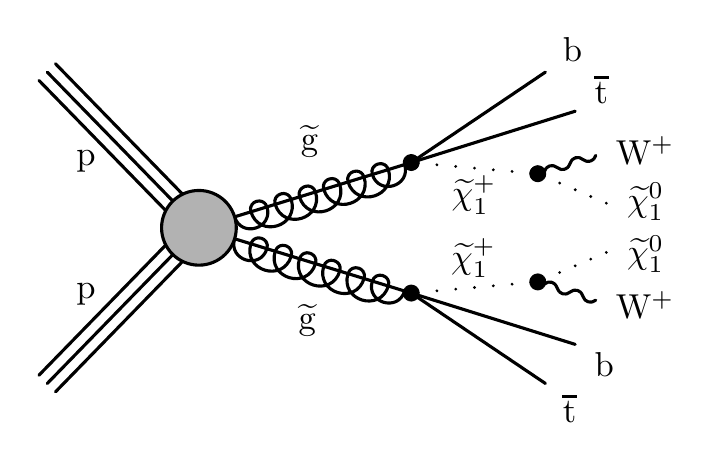}
\includegraphics[width=0.32\linewidth]{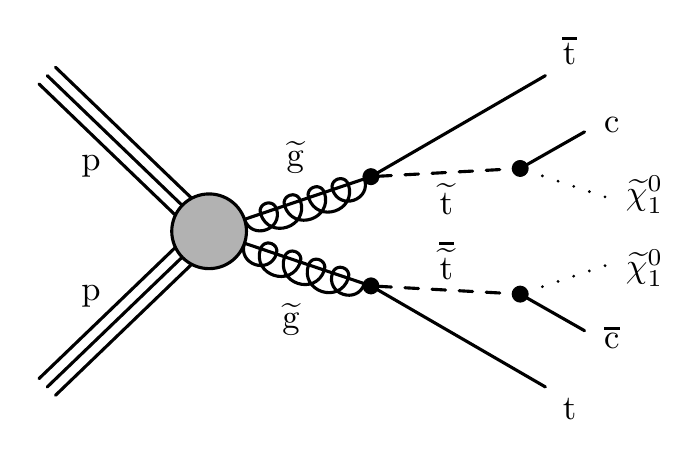}
\caption{
  Diagrams for the direct gluino production scenarios considered in this study:
  the T1tttt (left), T1ttbb (middle), and T5ttcc (right)
  simplified models.
}
\label{fig:gg-event-diagrams}
\end{figure*}

For gluino pair production, the models shown in Fig.~\ref{fig:gg-event-diagrams} are considered.
In the model denoted ``T1tttt'', each of the pair-produced gluinos decays to an off-shell top squark and an on-shell top quark.
The off-shell top squark decays to a top quark and the LSP.
The gluino decay is thus ${\PSg \to \ttbar\lsp}$.
In the ``T1ttbb'' model,
pair-produced gluinos each decay
via an off-shell top or bottom squark, which decay in turn, yielding
${\PSg \to\ttbar\lsp}$ (25\%),
${\PSg \to\PAQt\PQb\charginop}$ or its charge conjugate (50\%),
or ${\PSg \to\bbbar\lsp}$ (25\%).
The mass difference between \charginopm
and \lsp is taken to be ${\mchargino-\mlsp = 5\GeV}$, as in the T2tb model.
The \charginopm subsequently decays to \lsp and an off-shell \PW boson.
Search bins with multiple bottom quark, top quark, and \PW boson candidates enhance the sensitivity to the final states expected from these models.

In the ``T5ttcc'' model, pair-produced gluinos each decay to a top quark and an on-shell top squark, and subsequently
the top squark decays to a charm quark and the LSP.
For this model, ${\dm = 20\GeV}$ is assumed, so the decay of the top squark to a top quark and the LSP is kinematically forbidden.
In such cases, the top squark decay ${\stopq\to\PQc\lsp}$ is expected to be one of the dominant decay modes as discussed above.
The value of \dm has little effect on the final results for the T5ttcc model
when it remains below~\mW.
The T5ttcc model provides
sensitivity to scenarios in which the top squark is
kinematically unable to decay to an on-shell top quark
and can cover the scenarios of high top squark masses beyond the reach of T2cc through the cascade decays of gluinos.

\section{The CMS detector}
\label{sec:CMSdetector}

The CMS detector is a general-purpose particle detector surrounding the luminous region where protons from the LHC beams interact.
A 3.8\unit{T} magnetic field is produced by a solenoid of 6\unit{m} internal diameter, within which are a silicon pixel and silicon strip tracking detector, a lead tungstate crystal electromagnetic calorimeter (ECAL), and a brass and scintillator hadronic calorimeter (HCAL).
Each of these parts of the detector is composed of a cylindrical barrel section and two endcap sections.
The pseudorapidity $\eta$ coverage of the barrel and endcap detectors is extended by forward calorimeters which lie very close to the LHC beam line.
Outside the solenoid, returning magnetic flux is guided through a steel return yoke.
Gas-ionization detectors are sandwiched in between the layers of the return yoke and are used to detect muons.
The events used in the search were collected in 2016--2018 using a two-tier trigger system: a hardware-based level-1 trigger and a software-based high-level trigger~\cite{CMS_trigger}.
The integrated luminosities recorded in 2016, 2017, and 2018 are measured with uncertainties in the range of 2.3--2.5\%~\cite{CMS-PAS-LUM-17-001,CMS-PAS-LUM-17-004,CMS-PAS-LUM-18-002}.
The uncertainties in these measurements are mostly uncorrelated from year to year, resulting in a smaller uncertainty, 1.8\%, in the total integrated luminosity, 137\fbinv.
The CMS detector is described in more detail, along with the coordinate system and basic kinematic variables, in Ref.~\cite{CMS}.

\section{Simulated event samples}
\label{sec:simulation}

Samples of simulated events produced via the Monte Carlo (MC) method are used to optimize selection criteria, estimate signal acceptances, and develop background estimation techniques.

Simulated signal events are generated using {\MGvATNLO}~\cite{Alwall:2014hca} (versions 2.2.2 and 2.4.2)
with leading order (LO) predictions
including up to two additional partons in the matrix element calculations.
Version 2.2.2 of {\MGvATNLO} is used for 2016 simulation while version 2.4.2 is used for 2017 and 2018 simulation.
The production cross sections are
determined with approximate next-to-next-to-leading order (NNLO) plus next-to-next-to-leading logarithmic (NNLL)
accuracy~\cite{Beenakker:2016lwe,
bib-nlo-nll-01,bib-nlo-nll-02,bib-nlo-nll-03,bib-nlo-nll-04,
Beenakker:2011sf,Beenakker:2013mva,Beenakker:2014sma,
Beenakker:1997ut,Beenakker:2010nq,Beenakker:2016gmf}.

Events arising from SM processes are simulated using a number of MC event generators.
Samples of \ttbar events, \wjets events, \zjets events with \znn, $\PZ/\HepParticle{\PGg}{}{\ast} (\to\Pell^+\Pell^-) + \text{jets}$ events (\dyjets), \gjets events,
and quantum chromodynamics (QCD) multijet events containing solely jets produced through the strong interaction
are simulated using the {\MGvATNLO} event generator at LO (versions 2.2.2 and 2.4.2).
The \ttbar events are generated with
up to three additional partons in the matrix element calculations, while
the \wjets, \zjets, \dyjets, and \gjets events are generated
with up to four additional partons.
Events containing a single top quark produced through the $s$ channel, events containing a \ttbar pair produced in association with a \PZ boson, a \PW boson, or a photon, and rare events such as those containing multiple electroweak or Higgs bosons (\PW, \PZ, \PGg, and \PH)
are generated with {\MGvATNLO} (versions 2.2.2 and 2.4.2)
at next-to-leading order (NLO)~\cite{Frederix:2012ps}.
The {\POWHEG}\,v1.0~(v2.0)~\cite{Nason:2004rx,Frixione:2007vw,Alioli:2010xd,Alioli:2009je,Re:2010bp,Melia:2011tj,Nason:2013ydw,Hartanto:2015uka} program
is used to simulate events at NLO from 2016 (2017 and 2018) containing a single top quark produced through the $t$ and $\PQt\PW$ channels, as well as \WW and \ttH events.
Events containing \ZZ are generated at NLO with either {\POWHEG} or {\MGvATNLO} depending on the decay mode,
and \WZ production is simulated with {\PYTHIA}\,8.226~(8.230)~\cite{Sjostrand:2014zea} for 2016 (2017 and 2018) at LO.
Normalization of the simulated background samples is performed using
the most accurate cross section calculations
available~\cite{
Alioli:2009je,
Re:2010bp,Alwall:2014hca,
Gavin:2012sy,Gavin:2010az,Li:2012wna,
Gehrmann:2014fva,Campbell:2011bn,
Beneke:2011mq, Cacciari:2011hy, Baernreuther:2012ws,Czakon:2012zr,Czakon:2012pz,Czakon:2013goa,Czakon:2011xx},
which typically correspond to NLO or NNLO accuracy.

All simulated samples make use of the
{\PYTHIA}\,8.226~(8.230) program for 2016 (2017 and 2018)
to describe parton showering and hadronization.
Samples that are simulated at NLO with {\MGvATNLO} adopt the FxFx~\cite{Frederix:2012ps} scheme for matching partons from the matrix-element calculation to those from parton showers.
Samples simulated at LO adopt the MLM~\cite{Mangano:2006rw} scheme for the same purpose.
The CUETP8M1~\cite{Khachatryan:2015pea} {\PYTHIA}\,8.226 tune is used to produce the SM background and signal samples for the analysis of the 2016 data.
For the analysis of the 2017 and 2018 data,
the CP5 and CP2~\cite{Sirunyan:2019dfx} tunes are used for the SM background samples and signal samples, respectively.
Simulated samples generated at LO or NLO with the CUETP8M1 tune use the
\textsc{nnpdf2.3lo} or \textsc{nnpdf2.3nlo}~\cite{Ball:2013hta}
parton distribution functions (PDFs), respectively.
The samples using the CP2 or CP5 tune use the \textsc{nnpdf3.1lo} or \textsc{nnpdf3.1nnlo}~\cite{Ball:2017nwa} PDFs, respectively.

Simulated SM events are processed through a \GEANTfour-based~\cite{Agostinelli:2002hh} simulation of the CMS detector.
In order to keep the computational processing time manageable,
simulated signal events are processed through the CMS fast simulation program~\cite{Abdullin:2011zz,Giammanco:2014bza},
which yields results that are generally
consistent with the {\GEANTfour}-based simulation.
The simulated events are generated with nominal distributions
of additional $\Pp\Pp$ interactions per bunch crossing, referred to as pileup.
They are reweighted to match the corresponding pileup distribution measured in data.

In order to improve the modeling of additional jet multiplicities originating from radiation in events containing \ttbar,
the \MGvATNLO prediction is compared to data in a \ttbar{}-enriched data set.
The events in this data set are required to contain two reconstructed charged leptons ($\Pe\Pe$, $\PGm\PGm$, or $\Pe\PGm$)
and two jets that are identified as originating from a bottom quark.
A correction factor is derived from this comparison.
This correction factor is applied to the simulated \ttbar events for 2016, which use the CUETP8M1 tune, and to all the simulated signal events, which use the CUETP8M1 and CP2 tunes.
The correction factor is not applied to the simulated \ttbar events for 2017 and 2018, which use the CP5 tune,
because these simulated event samples already show a reasonable agreement with the data before the correction.
In addition, simulated \ttbar events for all three years are corrected for the observed mismatch
in the top quark \pt spectrum between data and simulation according to the results presented in Ref.~\cite{Khachatryan:2016mnb}.

\section{Event reconstruction}
\label{sec:reconstruction}

Events are reconstructed using the particle-flow (PF) algorithm~\cite{PF}, which uses information from all of the subdetectors to reconstruct candidates (PF candidates) of charged hadrons, neutral hadrons, photons, electrons, and muons.
Combinations of these PF candidates are used to reconstruct higher-level objects such as the missing transverse momentum (\ptvecmiss).
The \ptvecmiss is defined as the negative vector sum of the transverse momentum \ptvec of all PF candidates in the event, and its magnitude is denoted as \ptmiss~\cite{CMS_MET}.

We use only events with at least one reconstructed vertex.
The primary $\Pp\Pp$ interaction vertex (PV) is taken to be the one with the largest value of the summed $\pt^2$, summing over jets and the associated \ptmiss.
Jets are reconstructed from tracks assigned to the vertex using the anti-\kt jet finding algorithm~\cite{Cacciari:2008gp,Cacciari:2011ma} and the associated \ptmiss used for the PV identification is defined based on these track jets.

The primary set of jets used to define the data set for this search is reconstructed by clustering charged and neutral PF candidates using the anti-\kt algorithm~\cite{Cacciari:2008gp,Cacciari:2011ma} with a distance parameter of 0.4 (AK4 jets).
Only those charged PF candidates identified as originating from the PV are considered; any charged PF candidates originating from pileup vertices are ignored.
Jet quality criteria~\cite{CMS-PU-mitigation} are imposed to eliminate jets from spurious sources such as electronics noise.
The energies of jets are corrected for the presence of particles from pileup interactions~\cite{Cacciari:2007fd} as well as for the response of the detector as a function of \pt and $\eta$~\cite{CMS_JES}.
We count jets (\Nj) with ${\pt>30\GeV}$ and ${\abs{\eta} < 2.4}$.

Because the final states of the signal processes generally include at least one bottom quark, the identification of jets originating from a bottom quark plays an important role in this search.
Bottom quark jets (\PQb jets) are identified by applying a version of the combined secondary vertex algorithm based on a deep neural network~(DeepCSV)~\cite{deepCSV}.
The ``medium working point'' of this algorithm is used, which provides a tagging efficiency for \PQb jets (in the \pt range typical of \PQb quarks from top quark decay) of 68\%~\cite{deepCSV}.
The corresponding misidentification probability for light-flavor jets originating from gluons and up, down, and strange quarks is 1\%,
while that for charm quark jets is 12\%~\cite{deepCSV}.
We count \PQb jets (\Nb) with ${\pt>20\GeV}$ and ${\abs{\eta}<2.4}$.
The minimum \pt threshold for counting \Nb is set to 20 instead of 30\GeV, as used
for \Nj, in order to improve the sensitivity to top squark signal models, particularly those with compressed mass spectra, which yield low-\pt \PQb quarks.
Even without the use of a dedicated charm quark tagger, we find that we have adequate sensitivity to models containing charm quarks in the final state.

A large fraction of signal events from models with compressed mass spectra, \eg, events expected from the T2ttC and T2bWC models, contain \PQb~quarks with \pt below the 20\GeV \PQb jet \pt threshold, which would fail to be reconstructed as \PQb jets.
Identification of these soft \PQb quarks improves our ability to separate potential signal events from the SM background.
We therefore identify soft \PQb quarks based on the presence of a secondary vertex (SV) reconstructed using the inclusive vertex finder algorithm~\cite{IVF}.
Additional requirements on SV observables and the distance between the SV and PV are applied to suppress the background originating from light-flavor jets, as done in the previous top squark search~\cite{SUS-16-049}.
These requirements result in an efficiency of 40--55\% for correctly identifying soft \PQb quarks,
and a misidentification rate of ${\approx}2$--5\% for objects originating from light-flavor hadrons~\cite{SUS-19-009}.
To maintain orthogonality to \PQb tagging, SVs are further required to be separated by ${\DR > 0.4}$ from any jets with ${\pt > 20\GeV}$.
The selected SVs are counted (\Nsv).

As discussed in Section~\ref{sec:signal_models}, signal events with small \dm generally leave little visible energy in the detector, making such events difficult to identify.
They can still be detected if the top squark pair recoils against a high-\pt jet arising from ISR.
The ISR jet gives a transverse boost to the top squark pair and its decay products, including two LSPs, which can result in greater \ptmiss than in comparable events without a high-\pt ISR jet.
Jets clustered using the anti-\kt algorithm with a size parameter of 0.8 (AK8 jets), instead of 0.4 as used for \Nj counting, are used to identify ISR jet candidates as well as boosted high-\pt top quark and \PW boson candidates.
The identification of Lorentz-boosted top quarks and \PW bosons is discussed in detail in Section~\ref{sec:deepAK8}.

Pileup contributions to AK8 jets are statistically subtracted using the ``pileup per particle identification''~\cite{PUPPI,CMS-PU-mitigation} method, by which each charged and neutral particle is weighted by a factor representing its probability to originate from the PV before the clustering is performed.
The use of AK8 jets improves the ISR jet identification by capturing ISR gluons which often radiate additional gluons and result in large size jets.
The AK8 jet with the largest \pt among the AK8 jets with ${\pt>200\GeV}$ in the event is considered an ISR jet candidate.
The ISR jet is required to fail the DeepCSV \PQb-tag identification defined at the ``loose working point''~\cite{deepCSV}, which is characterized by a tagging efficiency of about 80\% and a misidentification rate of about 10\% for light-flavor jets.
The ISR jet may not be tagged as a top or \PW jet as defined in Section~\ref{sec:deepAK8}.

In order to obtain a sample of fully-hadronic events, all events with charged leptons, including electron and muon candidates, hadronically decaying tau lepton candidates, and isolated tracks, are removed from the search data set.
Electron and muon candidates are also used to define control samples of events with one or two isolated leptons, which are used for background estimation as discussed in Sections~\ref{subsec:lostlepton} and~\ref{subsec:Zinvisible}.

Electron candidates are reconstructed starting from clusters of energy deposited in the ECAL that are then matched to a track in the silicon tracker~\cite{CMS_electron}.
Electron candidates are required to have ${\abs{\eta}<2.5}$.
Muon candidates are reconstructed by matching tracks in the muon detectors to compatible track segments in the silicon tracker~\cite{CMS_muon} and are required to be within the muon detector fiducial region of ${\abs{\eta}<2.4}$.
Electron and muon candidates are further required to be isolated.

The isolation criterion for electron and muon candidates is based on the ``mini-isolation'' variable \miniiso, which is the scalar \pt sum of all charged-hadron, neutral-hadron, and photon PF candidates within a cone around the lepton candidate direction, where the radius \DR{} of the cone depends on the lepton candidate \pt.
For ${\pt < 50\GeV}$, ${\DR = 0.2}$; for ${\pt > 200\GeV}$, ${\DR = 0.05}$; and for ${50 < \pt < 200\GeV}$, ${\DR = 10\GeV / \pt}$.
The decrease in cone size with increasing \pt is motivated by the concomitant increase in collimation of the lepton decay products.
It reduces the rate of accidental overlap between the lepton and jets in high multiplicity or highly boosted events, particularly overlap between \PQb jets and leptons originating from a boosted top quark.
The mini-isolation variable is corrected for contributions from pileup using an estimate of the pileup energy inside the cone~\cite{CMS_muon,CMS_electron}.
The isolation requirement is ${\miniiso/\pt < 0.1}$ for electron candidates and ${\miniiso/\pt < 0.2}$ for muon candidates.

Hadronically decaying tau lepton candidates \tauh are reconstructed by the hadron-plus-strips algorithm~\cite{CMS_tau_1,CMS_tau_2}.
The \tauh candidates are required to have ${\pt > 20\GeV}$, ${\abs{\eta} < 2.4}$, and transverse mass ${\mT = \sqrt{\smash[b]{2 \pt \ptmiss \left(1 - \cos \dphi\right)}} < 100\GeV}$, where \dphi is the azimuthal separation between the candidate \ptvec and \ptvecmiss.
The goal of the transverse mass requirement is to suppress the background with $\PW\to\tauh\nu$ decays and no other source of \ptmiss.

Some electrons, muons, and tau leptons that do not satisfy the above criteria are still reconstructed as electron, muon, or charged-hadron PF candidates.
Electron, muon, and charged-hadron PF candidates are identified as electron tracks, muon tracks, and charged-hadron tracks, respectively, provided they satisfy criteria on the PF candidate \pt and $\eta$, the transverse mass \mT, and the isolation,
and are collectively referred to as isolated tracks.
Isolation is defined based on the scalar \pt sum \ptsum of all other charged PF candidates lying within a cone ${\DR < 0.3}$ around the PF candidate.
Isolated electron and muon tracks are required to have ${\pt > 5\GeV}$ and ${\ptsum < 0.2\,\pt}$, while charged-hadron tracks are required to have ${\pt > 10\GeV}$ as well as ${\ptsum < 0.1\,\pt}$.
Isolated tracks are all required to satisfy ${\abs{\eta} < 2.5}$ and ${\mT < 100\GeV}$.

Photon candidates, which are used in the estimation of some backgrounds, are reconstructed from clusters of energy deposited in the ECAL.
They must satisfy requirements on the cluster shape, the relative fraction of energy deposited in the HCAL behind the cluster in the ECAL, and the photon isolation~\cite{CMS_photon}.

Events expected from direct top squark production models with \dm larger than the top quark mass \mtop and from all the gluino-mediated top squark production models considered in this search produce on-shell top quarks and/or on-shell \PW bosons in the decay chain.
Thus, identification of hadronically decaying top quarks and \PW bosons plays a central role in this analysis.
Previous searches~\cite{SUS-16-049,SUS-16-050} already employed top quark and \PW boson tagging algorithms, but this search benefits from the improved tagging algorithms discussed below.
Because the top quarks and \PW bosons may have a wide range of \pt, we employ a combination of two tagging algorithms, which are optimized for different \pt ranges.

\subsection{Merged top quark and \texorpdfstring{\PW}{W} boson tagging algorithm}
\label{sec:deepAK8}

When a top quark or \PW boson is produced with high \pt, its decay products are often merged into a single AK8 jet.
Top quark and \PW boson candidates are selected from AK8 jets based on the jet \pt and soft-drop mass.
The soft-drop mass is a groomed jet mass calculated using the soft-drop algorithm~\cite{Larkoski:2014wba,Dasgupta:2013ihk} with an angular exponent ${\beta = 0}$ and soft cutoff threshold ${z_\text{cut} < 0.1}$.
The soft-drop algorithm recursively removes soft wide-angle radiation from a jet.
Top quark candidates are defined as those AK8 jets that have ${\pt > 400\GeV}$ and soft-drop mass above 105\GeV, while \PW boson candidates are required to have ${\pt > 200\GeV}$ and soft-drop mass between 65 and 105\GeV.

Final identification of top quark and \PW boson candidates is performed using the ``DeepAK8'' algorithm~\cite{JME-18-002}.
DeepAK8 is a multiclass classifier that identifies hadronically decaying particles as one of five main categories: \PW, \PZ, \PH, \PQt, and ``other''.
These categories are further subdivided into minor categories corresponding to the decay modes of each particle.
DeepAK8 uses a customized deep neural network architecture tailored to the jet classification task, which exploits PF information directly.
The neural network uses information about all of the PF candidates and all of the SVs associated with each AK8 jet.
A detailed description of the algorithm can be found in Ref.~\cite{JME-18-002}.
In the context of this analysis the output classes are combined to achieve ``top quark \vs QCD multijet'' and ``\PW boson \vs QCD multijet'' discrimination.
The other discriminators are not used.

\begin{figure}[thb]
\centering
\includegraphics[width=0.48\textwidth]{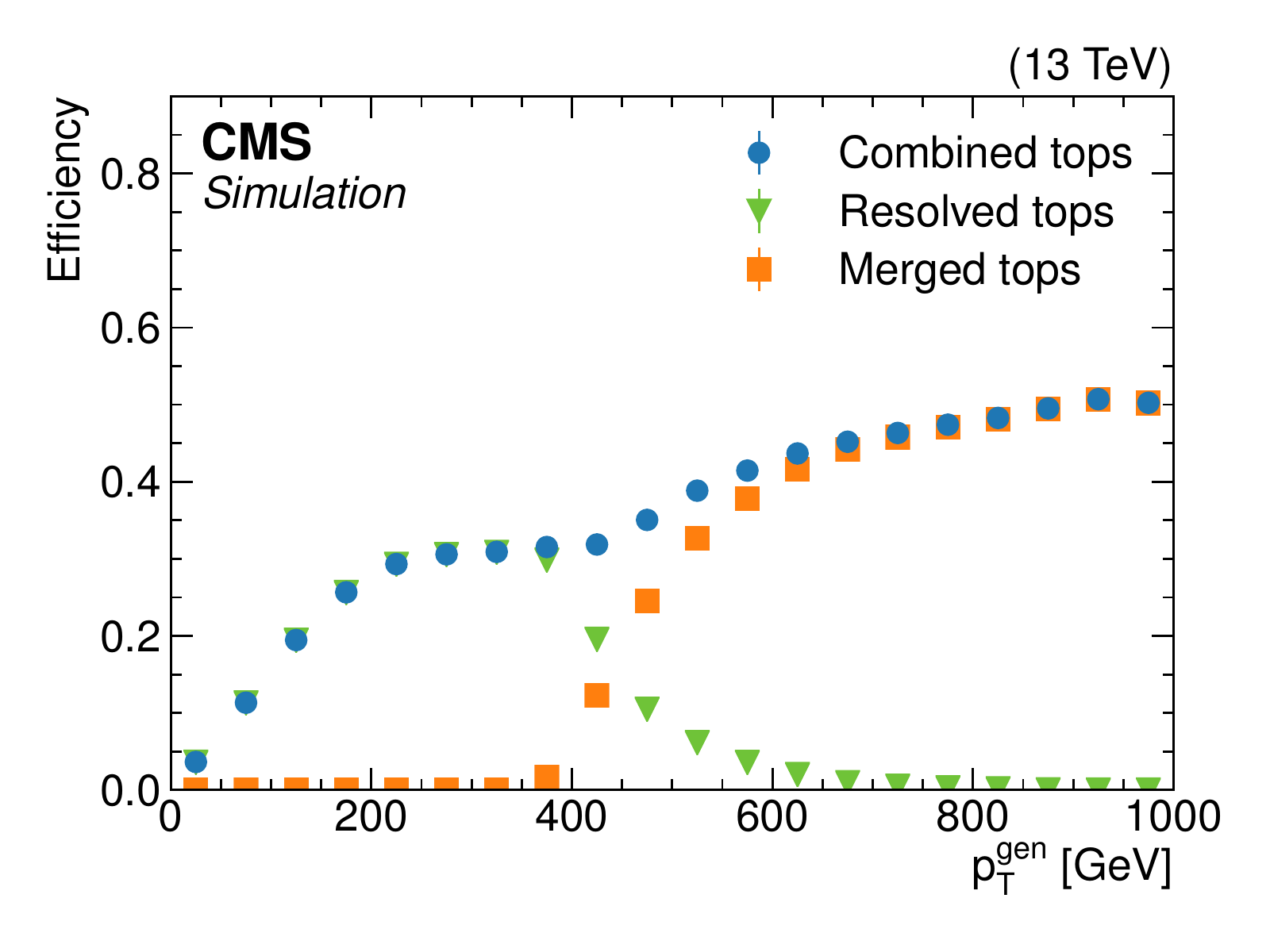}
\includegraphics[width=0.48\textwidth]{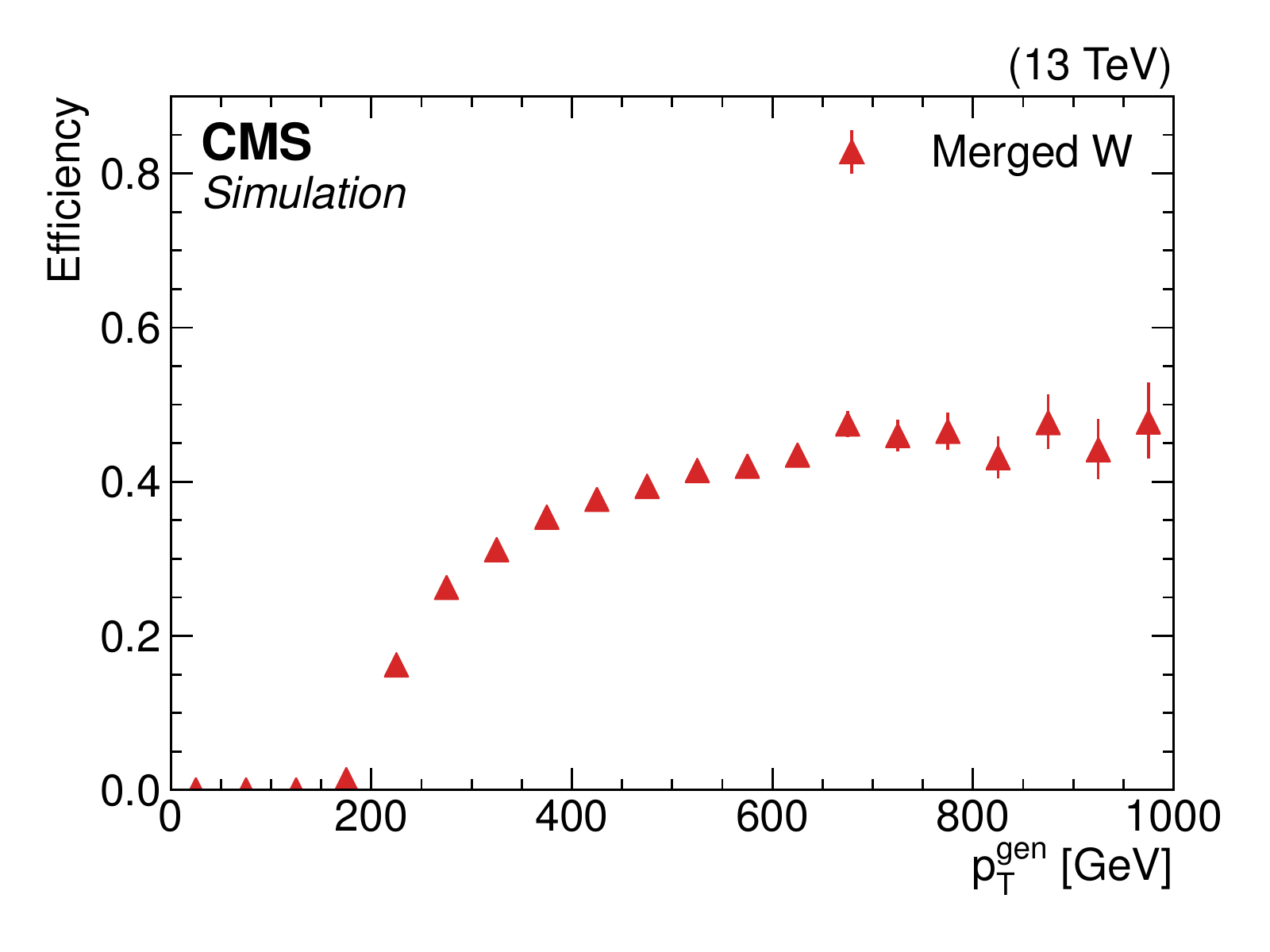}
\caption{
	Top quark and \PW boson tagging efficiencies are shown as a function of the generator-level top quark \pt and the generator-level \PW boson \pt, respectively, for the merged tagging algorithm described in Section~\ref{sec:deepAK8} and the resolved tagging algorithm described in Section~\ref{sec:resolvedtop}.  The \cmsLeft plot shows the efficiencies as calculated in a sample of simulated \ttbar events in which one top quark decays leptonically, while the other decays hadronically.  The \cmsRight plot shows the \PW boson tagging efficiency when calculated in a sample of simulated \WW events.  In addition to the individual algorithms shown as orange squares (boosted top quarks), green inverted triangles (resolved top quarks), and red triangles (boosted \PW bosons), the total top quark tagging efficiency (blue dots) is also shown.
}
\label{fig:topWEfficiency}
\end{figure}

Top quark candidates that satisfy a requirement on the value of the top quark \vs QCD multijet discriminator are considered tagged and are counted (\Nt).
The requirement used in this analysis yields a misidentification rate in QCD multijet events of 0.5\%, and a top quark tagging efficiency shown in Fig.~\ref{fig:topWEfficiency}.
Similarly, \PW boson candidates that satisfy a requirement on the value of the \PW boson \vs QCD multijet discriminator and are not already tagged as top quarks are considered tagged and are counted (\Nw).
The requirement used in this analysis yields the \PW{} boson tagging efficiency shown in Fig.~\ref{fig:topWEfficiency} and an average misidentification rate in QCD multijet events of 1\%.

\subsection{Resolved top quark tagging algorithm}
\label{sec:resolvedtop}

The DeepResolved algorithm~\cite{SUS-19-009} identifies top quarks with small boost, in the \pt range of roughly 100 to 500\GeV, whose decay products are too spread out to be contained inside a single AK8 jet.
Top quark candidates are formed from the combination of three AK4 jets with \pt of at least 40, 30, and 20\GeV, respectively.
The three jets of each top quark candidate must have an invariant mass between 100 and 250\GeV, no more than one of the jets can be identified as a \PQb jet using the DeepCSV medium working point, and the three jets must all lie within a cone of size ${\DR < 3.1}$ around the trijet centroid, the vector sum of the momenta of the three jets.

After this loose preselection, a feed-forward neural network with a single hidden layer is used to distinguish between trijet combinations whose three jets all match to a decay product of a top quark versus those that do not.
More complex neural network architectures did not result in improved discrimination power in our study.
The network uses high-level information such as the invariant mass of the trijet and individual dijet pairs, as well as information from each jet including the relativistic energy-momentum four-vector describing the jet, the DeepCSV heavy-flavor discriminator value, jet shape variables~\cite{CMS-PAS-JME-13-002},
the number of PF candidates associated with the jet, and variables describing the fraction of the jet energy carried by each of several categories of PF candidates.
The network is trained using simulated \ttbar events, simulated QCD multijet events, and events from a collision data set that is dominated by QCD multijet production.

The simulation is used to define which trijets are considered ``signal'' and ``background'' during neural network training.
Signal is defined as any trijet passing the preselection in which each jet is matched to a simulated decay product of a top quark within a cone of size ${\DR < 0.4}$ and the overall trijet system is matched to the simulated top quark within a cone of size ${\DR < 0.6}$.
Background is defined as any trijet combination that is not categorized as signal.
Background includes trijet combinations where some, but not all, of the jets are matched to top quark decay products, as well as trijet combinations in which the three top quark decay products to which the three jets are matched originate from two or more simulated top quarks.
Collision data that are highly enriched in QCD multijet events are included in the training.
These data are included using domain adaptation via gradient reversal~\cite{DomainAdaptation3,CMSDomainAdaptation} to discourage the network from learning features of the simulation that are not present in data.
With this method an additional output is added to the neural network, connected to the hidden layer in the same manner as the primary neural network output.
This additional output is tasked with distinguishing between trijet candidates from QCD multijet simulation and trijet candidates from collision data.
The primary neural network output is trained to minimize the ability to discriminate based on observables that are not well modeled in simulation.
This yields a network that is able to discriminate between signal and background almost as well as a network trained without domain adaptation but with minimal reliance on features that exist only in simulation.

Before the final selection of trijets as tagged top quarks can be made, any overlap between trijet candidates that share jets with another candidate must be resolved.
When considering any pair of overlapping trijets, the trijet that is more background-like according to the neural network is removed from further consideration.
Additionally, trijet candidates that overlap with top quark and \PW boson candidates identified by the DeepAK8 algorithm are removed.
A trijet overlaps a DeepAK8-tagged jet if any of the trijet constituents lies within a cone of size ${\DR < 0.4}$ around one of the subjets (as identified by the soft-drop algorithm~\cite{JME-18-002}) of the AK8 jet.
Any remaining trijets with a neural network output greater than a threshold are considered tagged and are counted (\Nres).
This threshold is chosen to yield a misidentification rate in QCD multijet events of 2\%.

The overall efficiencies of the top quark and \PW boson tagging algorithms are shown in Fig.~\ref{fig:topWEfficiency}.
The efficiency for each object is defined as the fraction of simulated hadronically decaying top quarks or \PW bosons that are identified by the appropriate algorithm.
The simulated top quark or \PW boson is considered to have been identified by the DeepAK8 algorithm if all of its primary decay products lie within a cone of size ${\DR < 0.6}$ around the AK8 jet.
Similarly, a simulated top quark is considered to have been identified by the DeepResolved algorithm if at least two of its three primary decay products lie within a cone of size ${\DR < 0.4}$ around distinct constituents of the trijet and the simulated top quark lies within a cone of size ${\DR < 0.6}$ around the trijet centroid.
The sum of the DeepAK8 and DeepResolved efficiencies for tagging top quarks is also shown in Fig.~\ref{fig:topWEfficiency} and demonstrates the complementary nature of these two algorithms over the full range of relevant \pt.

The top quark and \PW boson taggers exhibit slightly different performance in data and in simulation.
We derive scale factors to correct the performance of the taggers in simulation to match their performance in data.
For both the DeepAK8 and DeepResolved algorithms, the tagging efficiency is estimated in both data and simulation using a dedicated sample of events containing a single charged lepton, selected to be enriched in top quark and \PW boson production.
For the DeepAK8 algorithm, the misidentification rate is estimated in a sample of events containing a single photon, which is selected to be depleted of top quark and \PW boson production.
For the DeepResolved algorithm, the misidentification rate is estimated in a sample of events containing jets but no charged leptons and small \ptmiss.
This sample is similarly depleted of top quark and \PW boson production.

For each category of tagged top quark or \PW boson, data-to-simulation scale factors are defined as the ratio of the performance (either tagging efficiency or misidentification rate) in data to the performance in simulation.
These scale factors are parameterized as a function of the \pt of the tagged top quark or \PW boson candidate and are used to reweight simulated events to more accurately describe the data.
The efficiency scale factors for the DeepResolved algorithm are within 6\% of unity while the misidentification scale factors are within 8\% of unity.
The DeepAK8 efficiency scale factors are within 8\% of unity for the top quark and \PW boson categories while the misidentification scale factors vary up to 20 (30)\% from unity for the top quark (\PW boson) categories.
The DeepAK8 scale factors are discussed in detail in Ref.~\cite{JME-18-002}.
The most important sources of uncertainty in the scale factors arise from the jet energy scale and resolution, parton shower modeling, choice of factorization and renormalization scales, and statistical uncertainties in the data and the simulation.

\section{Event selection and search regions}
\label{sec:selection}

Events used for the search regions in this analysis were collected with a trigger that requires both \ptmiss and \mht larger than a threshold that varied between 100 and 140\GeV
depending on the LHC instantaneous luminosity and data taking period, where \mht is the magnitude of the vector \pt sum of jets reconstructed at the trigger level.
The trigger efficiency is greater than 95\% after application of the event selection criteria described below, including the requirement of ${\ptmiss > 250\GeV}$.

All events are required to pass filters designed to remove detector- and beam-related noise and events that suffered from event reconstruction failures~\cite{CMS_MET}.
The data set used in this analysis is defined broadly by the exclusive presence of multiple jets of strongly interacting particles along with large \ptmiss.
Large \ptmiss in SM events generally arises from leptonic decays of \PW bosons, \PZ boson decays to neutrino-antineutrino pairs, or jet energy mis\-meas\-ure\-ments in QCD multijet events.
Events with isolated electrons or muons, as defined in Section~\ref{sec:reconstruction}, with ${\pt > 5\GeV}$ are removed from the search data set in order to suppress SM backgrounds with large \ptmiss from leptonic \PW boson decays.
This requirement provides a search data set that is orthogonal to the data set used for top squark searches performed using final states with a single lepton~\cite{SUS-19-009} or with two oppositely charged leptons~\cite{SUS-19-011},
which will enable the statistical combination of the results from these other searches.

In order to further suppress events with charged leptons from \PW boson decays, we remove events from the search data set that contain \tauh candidates, isolated electron tracks, isolated muon tracks, or isolated charged-hadron tracks, as defined in Section~\ref{sec:reconstruction}.

Large \ptmiss in QCD multijet events typically arises from undermeasured jet energies, which result in a small azimuthal separation between the undermeasured jet and \ptvecmiss.
This background is suppressed by removing events with small azimuthal separation between a high-\pt jet and \ptvecmiss.

The detailed ``baseline'' requirements that define the search data set are given in Table~\ref{tab:evtsel}.
The data set is further divided into two regions.
The low \dm region is designed to be sensitive to low \dm signal models while the high \dm region is designed to be sensitive to high \dm signal models.

\begin{table*}
    \renewcommand{\arraystretch}{1.05}
        \topcaption{\label{tab:evtsel}
      Summary of the preselection requirements (baseline selection)
      imposed on the reconstructed physics objects for this search, as well as the low \dm and high \dm baseline selections.
      Here $R$ is the distance parameter of the anti-\kt algorithm.
      Electron and muon candidates as well as~\tauh candidates and isolated tracks are as defined in Section~\ref{sec:reconstruction}.
      The $i$-th highest-\pt jet is denoted by~\jet{i}.
    }
    \centering
    \begin{scotch}{ll}
		\multicolumn{2}{l}{Baseline selection} \\ [\cmsTabSkip]

        Jets & $\Nj \geq 2$ ($R = 0.4$), $\pt > 30\GeV$, $\abs{\eta} < 2.4$ \\ [\cmsTabSkip]

        \HT & $\HT > 300\GeV$ \\ [\cmsTabSkip]

        \multirow{4}{*}{\ptmiss} & $\ptmiss > 250\GeV$\\
                               & $\dphi\left(\ptvecmiss, \jet{1}\right) > 0.5$\\
                               & $\dphi\left(\ptvecmiss, \jet{2}\right) > 0.15$\\
                               & $\dphi\left(\ptvecmiss, \jet{3}\right) > 0.15$ (when applicable) \\ [\cmsTabSkip]

        Veto electron & $\pt > 5\GeV$, $\abs{\eta} < 2.5$, $\ptsum < 0.1 \, \pt$  \\
        Veto muon & $\pt > 5\GeV$, $\abs{\eta} < 2.4$, $\ptsum < 0.2 \,\pt$ \\
        Veto \tauh & $\pt > 20\GeV$, $\abs{\eta} < 2.4$, $\mT < 100\GeV$ \\ [\cmsTabSkip]
        \multirow{3}{*}{Veto track}
        & PF charged candidates, $\abs{\eta} < 2.5$, $\mT < 100\GeV$ \\
        & $\pt > 5 \GeV$, $\ptsum < 0.2\, \pt$ for electron and muon tracks \\
        & $\pt > 10 \GeV$, $\ptsum < 0.1\, \pt$ for charged-hadron tracks \\ [2\cmsTabSkip]

		\multicolumn{2}{l}{Low \dm baseline selection} \\ [\cmsTabSkip]

        \Nt, \Nw, \Nres & $\Nt = \Nw = \Nres = 0$ \\ [\cmsTabSkip]
        \mTb & $\mTb < 175\GeV$ (for events with $\Nb \geq 1$) \\ [\cmsTabSkip]
        \multirow{2}{*}{ISR jet} & $\Nj(\text{ISR}) = 1$ ($R = 0.8$), $\ptISR > 200\GeV$, $\abs{\eta} < 2.4$ \\
                                 & $\dphi\left(\ptvecmiss, \jet{\text{ISR}}\right) > 2$ \\ [\cmsTabSkip]
        \ptmiss & $\metsig > 10\,\sqrt{\GeVns}$\\ [2\cmsTabSkip]

		\multicolumn{2}{l}{High \dm baseline selection} \\ [\cmsTabSkip]

        Jets & $\Nj \geq 5$ ($R = 0.4$), $\pt > 30\GeV$, $\abs{\eta} < 2.4$ \\ [\cmsTabSkip]
        \PQb tagging & $\Nb \geq 1$, $\pt > 20\GeV$ \\ [\cmsTabSkip]
        \ptmiss & $\dphi\left(\ptvecmiss, \jet{1,2,3,4}\right) > 0.5$ \\
    \end{scotch}
\end{table*}

\subsection{Low \texorpdfstring{\dm}{dm} search region}
\label{subsec:lowdm_region}

In order to enhance sensitivity to low \dm signal models, as discussed in Section~\ref{sec:signal_models}, the low \dm region requires an ISR jet candidate with ${\pt > 200\GeV}$, ${\abs{\eta} < 2.4}$, and ${\dphi\left(\ptvecmiss, \jet{\text{ISR}}\right) > 2}$.
In the low \dm signal models, the ISR jet recoils against the top squark pair, including the two LSPs from the top squark decay chains, and provides \ptmiss in the targeted signal events.
Because the low \dm signal models involve neither on-shell top quarks nor on-shell \PW bosons, events with tagged top quarks or tagged \PW bosons, as described in Sections~\ref{sec:deepAK8} and~\ref{sec:resolvedtop}, are vetoed in the low \dm region.
We also require ${\metsig > 10\,\sqrt{\GeVns}}$, which suppresses events with a large \ptmiss arising from jet energy mis\-meas\-ure\-ments.

For events from signal models with compressed mass spectra, such as T2ttC and T2bWC or T2tt and T2bW with small \dm, we expect low-\pt bottom quarks in the final state.
For events with ${\Nb \geq 1}$, the minimum transverse mass of all \PQb jets with respect to the \ptvecmiss (\mTb) is required to be less than 175\GeV because events from low \dm signal models tend to lie in this region.
In events with ${\Nb \geq 3}$, only the two jets with the highest DeepCSV discriminator value are considered in the calculation of \mTb.

The dominant source of SM events that satisfy these requirements is \zjets production in which the \PZ boson decays to a neutrino-antineutrino pair.

Events passing the low \dm baseline selection are further required to have ${\ptISR > 300\GeV}$ (while events with ${200 < \ptISR < 300\GeV}$ are still used for the validation of background estimation, as discussed below)
and are further divided into 53 disjoint search bins as shown in Table~\ref{tab:low_dm}.
Eight search bins require ${\Nb = \Nsv = 0}$ and are divided based on \Nj and \ptmiss.
These bins are designed to provide sensitivity primarily to the T2cc model.
The remaining 45 search bins target other low \dm signal models with \PQb quarks in the final state, \ie, the T2ttC and T2bWC models, require ${\Nb \geq 1}$ and/or ${\Nsv \geq 1}$, and are divided based on \Nj, \Nb, \Nsv, \ptISR, \ptb, and \ptmiss.
The variable \ptb is defined as the \pt of the \PQb jet for events with ${\Nb = 1}$ and as the scalar \pt sum of the leading two \PQb jets for events with ${\Nb \geq 2}$.

\begin{table*}
        \topcaption{
        Summary of the 53 search bins that mainly target low \dm signal models.
        For these search bins, events are required to pass the low \dm region selection discussed in Section~\ref{subsec:lowdm_region}.
        Within each row of this table, the edges of the bins as a function of \ptmiss are given, and the bin numbers increase with increasing \ptmiss.
        \NAdescription (\NA) indicates that no requirements are made.
    }
    \label{tab:low_dm}
    \centering
    \begin{scotch}{cccccccc}
        \Nj        & \Nb        & \Nsv       & \mTb [{\GeVns}] & \ptISR [{\GeVns}] & \ptb [{\GeVns}] & \ptmiss [{\GeVns}]             & Bin number \\ [\cmsTabSkip]
        \hline

        2--5       & 0          & 0          & \NA             & ${>} 500$         & \NA             & [450, 550, 650, 750, $\infty$] &  0--3      \\
        ${\geq} 6$ & 0          & 0          & \NA             & ${>} 500$         & \NA             & [450, 550, 650, 750, $\infty$] &  4--7      \\
        2--5       & 0          & ${\geq} 1$ & \NA             & ${>} 500$         & \NA             & [450, 550, 650, 750, $\infty$] &  8--11     \\
        ${\geq} 6$ & 0          & ${\geq} 1$ & \NA             & ${>} 500$         & \NA             & [450, 550, 650, 750, $\infty$] & 12--15     \\ [\cmsTabSkip]

        ${\geq} 2$ & 1          & 0          & ${<} 175$       & 300--500          & 20--40          & [300, 400, 500, 600, $\infty$] & 16--19     \\
        ${\geq} 2$ & 1          & 0          & ${<} 175$       & 300--500          & 40--70          & [300, 400, 500, 600, $\infty$] & 20--23     \\
        ${\geq} 2$ & 1          & 0          & ${<} 175$       & ${>} 500$         & 20--40          & [450, 550, 650, 750, $\infty$] & 24--27     \\
        ${\geq} 2$ & 1          & 0          & ${<} 175$       & ${>} 500$         & 40--70          & [450, 550, 650, 750, $\infty$] & 28--31     \\
        ${\geq} 2$ & 1          & ${\geq} 1$ & ${<} 175$       & ${>} 300$         & 20--40          & [300, 400, 500, $\infty$]      & 32--34     \\ [\cmsTabSkip]

        ${\geq} 2$ & ${\geq} 2$ & \NA        & ${<} 175$       & 300--500          & 40--80          & [300, 400, 500, $\infty$]      & 35--37     \\
        ${\geq} 2$ & ${\geq} 2$ & \NA        & ${<} 175$       & 300--500          & 80--140         & [300, 400, 500, $\infty$]      & 38--40     \\
        ${\geq} 7$ & ${\geq} 2$ & \NA        & ${<} 175$       & 300--500          & ${>} 140$       & [300, 400, 500, $\infty$]      & 41--43     \\
        ${\geq} 2$ & ${\geq} 2$ & \NA        & ${<} 175$       & ${>} 500$         & 40--80          & [450, 550, 650, $\infty$]      & 44--46     \\
        ${\geq} 2$ & ${\geq} 2$ & \NA        & ${<} 175$       & ${>} 500$         & 80--140         & [450, 550, 650, $\infty$]      & 47--49     \\
        ${\geq} 7$ & ${\geq} 2$ & \NA        & ${<} 175$       & ${>} 300$         & ${>} 140$       & [450, 550, 650, $\infty$]      & 50--52     \\
    \end{scotch}
\end{table*}

\begin{table*}
        \topcaption{
        Summary of the 130 search bins that mainly target high \dm signal models.
        For these search bins, events are required to pass the high \dm region selection discussed in Section~\ref{subsec:highdm_region}.
        Within each row of this table, the edges of the bins as a function of \ptmiss are given, and the bin numbers increase with increasing \ptmiss.
    }
    \label{tab:high_dm}
    \centering
    \cmsTable{
        \begin{scotch}{ccccccccc}
            \mTb [{\GeVns}] & \Nj        & \Nb        & \Nt        & \Nw        & \Nres            & \HT [{\GeVns}] & \ptmiss [{\GeVns}]                  & Bin number \\ [\cmsTabSkip]
            \hline

            ${<} 175$       & ${\geq} 7$ & 1          & ${\geq} 0$ & ${\geq} 0$ & ${\geq} 1$       & ${>} 300$      & [250, 300, 400, 500, $\infty$]      & 53--56     \\
            ${<} 175$       & ${\geq} 7$ & ${\geq} 2$ & ${\geq} 0$ & ${\geq} 0$ & ${\geq} 1$       & ${>} 300$      & [250, 300, 400, 500, $\infty$]      & 57--60     \\

            ${>} 175$       & ${\geq} 5$ & 1          & 0          & 0          & 0                & ${>} 1000$     & [250, 350, 450, 550, $\infty$]      & 61--64     \\
            ${>} 175$       & ${\geq} 5$ & ${\geq} 2$ & 0          & 0          & 0                & ${>} 1000$     & [250, 350, 450, 550, $\infty$]      & 65--68     \\

            ${>} 175$       & ${\geq} 5$ & 1          & ${\geq} 1$ & 0          & 0                & 300--1000      & [250, 550, 650, $\infty$]           & 69--71     \\
            ${>} 175$       & ${\geq} 5$ & 1          & ${\geq} 1$ & 0          & 0                & 1000--1500     & [250, 550, 650, $\infty$]           & 72--74     \\
            ${>} 175$       & ${\geq} 5$ & 1          & ${\geq} 1$ & 0          & 0                & ${>} 1500$     & [250, 550, 650, $\infty$]           & 75--77     \\
            ${>} 175$       & ${\geq} 5$ & 1          & 0          & ${\geq} 1$ & 0                & 300--1300      & [250, 350, 450, $\infty$]           & 78--80     \\
            ${>} 175$       & ${\geq} 5$ & 1          & 0          & ${\geq} 1$ & 0                & ${>} 1300$     & [250, 350, 450, $\infty$]           & 81--83     \\
            ${>} 175$       & ${\geq} 5$ & 1          & 0          & 0          & ${\geq} 1$       & 300--1000      & [250, 350, 450, 550, 650, $\infty$] & 84--88     \\
            ${>} 175$       & ${\geq} 5$ & 1          & 0          & 0          & ${\geq} 1$       & 1000--1500     & [250, 350, 450, 550, 650, $\infty$] & 89--93     \\
            ${>} 175$       & ${\geq} 5$ & 1          & 0          & 0          & ${\geq} 1$       & ${>} 1500$     & [250, 350, 450, 550, 650, $\infty$] & 94--98     \\
            ${>} 175$       & ${\geq} 5$ & 1          & ${\geq} 1$ & ${\geq} 1$ & 0                & ${>} 300$      & [250, 550, $\infty$]                & 99--100    \\
            ${>} 175$       & ${\geq} 5$ & 1          & ${\geq} 1$ & 0          & ${\geq} 1$       & ${>} 300$      & [250, 550, $\infty$]                & 101--102   \\
            ${>} 175$       & ${\geq} 5$ & 1          & 0          & ${\geq} 1$ & ${\geq} 1$       & ${>} 300$      & [250, 550, $\infty$]                & 103--104   \\

            ${>} 175$       & ${\geq} 5$ & 2          & 1          & 0          & 0                & 300--1000      & [250, 550, 650, $\infty$]           & 105--107   \\
            ${>} 175$       & ${\geq} 5$ & 2          & 1          & 0          & 0                & 1000--1500     & [250, 550, 650, $\infty$]           & 108--110   \\
            ${>} 175$       & ${\geq} 5$ & 2          & 1          & 0          & 0                & ${>} 1500$     & [250, 550, 650, $\infty$]           & 111--113   \\
            ${>} 175$       & ${\geq} 5$ & 2          & 0          & 1          & 0                & 300--1300      & [250, 350, 450, $\infty$]           & 114--116   \\
            ${>} 175$       & ${\geq} 5$ & 2          & 0          & 1          & 0                & ${>} 1300$     & [250, 350, 450, $\infty$]           & 117--119   \\
            ${>} 175$       & ${\geq} 5$ & 2          & 0          & 0          & 1                & 300--1000      & [250, 350, 450, 550, 650, $\infty$] & 120--124   \\
            ${>} 175$       & ${\geq} 5$ & 2          & 0          & 0          & 1                & 1000--1500     & [250, 350, 450, 550, 650, $\infty$] & 125--129   \\
            ${>} 175$       & ${\geq} 5$ & 2          & 0          & 0          & 1                & ${>} 1500$     & [250, 350, 450, 550, 650, $\infty$] & 130--134   \\
            ${>} 175$       & ${\geq} 5$ & 2          & 1          & 1          & 0                & ${>} 300$      & [250, 550, $\infty]$                & 135--136   \\
            ${>} 175$       & ${\geq} 5$ & 2          & 1          & 0          & 1                & 300--1300      & [250, 350, 450, $\infty$]           & 137--139   \\
            ${>} 175$       & ${\geq} 5$ & 2          & 1          & 0          & 1                & ${>} 1300$     & [250, 350, 450, $\infty$]           & 140--142   \\
            ${>} 175$       & ${\geq} 5$ & 2          & 0          & 1          & 1                & ${>} 300$      & [250, 550, $\infty$]                & 143--144   \\
            ${>} 175$       & ${\geq} 5$ & 2          & 2          & 0          & 0                & ${>} 300$      & [250, 450, $\infty$]                & 145--146   \\
            ${>} 175$       & ${\geq} 5$ & 2          & 0          & 2          & 0                & ${>} 300$      & ${>} 250$                           & 147        \\
            ${>} 175$       & ${\geq} 5$ & 2          & 0          & 0          & 2                & 300--1300      & [250, 450, $\infty$]                & 148--149   \\
            ${>} 175$       & ${\geq} 5$ & 2          & 0          & 0          & 2                & ${>} 1300$     & [250, 450, $\infty$]                & 150--151   \\
            ${>} 175$       & ${\geq} 5$ & 2          & \multicolumn{3}{c}{$\Nt+\Nw+\Nres \geq 3$} & ${>} 300$      & ${>} 250$                           & 152        \\

            ${>} 175$       & ${\geq} 5$ & ${\geq} 3$ & 1          & 0          & 0                & 300--1000      & [250, 350, 550, $\infty$]           & 153--155   \\
            ${>} 175$       & ${\geq} 5$ & ${\geq} 3$ & 1          & 0          & 0                & 1000--1500     & [250, 350, 550, $\infty$]           & 156--158   \\
            ${>} 175$       & ${\geq} 5$ & ${\geq} 3$ & 1          & 0          & 0                & ${>} 1500$     & [250, 350, 550, $\infty$]           & 159--161   \\
            ${>} 175$       & ${\geq} 5$ & ${\geq} 3$ & 0          & 1          & 0                & ${>} 300$      & [250, 350, 550, $\infty$]           & 162--164   \\
            ${>} 175$       & ${\geq} 5$ & ${\geq} 3$ & 0          & 0          & 1                & 300--1000      & [250, 350, 550, $\infty$]           & 165--167   \\
            ${>} 175$       & ${\geq} 5$ & ${\geq} 3$ & 0          & 0          & 1                & 1000--1500     & [250, 350, 550, $\infty$]           & 168--170   \\
            ${>} 175$       & ${\geq} 5$ & ${\geq} 3$ & 0          & 0          & 1                & ${>} 1500$     & [250, 350, 550, $\infty$]           & 171--173   \\
            ${>} 175$       & ${\geq} 5$ & ${\geq} 3$ & 1          & 1          & 0                & ${>} 300$      & ${>} 250$                           & 174        \\
            ${>} 175$       & ${\geq} 5$ & ${\geq} 3$ & 1          & 0          & 1                & ${>} 300$      & [250, 350, $\infty$]                & 175--176   \\
            ${>} 175$       & ${\geq} 5$ & ${\geq} 3$ & 0          & 1          & 1                & ${>} 300$      & ${>} 250$                           & 177        \\
            ${>} 175$       & ${\geq} 5$ & ${\geq} 3$ & 2          & 0          & 0                & ${>} 300$      & ${>} 250$                           & 178        \\
            ${>} 175$       & ${\geq} 5$ & ${\geq} 3$ & 0          & 2          & 0                & ${>} 300$      & ${>} 250$                           & 179        \\
            ${>} 175$       & ${\geq} 5$ & ${\geq} 3$ & 0          & 0          & 2                & ${>} 300$      & [250, 350, $\infty$]                & 180--181   \\
            ${>} 175$       & ${\geq} 5$ & ${\geq} 3$ & \multicolumn{3}{c}{$\Nt+\Nw+\Nres \geq 3$} & ${>} 300$      & ${>} 250$                           & 182        \\ [\cmsTabSkip]
        \end{scotch}
    }
\end{table*}

\subsection{High \texorpdfstring{\dm}{dm} search region}
\label{subsec:highdm_region}

The high \dm search region is optimized for those direct top squark production signal models with ${\dm > \mW}$ and for the gluino pair production models considered in this search,
all of which often produce events with a large number of jets, some of which originate from \PQb quarks.
Therefore, the high \dm region selection requires ${\Nj \geq 5}$ and ${\Nb \geq 1}$.
The additional requirement of ${\dphi\left(\jet{1,2,3,4},\ptvecmiss\right) > 0.5}$ further suppresses QCD multijet events with severe jet energy mis\-meas\-ure\-ments.
Events passing the high \dm region selection are further divided among 130 disjoint search bins, which are described in detail in Table~\ref{tab:high_dm}.

The dominant source of SM background events that satisfy these requirements is \ttbar production in which one of the \PW bosons decays leptonically.
In such events, the \PQb quark from the same top quark decay as the leptonically decaying \PW boson is expected to yield a low \mTb value, typically below the top quark mass.
The high \dm region is divided into two subcategories with ${\mTb < 175\GeV}$ and ${\mTb > 175\GeV}$.

The subcategory with ${\mTb < 175\GeV}$ suffers from large \ttbar background yields but provides sensitivity to signal models with moderate \dm.
Events from these signal models are likely to produce relatively low-\pt top quarks, which leads to large \Nj, so events in this subcategory are required to have ${\Nj \geq 7}$ and ${\Nres \geq 1}$, and are further divided based on \Nb and \ptmiss.

The subcategory with ${\mTb > 175\GeV}$ is further divided into search bins based on \Nb, \Nt, \Nres, \Nw, \ptmiss, and \HT.
These search bins help to provide sensitivity to signal models with a wide range of top squark, gluino, and LSP masses.
The search bins with $\Nb=1$ and 2 primarily provide sensitivity to the direct top squark pair production models T2tt, T2bW, and T2tb, as well as the gluino-mediated top squark production model T5ttcc.
The sensitivity to the T1tttt and T1ttbb models is driven primarily by search bins with $\Nb \geq 3$, and additional requirements on the top quark candidate multiplicity enhance the sensitivity further for T1tttt in particular.
The requirement of one or more merged top quark candidates enhances the sensitivity to signal events from models with large \mstop or \mgluino and low \mlsp, in which the top quarks from top squark or gluino decays have a high transverse boost,
while the requirement of one or more resolved top quark candidates plays a more important role for signal events from models with higher \mlsp, in which top quarks are expected to be less boosted.
The requirement of one or more \PW boson candidates enhances the sensitivity to the T2bW model.

\subsection{Validation regions}
\label{subsec:validation_regions}
In addition to the search region, two validation regions are defined and used to validate the background estimation methods.
These validation regions are kinematically similar to but disjoint from the search region, and are depleted in expected signal relative to the search region.
This allows the estimated background yields to be compared to data, as is done in Section~\ref{subsec:validation}, while maintaining a blind search.

The low \dm validation region is divided into 19 validation bins, analogous to the search bins, as shown in Table~\ref{tab:low_dm_validation}.
Bins 0--14 have identical baseline requirements to the low \dm search region, but require lower \ptmiss than any of the low \dm search bins.
Bins 15--18 are used to validate the background model in a higher \ptmiss range, and are made disjoint from the low \dm search region by altering the requirement on the alignment between jets and \ptvecmiss.
The low \dm search region requires ${\dphi\left(\ptvecmiss, \jet{1}\right) > 0.5}$, while these low \dm validation bins instead require ${0.15 < \dphi\left(\ptvecmiss, \jet{1}\right) < 0.5}$ (medium \dphi).

The high \dm validation region is divided into 24 validation bins as shown in Table~\ref{tab:high_dm_validation}.
These bins have identical requirements to the high \dm search region except for the requirement on the alignment between jets and \ptvecmiss.
The high \dm search region requires ${\dphi\left(\ptvecmiss, \jet{1,2,3,4}\right) > 0.5}$,
while the high \dm validation bins instead require ${\dphi\left(\ptvecmiss, \jet{1}\right) > 0.5}$, ${\dphi\left(\ptvecmiss, \jet{2,3}\right) > 0.15}$, and at least one of ${\dphi\left(\ptvecmiss, \jet{2,3,4}\right) < 0.5}$.

\begin{table*}
        \topcaption{Summary of the 19 validation bins for low \dm.
        Bins 0 to 14 use the normal low \dm region selection with lower \ptmiss requirements than any of the low \dm search bins.
        Bins 15--18 use a similar selection, but additionally require medium \dphi, as discussed in Section~\ref{subsec:validation_regions}.
        \NAdescription (\NA) indicates that no requirements are made.
    }
    \label{tab:low_dm_validation}
    \centering
    \begin{scotch}{cccccccc}
        \dphi        & \Nb        & \Nsv       & \ptISR [{\GeVns}] & \ptb [{\GeVns}] & \Nj        & \ptmiss [{\GeVns}] & Bin number \\ [\cmsTabSkip]
        \hline
        \NA          & 0          & 0          & ${>} 500$         & \NA             & 2--5       & 250--400           & 0          \\
        \NA          & 0          & 0          & ${>} 500$         & \NA             & ${\geq} 6$ & 250--400           & 1          \\
        \NA          & 0          & ${\geq} 1$ & ${>} 500$         & \NA             & 2--5       & 250--400           & 2          \\
        \NA          & 0          & ${\geq} 1$ & ${>} 500$         & \NA             & ${\geq} 6$ & 250--400           & 3          \\
        \NA          & 1          & 0          & 300--500          & ${<} 40$        & ${\geq} 2$ & 250--300           & 4          \\
        \NA          & 1          & 0          & 300--500          & 40--70          & ${\geq} 2$ & 250--300           & 5          \\
        \NA          & 1          & 0          & ${>} 500$         & ${<} 40$        & ${\geq} 2$ & 250--400           & 6          \\
        \NA          & 1          & 0          & ${>} 500$         & 40--70          & ${\geq} 2$ & 250--400           & 7          \\
        \NA          & 1          & ${\geq} 1$ & \NA               & ${<} 40$        & ${\geq} 2$ & 250--300           & 8          \\
        \NA          & ${\geq} 2$ & \NA        & 300--500          & ${<} 80$        & ${\geq} 2$ & 250--300           & 9          \\
        \NA          & ${\geq} 2$ & \NA        & 300--500          & 80--140         & ${\geq} 2$ & 250--300           & 10         \\
        \NA          & ${\geq} 2$ & \NA        & 300--500          & ${>} 140$       & ${\geq} 7$ & 250--300           & 11         \\
        \NA          & ${\geq} 2$ & \NA        & ${>} 500$         & ${<} 80$        & ${\geq} 2$ & 250--400           & 12         \\
        \NA          & ${\geq} 2$ & \NA        & ${>} 500$         & 80--140         & ${\geq} 2$ & 250--400           & 13         \\
        \NA          & ${\geq} 2$ & \NA        & ${>} 500$         & ${>} 140$       & ${\geq} 7$ & 250--400           & 14         \\ [\cmsTabSkip]

        medium \dphi & 0          & 0          & ${>} 200$         & \NA             & ${\geq} 2$ & ${>} 250$          & 15         \\
        medium \dphi & 0          & ${\geq} 1$ & ${>} 200$         & \NA             & ${\geq} 2$ & ${>} 250$          & 16         \\
		medium \dphi & ${\geq} 1$ & 0          & ${>} 200$         & ${>} 20$        & ${\geq} 2$ & ${>} 250$          & 17         \\
        medium \dphi & ${\geq} 1$ & ${\geq} 1$ & ${>} 200$         & ${>} 20$        & ${\geq} 2$ & ${>} 250$          & 18         \\
    \end{scotch}
\end{table*}

\begin{table*}
        \topcaption{Summary of the 24 validation bins for high \dm.
        These search bins are orthogonal to the high \dm search region because of the \dphi requirements discussed in Section~\ref{subsec:validation_regions}.
    }
    \label{tab:high_dm_validation}
    \centering
    \begin{scotch}{cccccccc}
        \mTb [{\GeVns}] & \Nb        & \Nj        & \Nt        & \Nw        & \Nres            & \ptmiss [{\GeVns}] & Bin number \\ [\cmsTabSkip]
        \hline

        ${<} 175$       & 1          & ${\geq} 7$ & ${\geq} 0$ & ${\geq} 0$ & ${\geq} 1$       & 250--400           & 19         \\
        ${<} 175$       & 1          & ${\geq} 7$ & ${\geq} 0$ & ${\geq} 0$ & ${\geq} 1$       & ${>} 400$          & 20         \\
        ${<} 175$       & ${\geq} 2$ & ${\geq} 7$ & ${\geq} 0$ & ${\geq} 0$ & ${\geq} 1$       & 250--400           & 21         \\
        ${<} 175$       & ${\geq} 2$ & ${\geq} 7$ & ${\geq} 0$ & ${\geq} 0$ & ${\geq} 1$       & ${>} 400$          & 22         \\ [\cmsTabSkip]

        ${>} 175$       & 1          & ${\geq} 5$ & 0          & 0          & 0                & 250--400           & 23         \\
        ${>} 175$       & 1          & ${\geq} 5$ & 0          & 0          & 0                & ${>} 400$          & 24         \\
        ${>} 175$       & ${\geq} 2$ & ${\geq} 5$ & 0          & 0          & 0                & 250--400           & 25         \\
        ${>} 175$       & ${\geq} 2$ & ${\geq} 5$ & 0          & 0          & 0                & ${>} 400$          & 26         \\ [\cmsTabSkip]

        ${>} 175$       & 1          & ${\geq} 5$ & 1          & 0          & 0                & 250--400           & 27         \\
        ${>} 175$       & 1          & ${\geq} 5$ & 1          & 0          & 0                & ${>} 400$          & 28         \\
        ${>} 175$       & 1          & ${\geq} 5$ & 0          & 1          & 0                & 250--400           & 29         \\
        ${>} 175$       & 1          & ${\geq} 5$ & 0          & 1          & 0                & ${>} 400$          & 30         \\
        ${>} 175$       & 1          & ${\geq} 5$ & 0          & 0          & 1                & 250--400           & 31         \\
        ${>} 175$       & 1          & ${\geq} 5$ & 0          & 0          & 1                & ${>} 400$          & 32         \\
        ${>} 175$       & 1          & ${\geq} 5$ & \multicolumn{3}{c}{$\Nt+\Nw+\Nres \geq 2$} & 250--400           & 33         \\
        ${>} 175$       & 1          & ${\geq} 5$ & \multicolumn{3}{c}{$\Nt+\Nw+\Nres \geq 2$} & ${>} 400$          & 34         \\ [\cmsTabSkip]

        ${>} 175$       & ${\geq} 2$ & ${\geq} 5$ & 1          & 0          & 0                & 250--400           & 35         \\
        ${>} 175$       & ${\geq} 2$ & ${\geq} 5$ & 1          & 0          & 0                & ${>} 400$          & 36         \\
        ${>} 175$       & ${\geq} 2$ & ${\geq} 5$ & 0          & 1          & 0                & 250--400           & 37         \\
        ${>} 175$       & ${\geq} 2$ & ${\geq} 5$ & 0          & 1          & 0                & ${>} 400$          & 38         \\
        ${>} 175$       & ${\geq} 2$ & ${\geq} 5$ & 0          & 0          & 1                & 250--400           & 39         \\
        ${>} 175$       & ${\geq} 2$ & ${\geq} 5$ & 0          & 0          & 1                & ${>} 400$          & 40         \\
        ${>} 175$       & ${\geq} 2$ & ${\geq} 5$ & \multicolumn{3}{c}{$\Nt+\Nw+\Nres \geq 2$} & 250--400           & 41         \\
        ${>} 175$       & ${\geq} 2$ & ${\geq} 5$ & \multicolumn{3}{c}{$\Nt+\Nw+\Nres \geq 2$} & ${>} 400$          & 42         \\
    \end{scotch}
\end{table*}

\section{Background estimation}
\label{sec:background}

The data set is expected to be dominated by events that do not contain top squarks (backgrounds), which arise from SM processes.
The contributions of the major backgrounds are estimated through measurements in dedicated ``control regions''.
This approach produces background estimates that are more precise and less affected by potential mismodeling than estimates taken purely from simulation.
The control regions are each disjoint from the search region and from each other, are enriched in background events from a particular source, and are expected to be depleted of signal events.
With the aid of simulation, the observations in these control regions are translated to predictions in the search region.
This strategy makes use of methods described in previous searches in similar final states~\cite{Sirunyan:2016jpr,SUS-16-049,SUS-16-050}.

Events with large \ptmiss and a charged lepton mainly arise from \ttbar production, electroweak production of a single top quark, and production of a \PW boson with additional jets in the final state.
These events enter the search region when the charged lepton is not identified in the detector.
The estimation of this background is described in Section~\ref{subsec:lostlepton}.

Events containing a \PZ boson that decays to a neutrino-antineutrino pair contain large \ptmiss.
These events enter the search region when jets are produced along with the \PZ boson.
This background is described in Section~\ref{subsec:Zinvisible}.

QCD multijet events have nearly zero \ptmiss; however, mismeasurement of the \pt of one or more of the jets in the final state can result in large \ptmiss.
These mismeasured events enter the search region and constitute another important source of background, which is described in Section~\ref{subsec:QCD}.

Lastly, a variety of rare processes contribute to the search region, including production of multiple electroweak bosons
and production of a top quark-antiquark pair in association with one or more electroweak or Higgs bosons.
The estimation of these backgrounds is described in Section~\ref{subsec:rare}.

\subsection{Background from \texorpdfstring{\ttbar}{ttbar}, single top quark, and \texorpdfstring{\wjets}{W+jets} events}
\label{subsec:lostlepton}

The background from events containing at least one top quark, top antiquark, or \PW boson, along with additional jets in the final state is dominated by events in which the \PW boson (either prompt or from the decay of a top quark) decays to a charged lepton and a neutrino.
The event selection for this search requires the number of reconstructed charged leptons and isolated tracks (which can be the result of a partially reconstructed charged lepton) to be zero, which substantially reduces this background.
These events still pass the event selection when the charged lepton lies outside the lepton acceptance or is not reconstructed, or is not isolated, and thus is not counted even as an isolated track.
Therefore, this source of SM background is referred to as the lost lepton (LL) background.

This background is estimated from a \ljets control region with ${\Pell = \Pe}$ or \PGm,
selected with the same high \dm and low \dm baseline selection criteria as discussed above,
except that we require exactly one rather than zero isolated leptons and we do not remove events containing isolated tracks.
The \mT of the lepton is required to be less than 100\GeV in order to select events containing a ${\PW\to\Pell\PGn}$ decay and to suppress possible SUSY signal contamination,
\ie, signal events that satisfy the requirements of the \ljets control region.

The LL background yield in each search bin \NP{\text{LL}} is estimated based on the event count in data in a corresponding bin in the \ljets control region \ND{1\Pell}.
This count is extrapolated to the search region to obtain a prediction by means of a transfer factor \TF{LL} obtained from simulation:
\begin{linenomath*}
\begin{equation*}
\label{eqn:LLTF}
\NP{\text{LL}} = \TF{LL} \ND{1\Pell}, \quad \TF{LL}=\frac{\NMC{0\Pell}}{\NMC{1\Pell}},
\end{equation*}
\end{linenomath*}
where \NMC{1\Pell} is the yield expected from simulation in the corresponding control region bin and \NMC{0\Pell} is the yield expected from simulation in the search bin.
These event yields include contributions from \ttbar, \wjets, and single top quark production, as well as smaller contributions from events with two or three electroweak gauge bosons,
denoted by ``multiboson'', and from events with a \ttbar pair produced in association with a \PGg, a \PH boson, a \PW boson, or a \PZ boson, denoted by $\ttbar\PX$.
A unique \TF{LL} is defined for each of the 183 search bins.
The simulated events used to estimate these yields include at least one simulated charged lepton.
The definition of the corresponding bin in the control region is identical to the definition of the search bin except for the requirements on \Nt, \Nw, and \Nres.

For the control region bins corresponding to the high \dm region, no requirement on \Nt, \Nw, or \Nres is made.
This improves the statistical uncertainty in the background estimation in the high \dm region.
Thus, in the high \dm region, the transfer factor \TF{LL} is
\begin{linenomath*}
\begin{equation*}\label{TFExtrapolation}
	\TF{LL} = \frac{\NMC{0\Pell}(\Nb,\ptmiss,\HT,\Nt,\Nres,\Nw)}{\NMC{1\Pell}(\Nb,\ptmiss,\HT)}.
\end{equation*}
\end{linenomath*}
The event yields expected from simulation include the application of data-to-simulation scale factors for the efficiency of the DeepAK8 and DeepResolved top quark and \PW boson taggers.

\begin{figure*}[!h]
	\centering
	\includegraphics[width=0.48\textwidth]{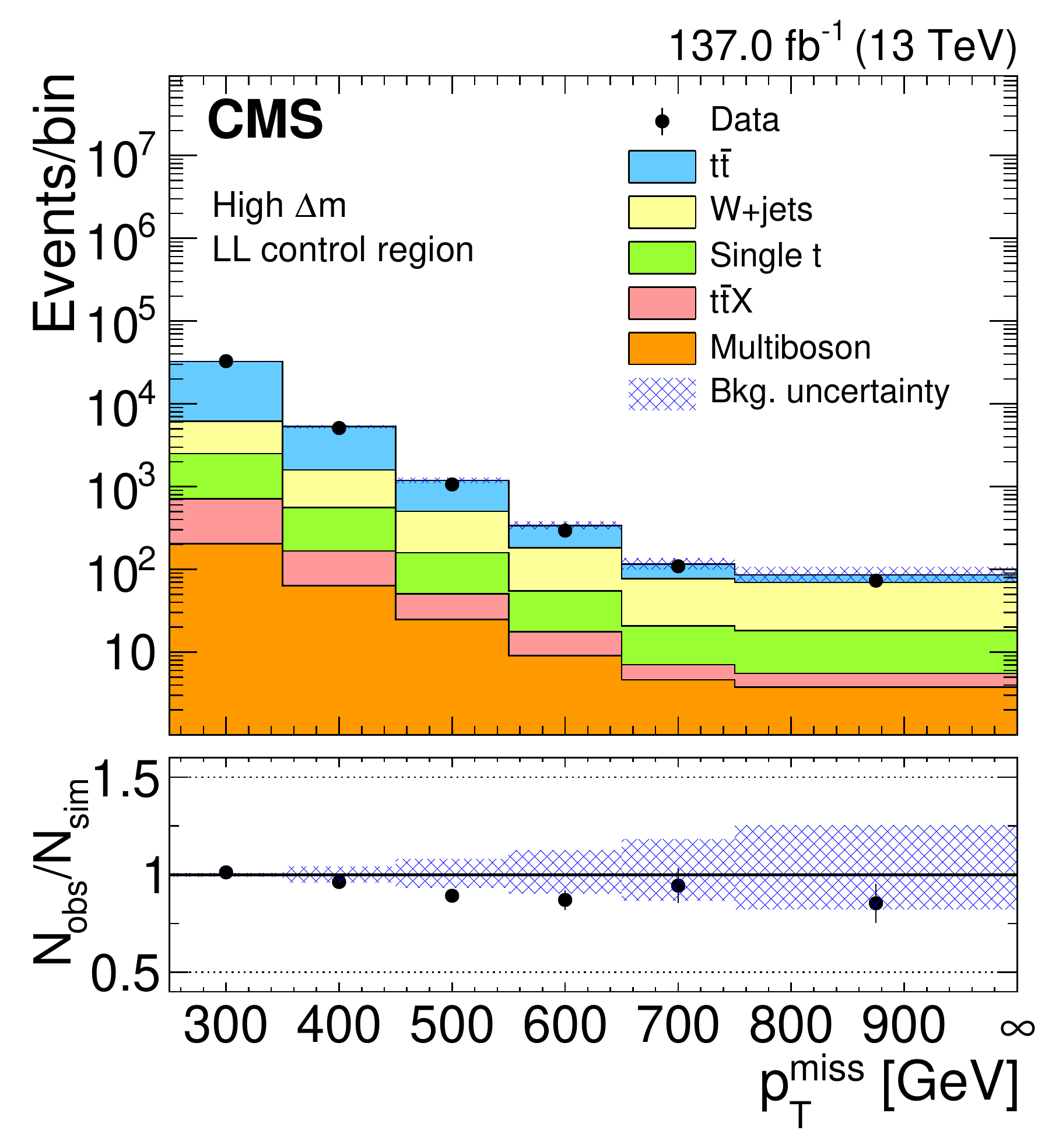}
	\includegraphics[width=0.48\textwidth]{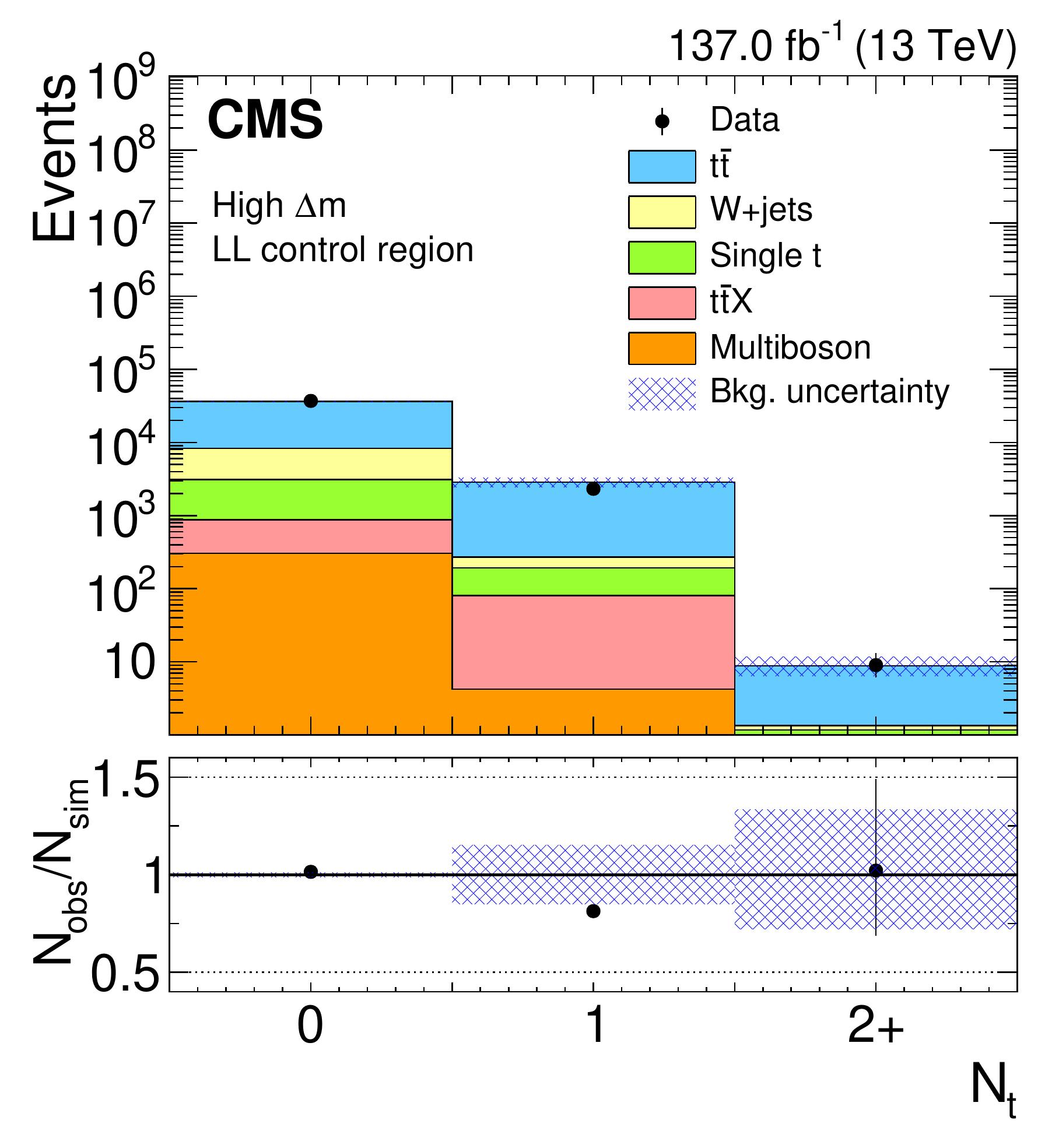} \\
	\includegraphics[width=0.48\textwidth]{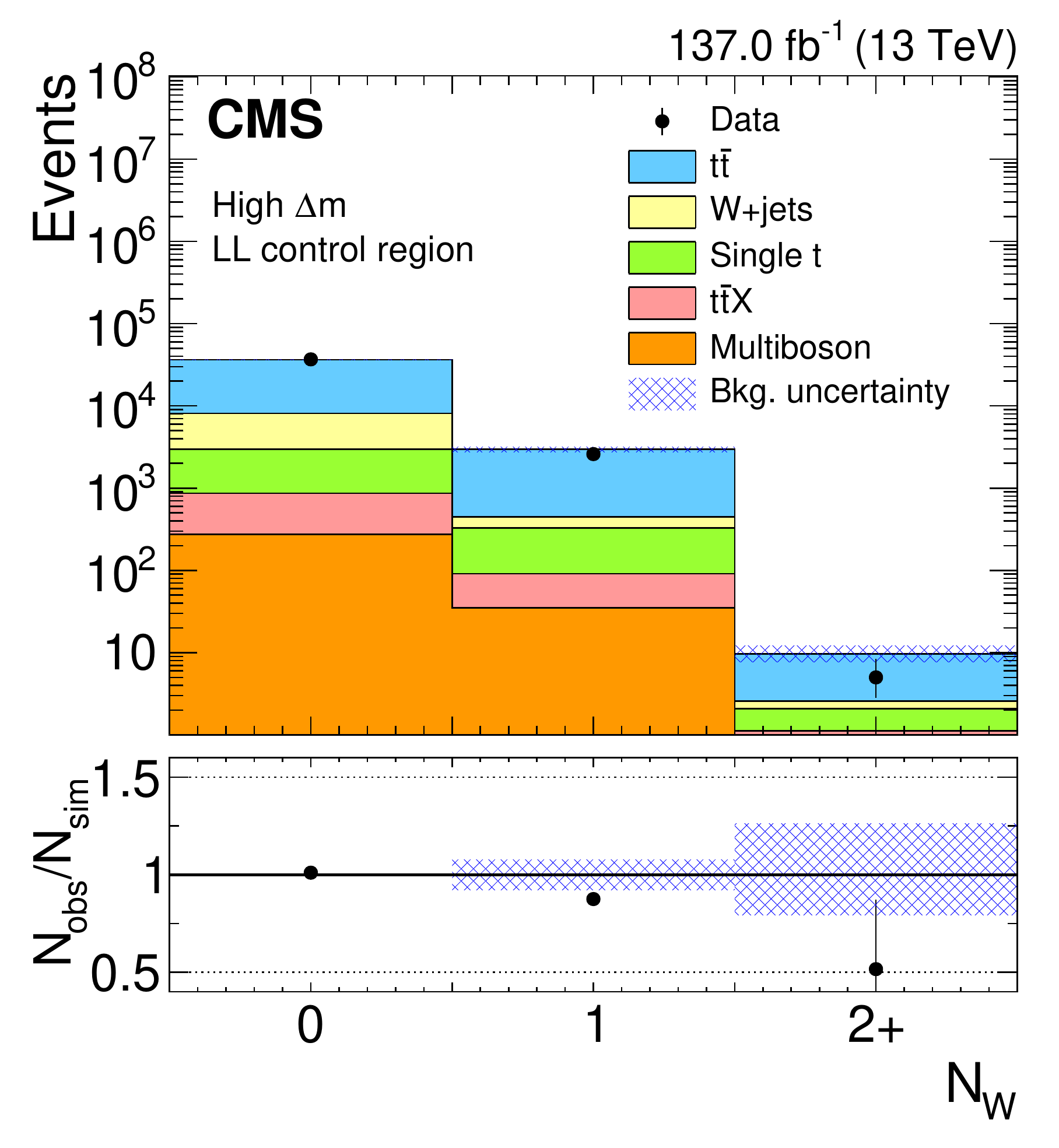}
	\includegraphics[width=0.48\textwidth]{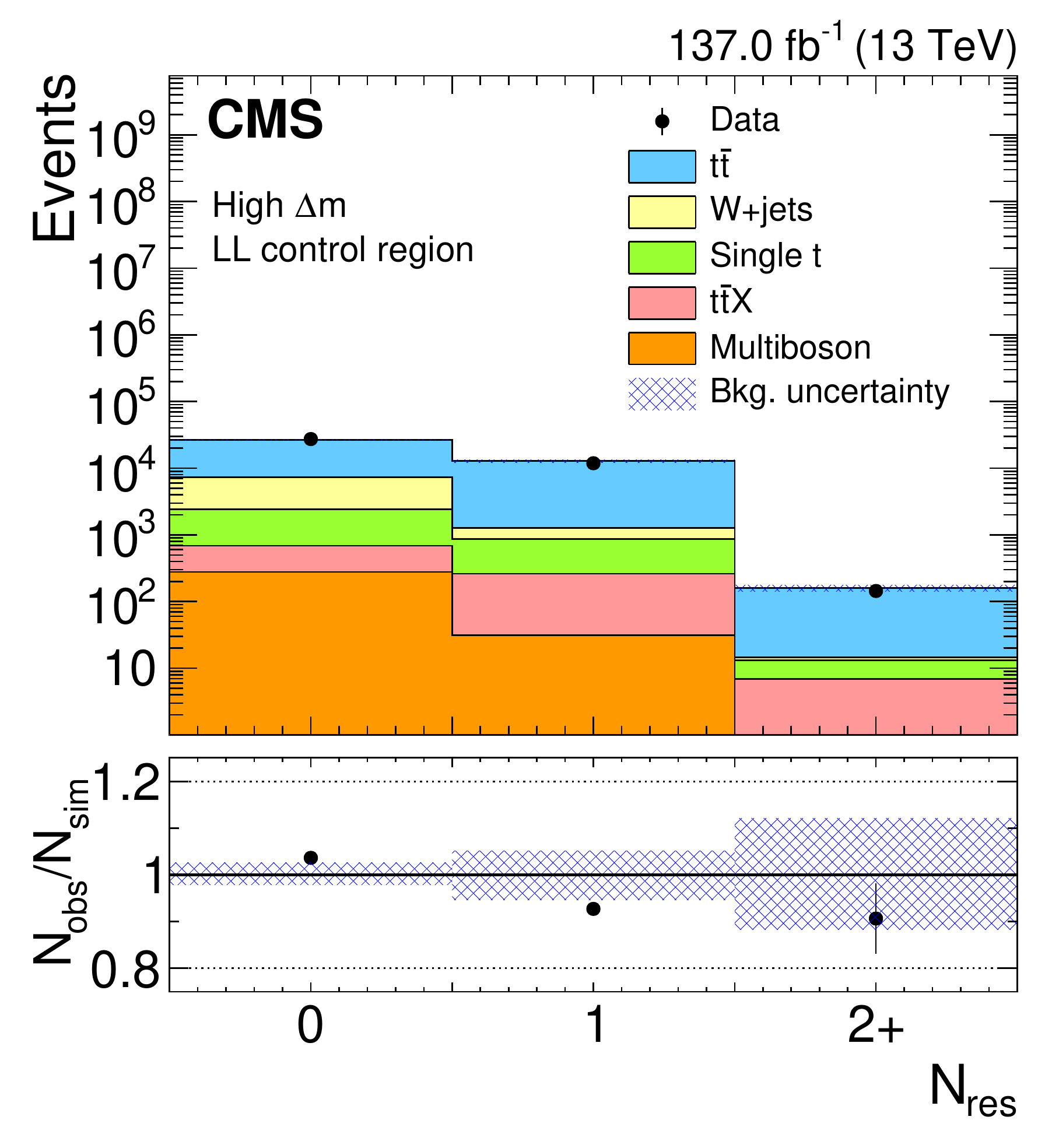}
	\caption[Heavy object comparisons in high \dm LL control region]{
				Comparison between data and simulation in the high \dm portion of the \ljets control region, as a function of \ptmiss (upper left), \Nt (upper right), \Nw (lower left), and \Nres (lower right) after scaling the simulation to match the total yield in data.
		The hatched region indicates the total shape uncertainty in the simulation.
		The lower panels display the ratios between the observed data and the simulation.
	}
	\label{fig:llcr-hm-heavy-objects}
\end{figure*}

Figure~\ref{fig:llcr-hm-heavy-objects} shows that the background model describes the data in the high \dm region of the \ljets control region well as a function of \Nt, \Nw, and \Nres.
The total yield of the background model is scaled to exactly match the total yield observed in collision data for the purpose of this comparison only, while the prediction for the LL background in the search region does not include this scaling.
This demonstrates that the transfer factor method described above will correctly describe the data as a function of \Nt, \Nw, and \Nres.
The background model includes the top quark \pt reweighting mentioned in Section~\ref{sec:simulation}.

One of the most important sources of systematic uncertainty in the LL background estimation arises from the data-to-simulation scale factor measurements for the merged top quark and \PW boson identification and resolved top quark identification.
This leads to an uncertainty in the estimated LL background yield of up to 8\% from the DeepAK8 top quark scale factor, 17\% from the DeepAK8 \PW boson scale factor, and up to 5\% from the DeepResolved top quark scale factor for some high \dm search bins that require one or more of these tagged objects.
Another significant source of systematic uncertainty arises from the top quark \pt reweighting, which improves the agreement between data and simulation.
The uncertainty associated with this reweighting is up to 15\%, depending on the search bin.
Other sources of systematic uncertainty include
the statistical uncertainties due to the control region data (up to 80\%)
and the simulated event samples (up to 43\%),
the electron and muon identification and isolation efficiencies (2--6\%),
the \tauh veto efficiency (1--7\%),
the jet energy scale (1--12\%),
the \ptmiss energy resolution (1--8\%),
the \PQb tagging efficiency (2--5\%),
the PDF uncertainty (2--17\%),
the pile-up uncertainty (1--10\%),
and the \ttbar and \wjets production cross section uncertainties (4--6\%),
depending on the search bin.

\subsection{Background from \texorpdfstring{\zparennunujets}{Z(nunu)+jets} events}
\label{subsec:Zinvisible}

In previous searches~\cite{Sirunyan:2016jpr,SUS-16-049,SUS-16-050}, two different methods have been used to estimate the background from \zjets events with \znn decay (\zparennunujets events).
One method uses \zjets events in which the \PZ boson decays to $\Pell^+\Pell^-$ (\EE or \MM).
In these events, the \PZ bosons have very similar kinematic properties to those of the \zparennunujets events in the search region, but this method is statistically limited because of the small $\PZ\to\Pell^{+}\Pell^{-}$ branching fraction.
Another method uses \gjets events, which feature a cross section that is larger than the \zparennunujets cross section by roughly a factor of five in the range of \PZ boson \pt that is relevant for this search.
The LO Feynman diagrams involved in \gjets production are similar to those for \zjets production, but differ in the quark-boson couplings and in the fact that the \PZ boson is massive.
These differences are generally described well by simulation and are accounted for.
Taking into account these considerations we use a hybrid method to estimate the \zparennunujets background that makes use of both procedures.
The method is discussed in more detail in Refs.~\cite{Sirunyan:2016jpr,SUS-16-049}.

Two control regions are used.
The \zparenll control region requires two same-flavor, opposite-charge reconstructed leptons (\EE or \MM) and is enriched in \zjets events in which the \PZ boson decays to $\Pell^+\Pell^-$.
The \gjets control region requires a single reconstructed photon and is therefore enriched in \gjets events.

The predicted yield of \zparennunujets is
\begin{linenomath*}
\begin{equation*}
\NP{\zparennunujets} = \RZ \SGamma \NMC{\zparennunujets},
\end{equation*}
\end{linenomath*}
where \NP{\zparennunujets} and \NMC{\zparennunujets} are respectively the predicted number of \zparennunujets events and the number of simulated \zparennunujets events in each search bin,
\RZ is a flavor-dependent \zjets normalization factor measured in the \zparenll control region, and \SGamma is a shape correction factor measured in the \gjets control region.

The normalization factor \RZ is derived from a fit of the simulation to the data in the \zparenll control region as a function of the measured dilepton mass \mll.
The \zparenll control region is selected from single-electron and single-muon triggers,
and further requires offline ${\pt > 40~(50)\GeV}$ for the leading electron (muon) candidates and
${\pt > 20\GeV}$ for the subleading electron and muon candidates.
The quality, isolation, and $\eta$ selection criteria discussed in Section~\ref{sec:reconstruction} are also required.
Jets matched to these selected leptons are removed from the calculation of search variables, and
the dilepton \ptvec is added to \ptvecmiss to emulate the \ptmiss expected in \zparennunujets events.

Events in the \zparenll control region are required to pass the same low \dm and high \dm baseline selections shown in Table~\ref{tab:evtsel},
with the exception of the lepton and isolated track vetoes, and with the additional requirement that ${\pt(\Pell^+\Pell^-) > 200\GeV}$.
The \RZ factor is measured from events in the dilepton mass window ${81<\mll<101\GeV}$, while
events in the ranges ${50<\mll<81\GeV}$ and ${\mll>101\GeV}$ are used to measure the rate of nonresonant background contributions, which are primarily \ttbar events in which both \PW bosons decay leptonically.
Approximately 97\% of the events in the low \dm \zparenll control region and 79\% of the events in the high \dm \zparenll control region are expected to be \dyjets events.
Minor contributions from, \eg, \ZZ production are counted with \zjets events in the extraction of \RZ, and minor contributions from, \eg, single top processes are counted with \ttbar events when measuring the rate of nonresonant backgrounds.
The \RZ factor is measured separately in different ranges of \Nb and \Nsv as well as separately in the low \dm and high \dm regions, allowing it to account for dependence on heavy-flavor production.
This results in five distinct \RZ values in the low \dm region, as can be seen from Table~\ref{tab:low_dm}, which includes five unique combinations of \Nb and \Nsv requirements.
In the high \dm region, this results in two distinct \RZ values: one with a requirement of ${\Nb = 1}$ and one with a requirement of ${\Nb \geq 2}$.
The \RZ factor ranges from 0.71 to 1.05 for low \dm search bins and from 1.20 to 1.27 for high \dm search bins and has uncertainties of 4--14\%, which are propagated to the \zparennunujets predictions in the search regions.

The shape correction factor \SGamma is derived from the \gjets control region and corrects for any mismodeling of the search variable distributions by the simulation.
The \gjets control region is selected from single-photon triggers that require a photon candidate with a \pt threshold of 175--200\GeV, depending on the data collection period.
Offline photons are required to have ${\pt > 220\GeV}$ and ${\abs{\eta} < 1.44}$ or ${1.57 < \abs{\eta} < 2.5}$, avoiding the gap between the ECAL barrel and endcap detectors.
Similarly to what is done for dilepton events,
jets matched to selected photons are removed from the calculation of search variables, and
the photon four-vector is added to \ptvecmiss to emulate the \ptmiss expected in \zparennunujets events.
The \ptmiss prior to this addition is required to be less than 250\GeV to make the \gjets control region orthogonal to the search region.
Approximately 87\% of the events in the low \dm \gjets control region and 76\% of the events in the high \dm control region are expected to be \gjets events.

The \SGamma factor is not used to correct the estimated overall rate of \zparennunujets events, but only to correct the distribution of those events.
The yield of simulated events in the \gjets control region is scaled to the corresponding event yields from collision data separately in the low \dm and high \dm regions and in different ranges of \Nb and \Nj,
and then simulated events are compared to collision data as a function of all search bin variables except \Nt, \Nw, and \Nres to extract \SGamma.
The high \dm \gjets control region bins make no requirements on \Nt, \Nw, or \Nres, yielding 112 control region bins, with a distinct \SGamma value for each control region bin.
The \zparennunujets simulation provides an improved description of the distributions of \Nt, \Nw, and \Nres in the search region after the \RZ and \SGamma correction factors have been applied.
The \SGamma factor measured in the \gjets control region is validated in the \zparenll control region.
The observed differences between \SGamma calculated in the \gjets control region and \SGamma calculated in the \zparenll control region as a function of \ptmiss (up to 16\%) are treated as systematic uncertainties.

In addition to the uncertainties in the \RZ normalization factor obtained from the \zparenll control region and in the \SGamma shape correction factors discussed above, we consider several sources of uncertainty in the estimation of the \zparennunujets background, including
the statistical uncertainties in the photon control region data (up to 100\%) and simulated event samples (up to 110\%),
the photon identification efficiencies (5--13\%),
the photon trigger efficiency (up to 2\%),
the pileup reweighting (up to 40\%),
the jet energy scale corrections (up to 41\%),
the \ptmiss energy resolution (up to 35\%),
the PDF uncertainty (up to 59\%),
the \PQb tagging efficiencies for heavy-flavor jets (up to 5\%) and misidentification rates for light-flavor jets (up to 16\%),
the soft \PQb tagging efficiencies (up to 1\%), and
the top quark and \PW boson misidentification rates (up to 34\%).

\subsection{Background from QCD multijet events}
\label{subsec:QCD}

The QCD multijet background originates from mismeasurement of the energy of one or more jets in a QCD multijet event.
When that happens, large amounts of spurious \ptmiss can be present in the reconstructed event, causing it to satisfy the selection requirements.
The probability to produce such an event, including misidentified \PQb jets and top quarks, is very low, but the high QCD multijet production cross section makes them very numerous and therefore their contribution to the search bins must be estimated.

The QCD multijet control region requires that at least one of the three leading jets is close to the \ptvecmiss, that is, ${\dphi\left(\ptvecmiss, \jet{1,2,3}\right) < 0.1}$.
This control region definition otherwise requires the baseline selection described in Section~\ref{sec:selection}, including the low \dm and high \dm regions (with the exception of the \dphi requirements).
These requirements produce a control region in which QCD multijet events are expected to make up 84\% of the total yield.

The QCD multijet control region is divided into bins based on \ptmiss, \HT, \Nj, \mTb, \ptb, \ptISR, \Nb, and \Nsv, similarly to the search bins described in Section~\ref{sec:selection} and Tables~\ref{tab:low_dm} and \ref{tab:high_dm}, with the exception that the QCD multijet control region is not binned in \Nt, \Nw, or \Nres.
This allows us to maintain adequately small statistical uncertainties in each bin of the control region.

The yield of QCD multijet events in a search bin is extrapolated from the corresponding bin in the QCD multijet control region.
The ratio of the QCD multijet yield predicted by simulation in a search bin to the QCD multijet yield predicted in the corresponding control region bin, \TF{QCD}, is taken from simulation, and then the QCD multijet background yield \NP{\text{QCD}} is estimated as
\begin{linenomath*}
\[\NP{\text{QCD}} = \TF{QCD} \left(\ND{} - \NMC{\text{non-QCD}}\right)\]
\end{linenomath*}
where \ND{} is the number of events in the QCD multijet control region bin and \NMC{\text{non-QCD}} is the number of events from all other backgrounds in the same bin as predicted by simulation.

To improve the statistical power of the QCD multijet simulation, we employ a ``smearing'' procedure, which involves resampling the \pt of the leading jets from the expected jet response distribution.
This simulates the effects of jet \pt mismeasurement, and allows simulated events with low \ptmiss to be used.
A correction scale factor is applied to each simulated event to correct for any mismodeling of the jet response distribution.
This scale factor varies as a function of the ratio of the reconstructed jet \pt to the simulated jet \pt, and is derived by fitting the simulated events to the data in the QCD multijet control region as a function of a proxy variable, ${\pt^\text{reco} / (\pt^\text{reco} + \ptmiss)}$, where $\pt^\text{reco}$ is the reconstructed jet \pt.

\begin{linenomath*}
We account for the effects of a number of sources of systematic uncertainty in the estimate of the QCD multijet background yield,
including the statistical uncertainties due to the control region data (1--260\%) and simulated event samples (4--130\%),
the \PQb tagging efficiencies for heavy-flavor jets (up to 18\%) and misidentification rates for light-flavor jets (up to 9\%),
the soft \PQb tagging efficiency (up to 8\%),
the trigger efficiency (up to 40\%),
the pileup reweighting (up to 50\%),
the jet energy scale corrections (up to 63\%),
the \ptmiss energy resolution (up to 64\%),
the top quark and \PW boson misidentification rates (up to 36\%),
the top quark \pt reweighting (up to 39\%),
the PDF uncertainty (up to 67\%),
the smearing procedure (up to 41\%),
the jet response correction (up to 42\%), and
residual bias in the \ptmiss distribution (up to 20\%).
\end{linenomath*}

\subsection{Background from rare processes}
\label{subsec:rare}

Besides the backgrounds discussed above, other SM processes with small production cross sections are also considered in this analysis.
These include the diboson (\WW, \WZ, and \ZZ) processes, multiboson ($\PW\PW\PW$, $\PW\PW\PZ$, $\PW\PZ\PZ$, and $\PZ\PZ\PZ$) processes, associated production with a top quark-antiquark pair (\ttH, $\ttbar\PGg$, $\ttbar\PW$, and \ttZ), and other processes ($\PQt\PW\PZ$, $\PW\PZ\PGg$, and $\PW\PW\PGg$).
Of these, the most important is the \ttZ background because, in the case where the \PZ boson decays to $\PGn\PAGn$, this background is irreducible.

Simulated events are used to estimate the background, and the total yield is given by the product of the luminosity and the theoretical cross section, with the exception of the \ttZ background.
The \ttZ cross section is taken from a recent measurement using CMS data~\cite{CMS:2019too}.
These backgrounds, other than \ttZ, are not estimated from data because they are sufficiently rare that the estimate based on the theoretical cross sections is more precise than an estimate based on data.
The LL background estimation procedure already accounts for the portion of these backgrounds that include one or more charged leptons, and therefore simulated events that include generated charged leptons are not part of this prediction.

The uncertainties for the rare backgrounds are determined individually for each search bin and arise from
the statistical uncertainty in the simulated event samples (up to 100\%),
the integrated luminosity (1.8\%),
the \PQb tagging efficiency for heavy-flavor jets (up to 7\%) and
misidentification rates for light-flavor jets (up to 14\%),
the soft \PQb tagging efficiency (up to 5\%),
the trigger efficiency (less than 1\%),
the renormalization and factorization scales (up to 35\%),
the pileup reweighting (up to 48\%),
the jet energy scale corrections (up to 39\%),
the \ptmiss energy resolution (up to 23\%),
the PDF uncertainty (up to 15\%),
the merged top quark and \PW boson reconstruction efficiencies (up to 19\%),
the resolved top quark reconstruction efficiencies (up to 17\%), and
the \ttZ scale factor derived from comparison to data in the 3- and 4-lepton channels (8\%).

\subsection{Validation of the background estimation}
\label{subsec:validation}

\begin{figure}[htb!p]
	\includegraphics[width=0.48\textwidth]{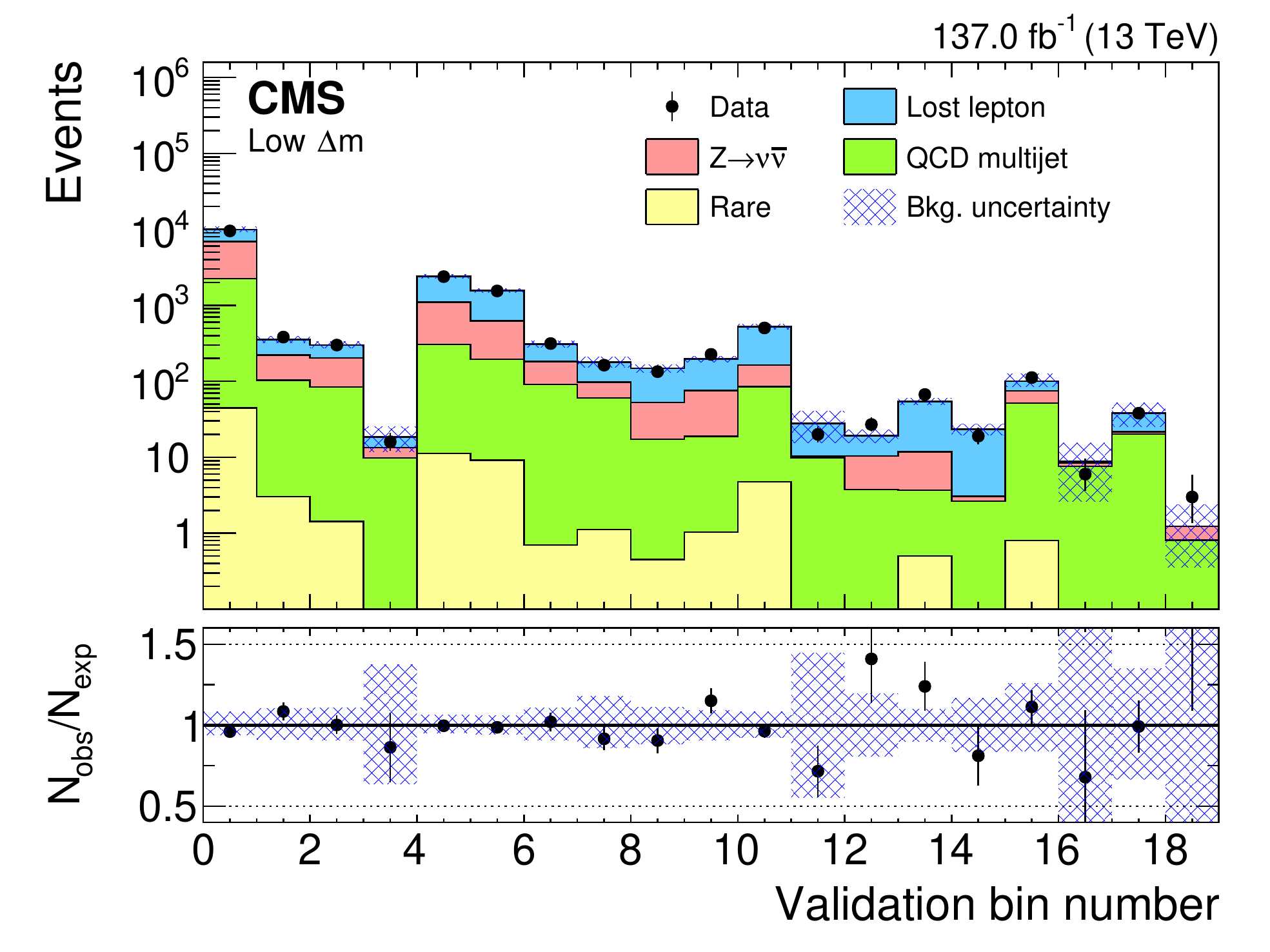}
	\includegraphics[width=0.48\textwidth]{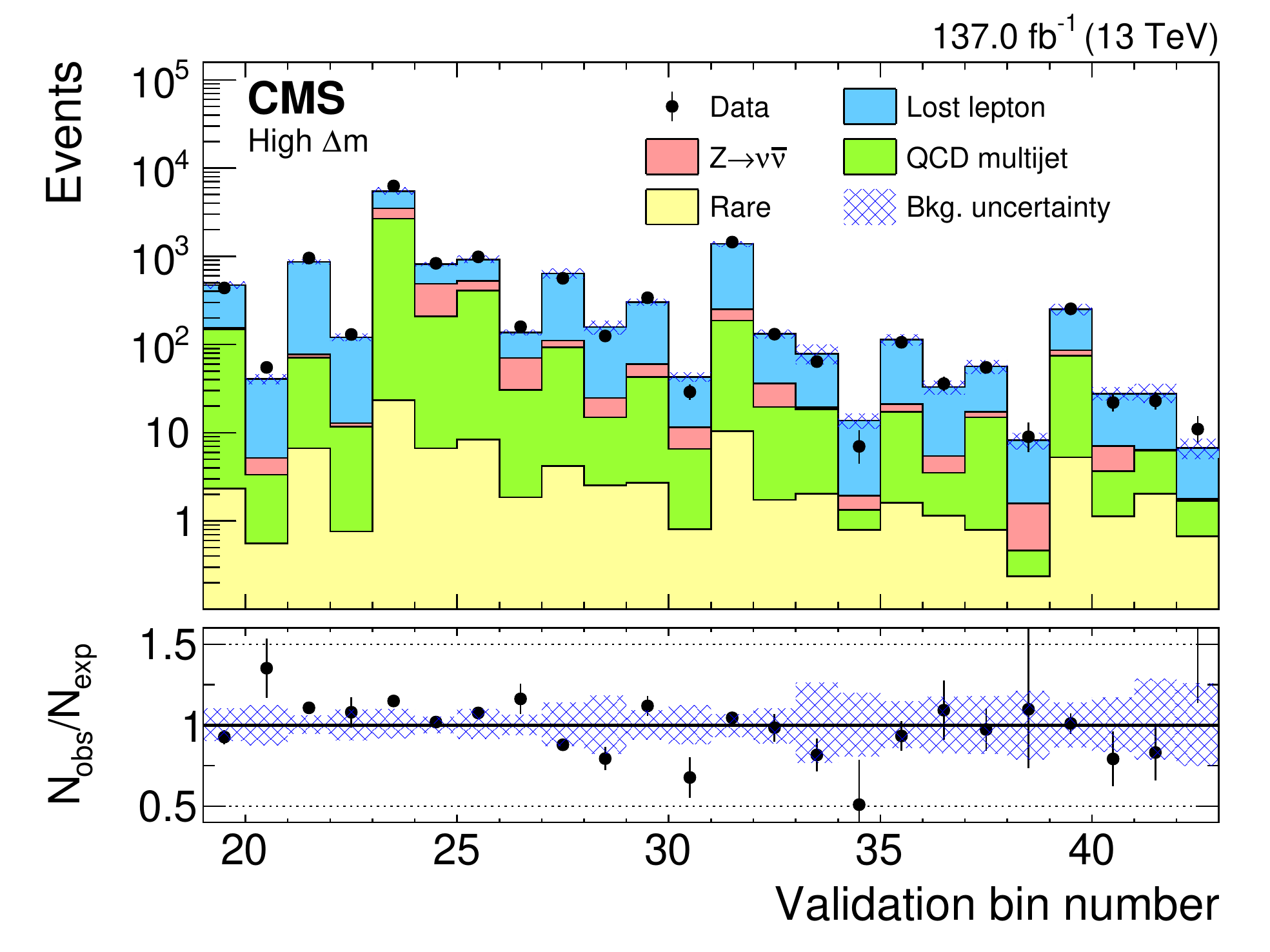}
	\caption{\label{fig:validation_bins}
          The observed numbers of events and the SM background predictions for the low \dm validation bins (\cmsLeft)
          and for the high \dm validation bins (\cmsRight).
          The hatched region indicates the total uncertainty in the background predictions.
          The lower panels display the ratios between the data and the SM predictions.
	}
\end{figure}

To validate our background model, we compare the model to data in the validation regions described in Section~\ref{subsec:validation_regions}.
The validation regions are kinematically very similar to the search region, but do not overlap with it and are not expected to contain any appreciable yield from any of the signal models.
The low \dm validation region is described in Table~\ref{tab:low_dm_validation} and the high \dm validation region is described in Table~\ref{tab:high_dm_validation}.

The background predictions in these validation bins are calculated as described in the preceding sections and compared to data, as shown in Fig.~\ref{fig:validation_bins}.
The background prediction is compatible within uncertainties with the observed data.
This compatibility demonstrates that the background model adequately describes the backgrounds present in the data and can be used to describe the backgrounds in the search region.

\section{Results and interpretation}
\label{sec:results}

\begin{figure*}[p!htb]
  \centering
  \includegraphics[width=0.80\linewidth]{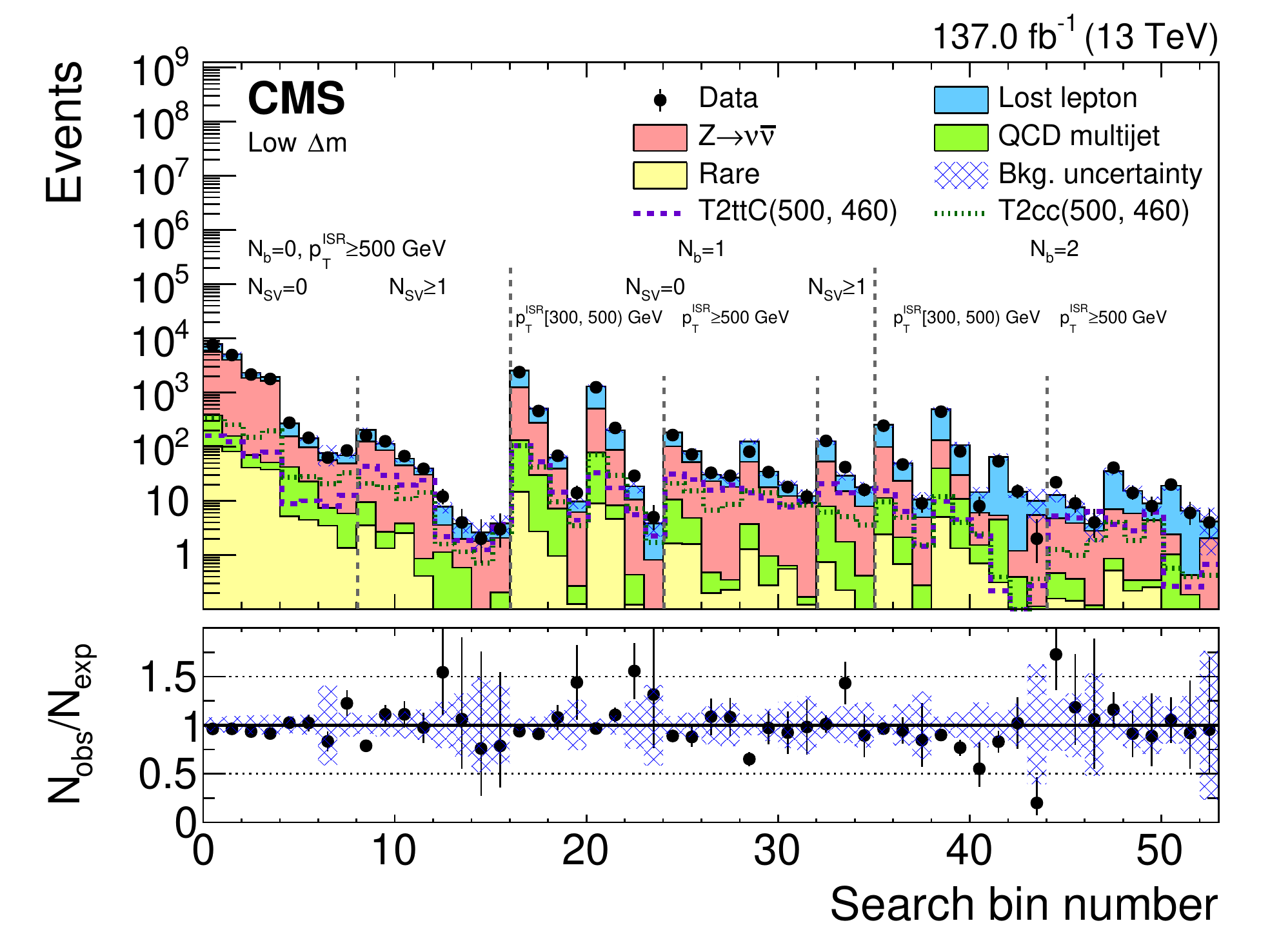}
  \includegraphics[width=0.80\linewidth]{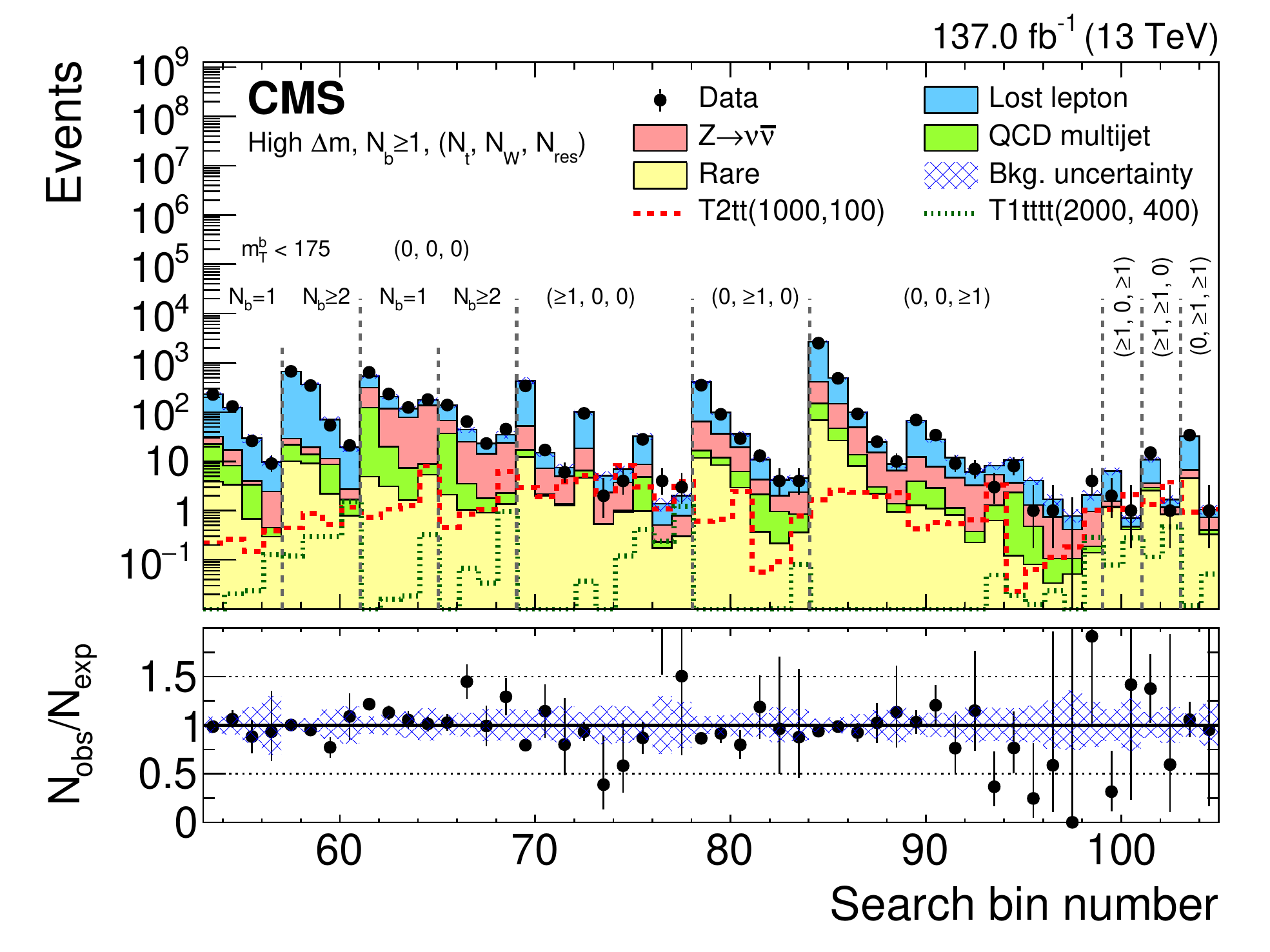}
\caption{Observed event yields in data (black points) and predicted SM background (filled histograms) for the low \dm search bins 0--52 (upper),
         and for the high \dm search bins 53--104 (lower).
         The bracketed numbers in the lower plot
         represent the respective \Nt, \Nw, and \Nres requirements used in that region.
         The signal models are denoted in the legend with the masses in \GeVns of the SUSY particles in parentheses: $(\mstop, \mlsp)$ or $(\mgluino, \mlsp)$ for the T2 or T1 signal models, respectively.
         For both plots, the lower panel shows the ratio of the data to the total background prediction.
         The hatched bands correspond to the total uncertainty in the background prediction.
         The (unstacked) distributions for two example signal models are also shown in both plots.
        }
  \label{fig:pred-lm&pred-hm-nb1}
\end{figure*}

\begin{figure*}[p!htb]
  \centering
  \includegraphics[width=0.80\linewidth]{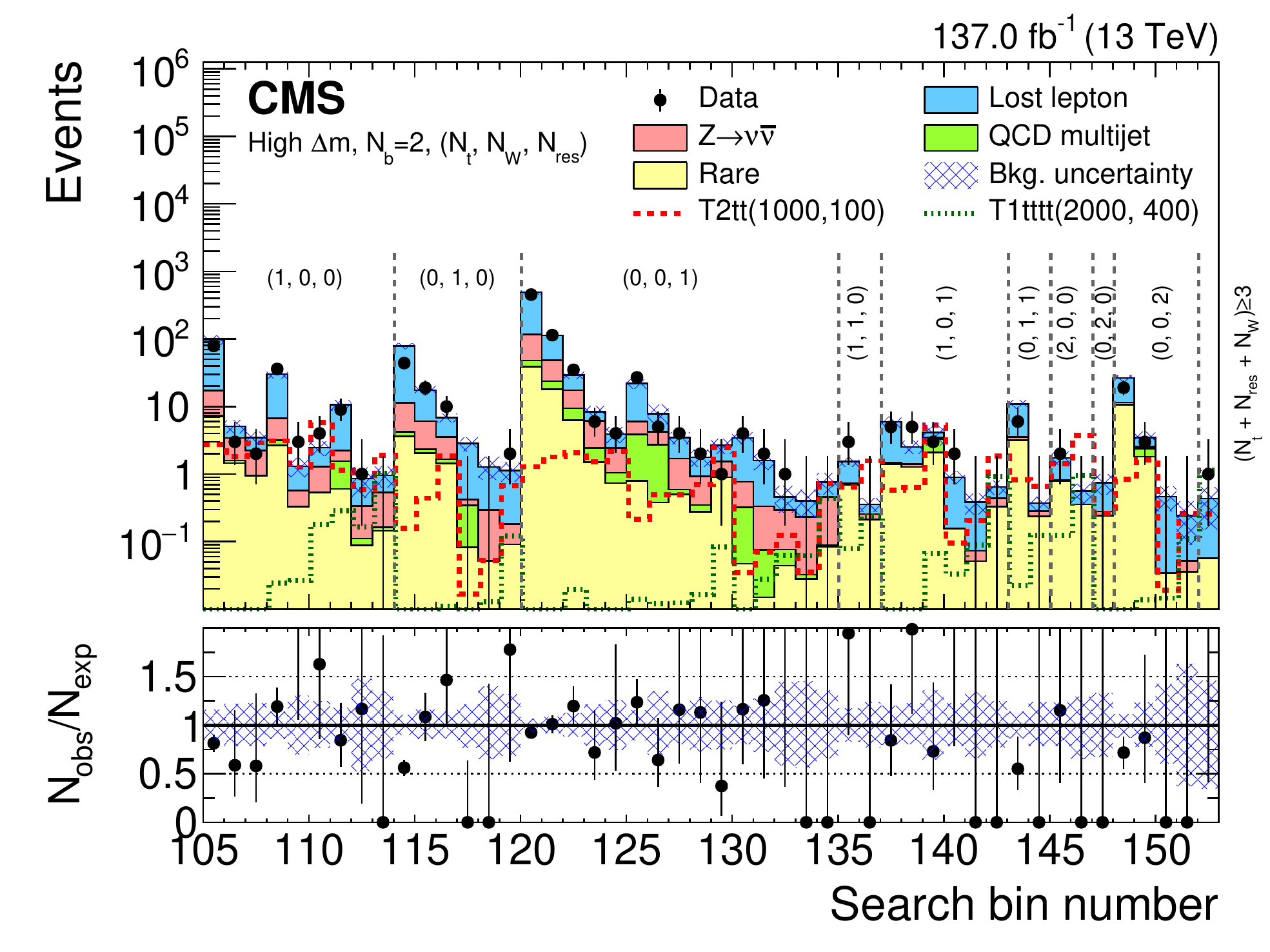}
  \includegraphics[width=0.80\linewidth]{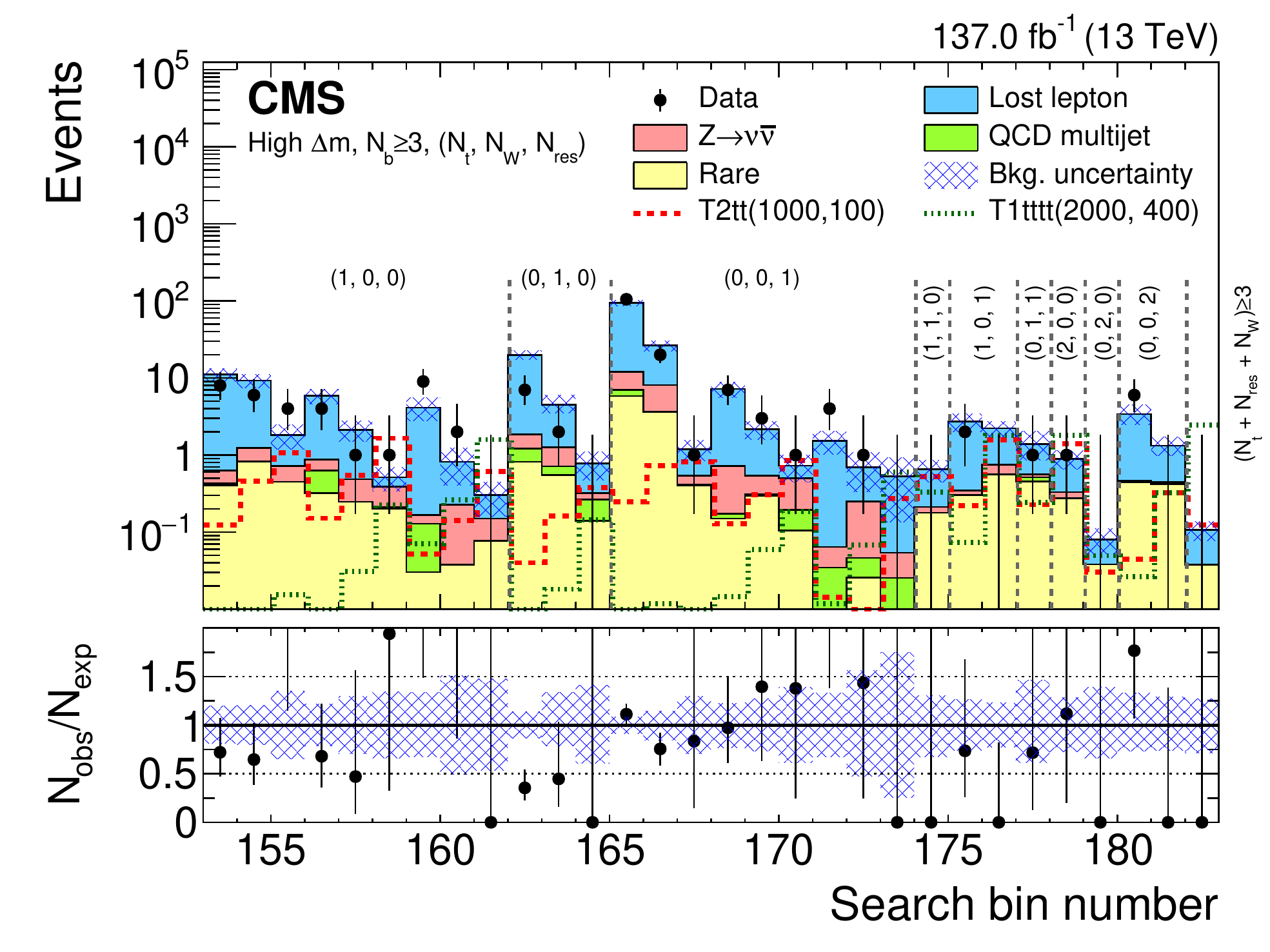}
\caption{Observed event yields in data (black points) and predicted SM background (filled histograms) for the high \dm search bins 105--152 with ${\Nb = 2}$ (upper),
         and for the high \dm search bins 153--182 with ${\Nb \geq 3}$ (lower). The bracketed numbers in each plot
         represent the respective \Nt, \Nw, and \Nres requirements used in that region.
         The signal models are denoted in the legend with the masses in \GeVns of the SUSY particles in parentheses: $(\mstop, \mlsp)$ or $(\mgluino, \mlsp)$ for the T2 or T1 signal models, respectively.
         For both plots, the lower panel shows the ratio of the data to the total background prediction.
         The hatched bands correspond to the total uncertainty in the background prediction.
         The (unstacked) distributions for two example signal models are also shown in both plots.
        }
  \label{fig:pred-hm-nb2&pred-hm-nb3}
\end{figure*}

The predicted and observed yields in the 183 search bins defined in Section~\ref{sec:selection} are summarized in Figs.~\ref{fig:pred-lm&pred-hm-nb1} and \ref{fig:pred-hm-nb2&pred-hm-nb3}, and numerical values are presented in Tables~\ref{tab:pred-0}--\ref{tab:pred-6} of Appendix~\ref{appdx}.
No statistically significant excess of events is observed relative to the expectation from the SM.
All but six out of 183 search bins have agreement within two standard deviations, and all search bins have agreement within three standard deviations.
A goodness-of-fit test under the background-only hypothesis yields a $p$-value of 0.66, indicating good agreement with the SM expectation.
The observations are interpreted in the context of the models described in Section~\ref{sec:signal_models} as upper limits on the cross section for production of top squarks as a function of the masses of the top squark and LSP or the gluino and LSP.

Upper limits on the direct top squark pair production cross section or gluino-mediated top squark production cross section are derived via a modified frequentist method using the \CLs criterion in an asymptotic formulation~\cite{Read_2002,Junk_1999,Cowan}.
The observed and predicted yields in the search bins as well as all of the control region bins are included in the limit calculation.
To implement the background estimation procedures based on data in control regions described in Section~\ref{sec:background}, the yields of the relevant backgrounds in the search region bins and the corresponding control region bins are taken directly from the simulation, but are scaled by nuisance parameters with no a priori constraint.
These a priori unconstrained nuisance parameters are constrained in the fitting procedure by the observed yield in the data in the search region as well as the data in the control regions.
Systematic uncertainties are also implemented using nuisance parameters with log-normal a priori constraints.
When computing the limits, the signal yields are corrected to account for the expected signal contamination of the data control regions used to estimate the SM background.
These corrections are typically below 20\%.

The uncertainties in the signal modeling are determined
individually for each search bin and arise from
the statistical uncertainty in the simulated event samples (up to 100\%),
the integrated luminosity (1.8\%),
the charged lepton veto efficiencies (up to 10\%),
the \PQb tagging efficiency for heavy-flavor jets (up to 11\%) and misidentification rates for light-flavor jets (up to 14\%),
the soft \PQb tagging efficiency (up to 5\%),
the trigger efficiency (less than 1\%),
the pileup reweighting (up to 15\%),
the renormalization and factorization scales (up to 7\%),
the ISR modeling (up to 37\%),
the jet energy scale corrections (up to 26\%),
the \ptmiss energy resolution (up to 12\%),
the merged top quark and \PW boson reconstruction efficiencies (up to 17\%), and
the resolved top quark reconstruction efficiencies (up to 20\%),
where the systematic uncertainty upper range is defined as the 95\% upper quantile to indicate the typical ranges.
Because SUSY signal events are simulated using the CMS fast simulation program, additional uncertainties are assigned to the correction of the
\PQb tagging, soft \PQb, merged top quark and \PW boson, and resolved top quark reconstruction efficiencies, as well as to cover differences in \ptmiss
between the fast simulation and the full \GEANTfour-based model of the CMS detector, which lead to uncertainties of up to about 40\%, depending on the search bin.
All uncertainties except those from the statistical precision of the simulation
are treated as fully correlated among search bins.
The statistical uncertainties from the simulated signal events as well as those from the simulated SM background events
are incorporated into the limit calculation via the approach described in Ref.~\cite{Barlow-Beeston}.

\begin{figure*}[p!htb]
\centering
\includegraphics[width=0.48\linewidth]{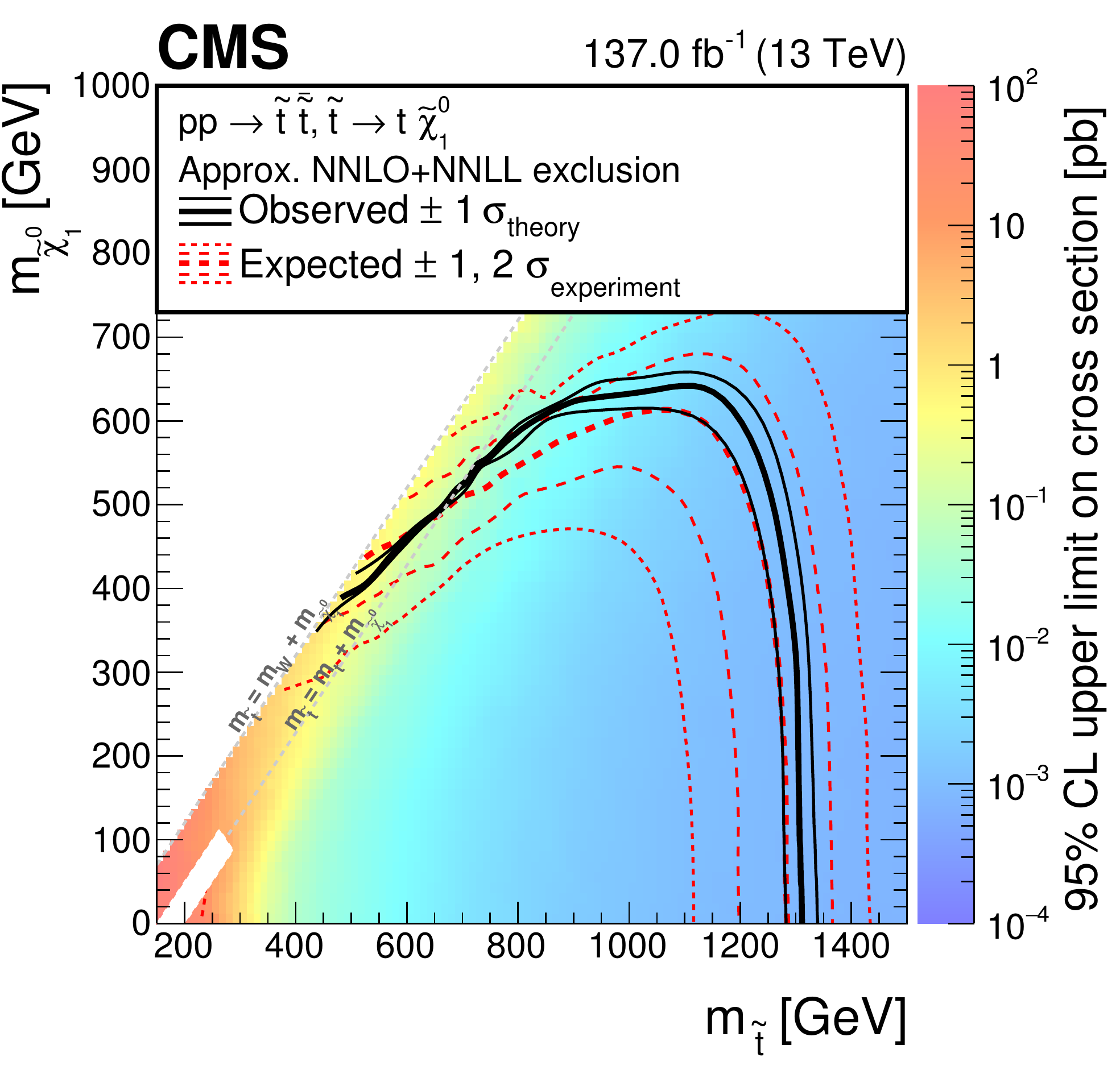}
\includegraphics[width=0.48\linewidth]{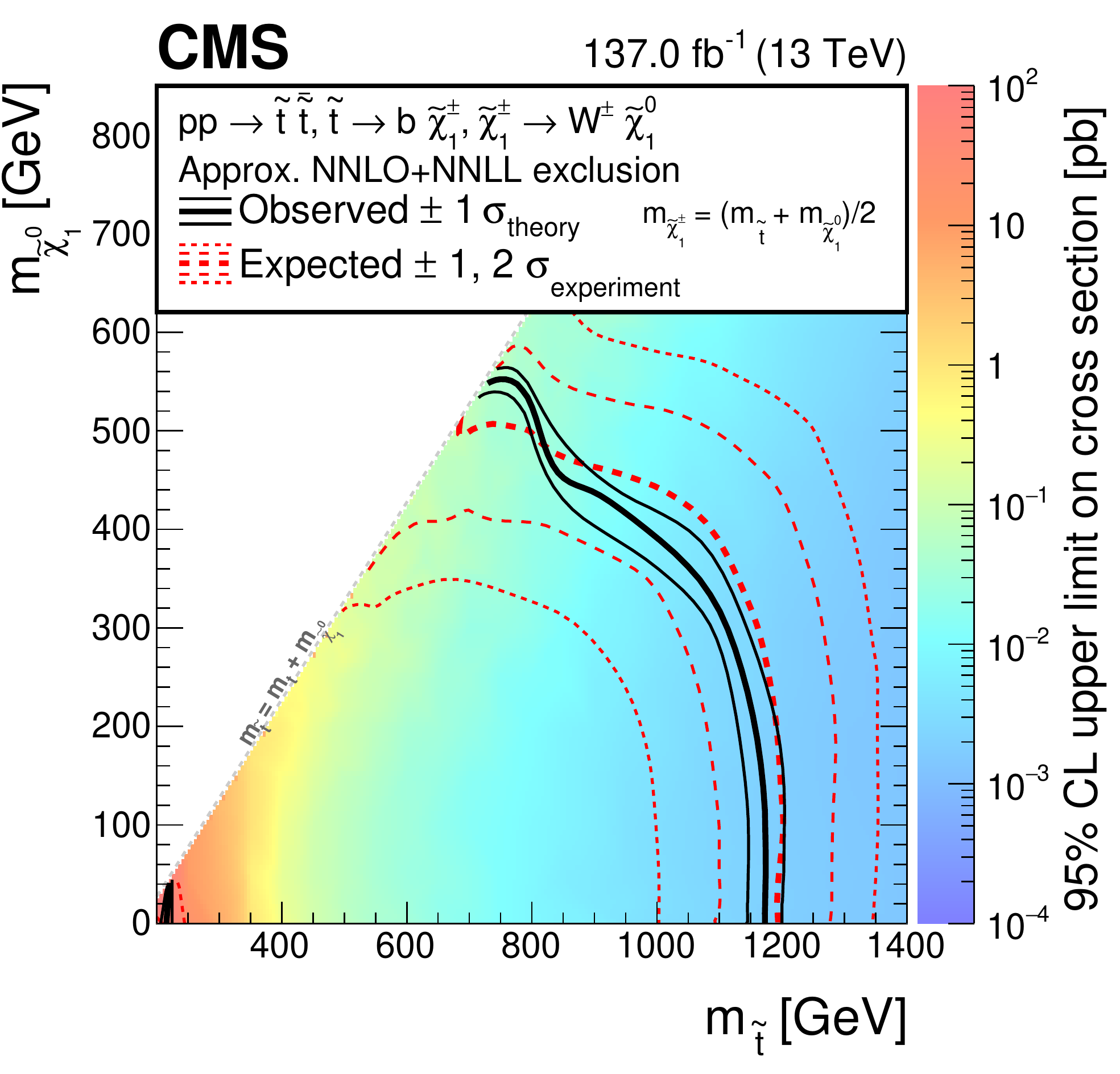}
\includegraphics[width=0.48\linewidth]{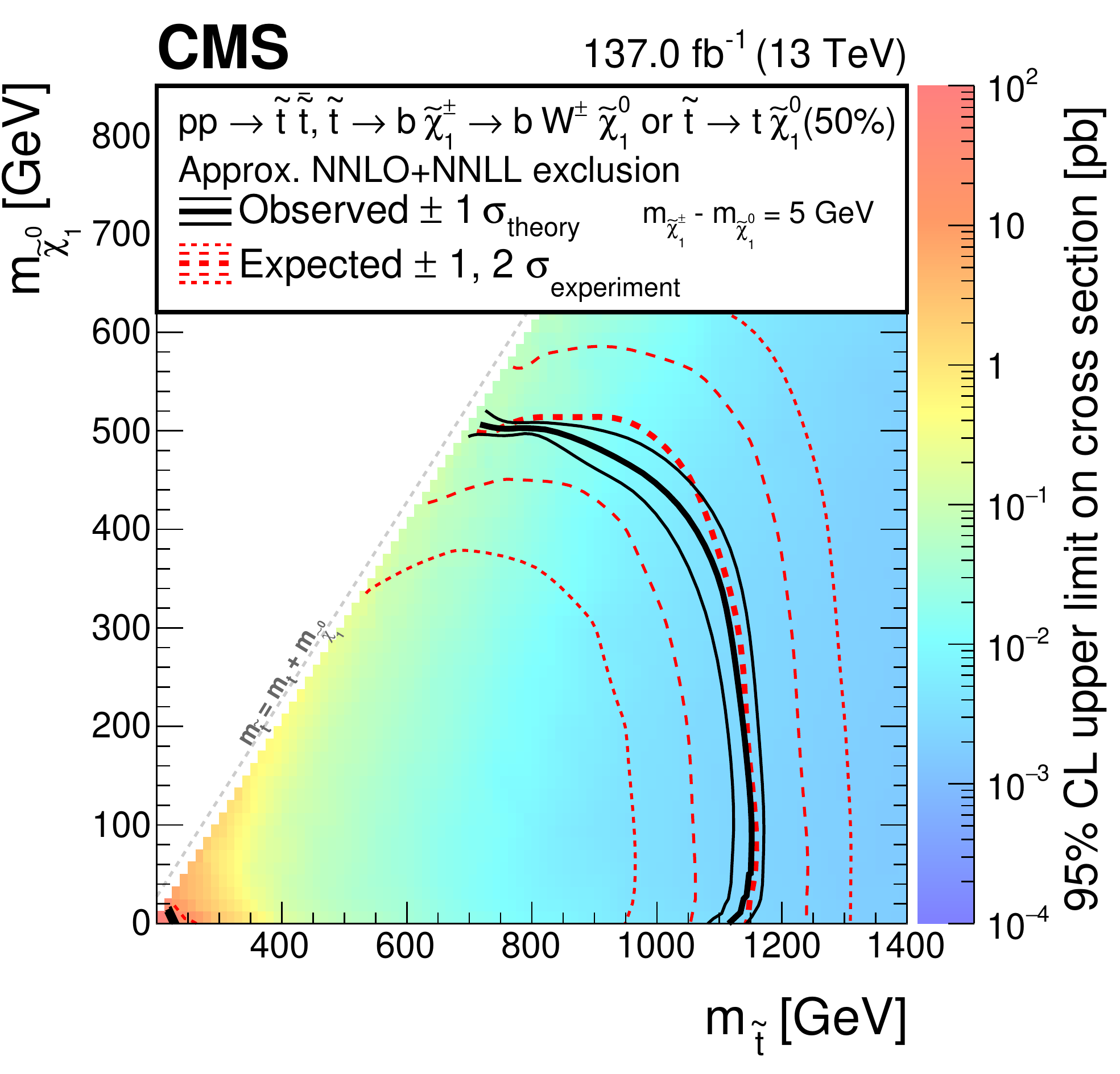}
\caption{
The 95\% \CL upper limit on the production cross section of the
T2tt (upper left),
T2bW (upper right), and
T2tb (lower)
simplified models as a function of the top squark and LSP masses.
The solid black curves represent the observed exclusion contour
with respect to approximate NNLO+NNLL signal cross sections
and the change in this contour due to variation of these
cross sections within their theoretical uncertainties ($\sigma_{\text{theory}}$)~\cite{Beenakker:2016lwe,
bib-nlo-nll-01,bib-nlo-nll-02,bib-nlo-nll-03,bib-nlo-nll-04,
Beenakker:2011sf,Beenakker:2013mva,Beenakker:2014sma,
Beenakker:1997ut,Beenakker:2010nq,Beenakker:2016gmf}.
The dashed red curves indicate the mean expected exclusion contour
and the region containing 68 and 95\% ($\pm 1$ and $2\,\sigma_{\text{experiment}}$) of the distribution of expected
exclusion limits under the background-only hypothesis.
For T2tt, no interpretation is provided for signal models for which
${\abs{\mstop - \mlsp - \mtop} < 25\GeV}$ and ${\mstop < 275\GeV}$ as described in the text.
}
\label{fig:limits_T2_highdm}
\end{figure*}

Figure~\ref{fig:limits_T2_highdm} shows the 95\% confidence level (\CL) upper limits on the direct top squark pair production cross section in the context of the T2tt, T2bW, and T2tb models.
The upper limits are used to derive an exclusion region in the \mstop-\mlsp plane by comparison to the theoretical cross sections calculated at approximate NNLO+NNLL in Refs.~\cite{bib-sms-3,bib-sms-2,bib-sms-4,CMS-SMS}.
For the T2tt model, top squark masses up to 1310\GeV and LSP masses up to 640\GeV are excluded.
For the T2tt model we do not present cross section upper limits in the region of $\abs{\mstop-\mtop-\mlsp} < 25\GeV$ and $\mstop < 275\GeV$ as shown in Fig.~\ref{fig:limits_T2_highdm} (upper left).
In this region, signal events become similar to SM \ttbar events and the signal acceptance changes rapidly and is very sensitive to the details of the simulation, so no interpretation is presented~\cite{Sirunyan:2016jpr}.
To constrain the top squark pair production cross section in this region, a dedicated search could be performed~\cite{Sirunyan:2019zyu}, or a measurement of spin correlations in the \ttbar dileptonic decay system~\cite{Aaboud:2019hwz} could provide some constraint.
For the T2bW and T2tb models, top squark masses up to 1170 and 1150\GeV and LSP masses up to 550 and 500\GeV are excluded, respectively.
For the T2bW and T2tb models, there are regions of low \mstop and \mlsp where signals are not excluded by the observed 95\% \CL limits.
The sensitivity is reduced in this region because the top squark decay products have low \pt and the signal acceptance becomes smaller.

\begin{figure*}[p!htb]
\centering
\includegraphics[width=0.48\linewidth]{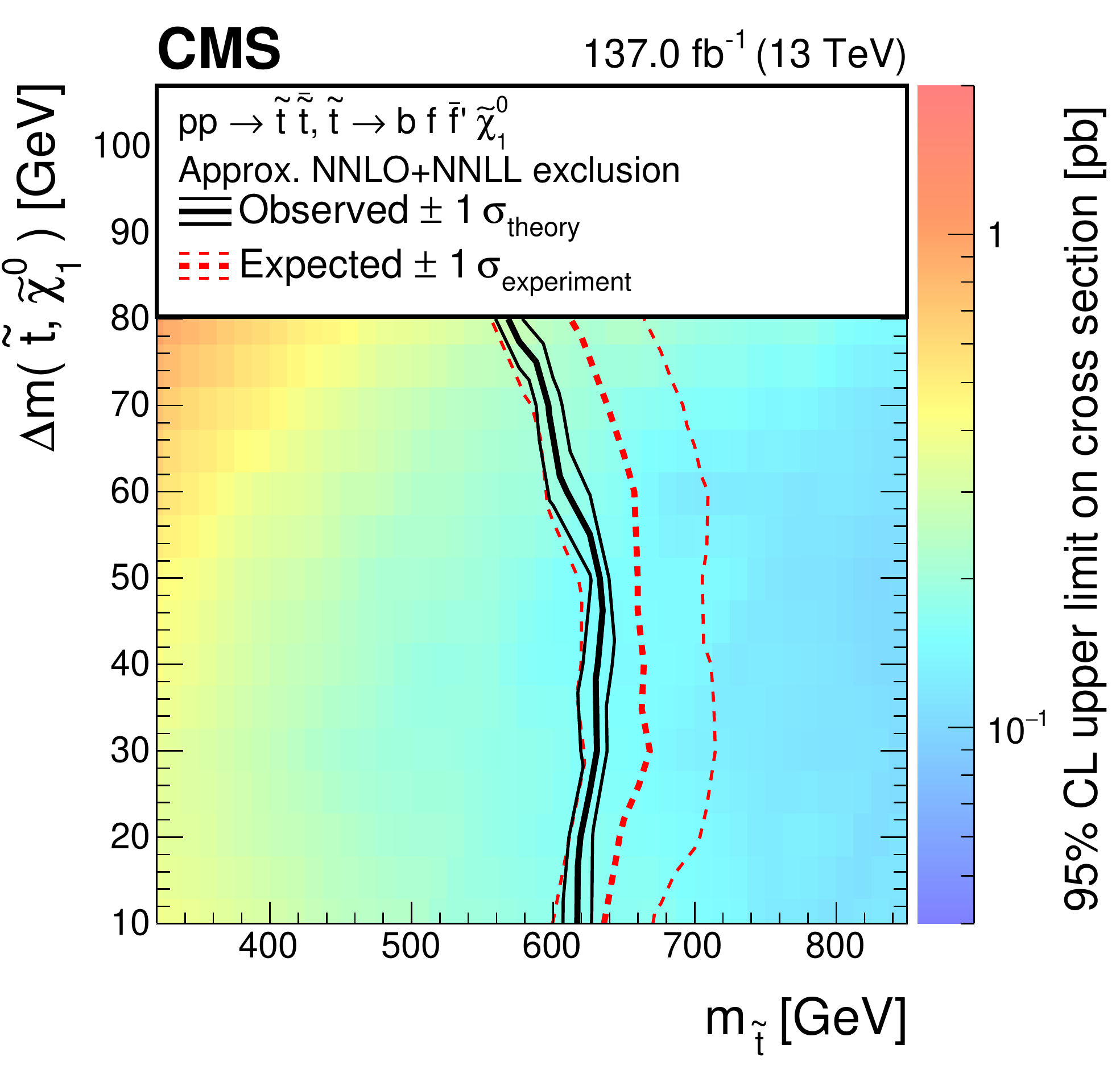}
\includegraphics[width=0.48\linewidth]{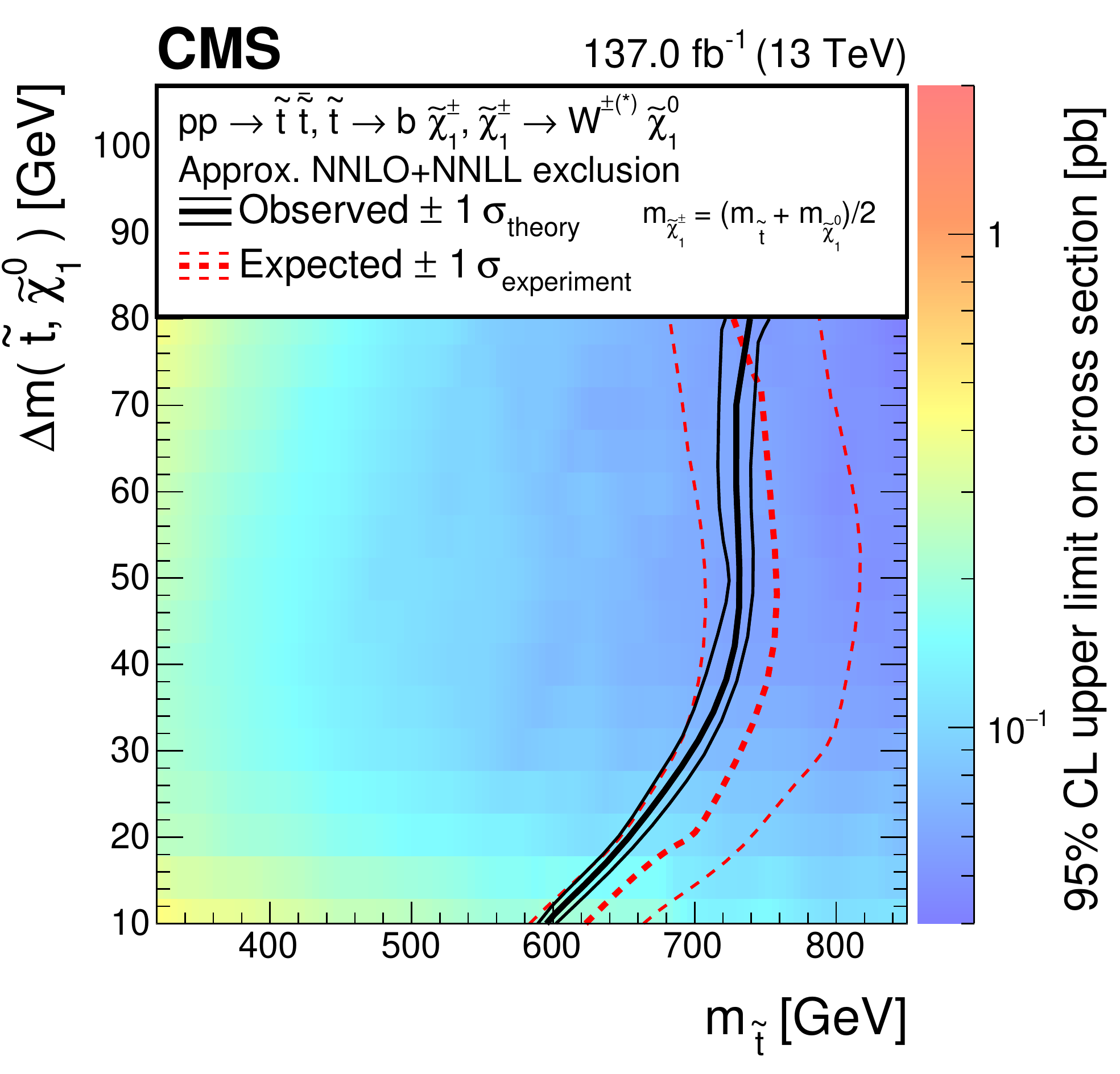}
\includegraphics[width=0.48\linewidth]{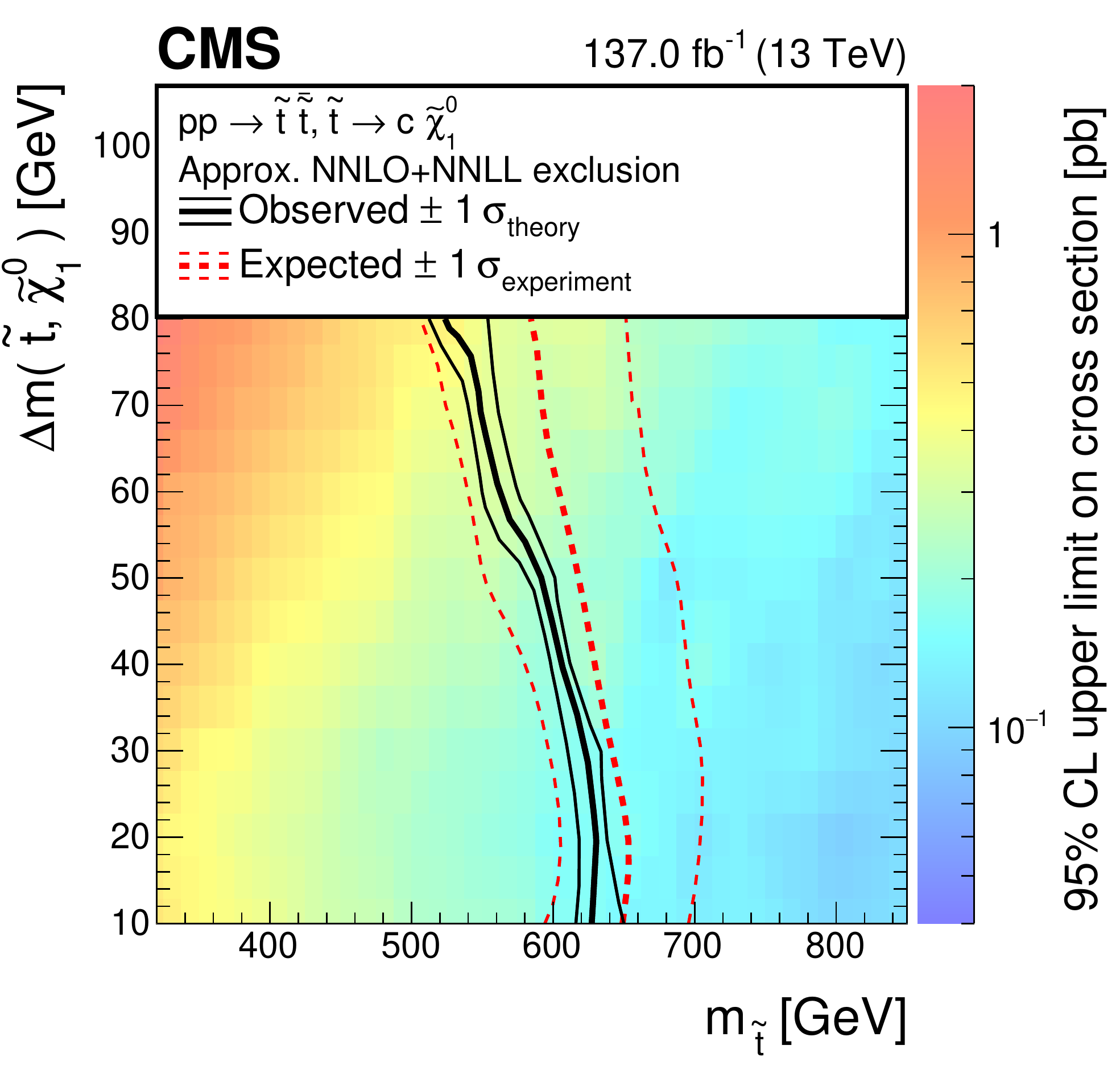}
\caption{
The 95\% \CL upper limit on the production cross section of the
T2ttC (upper left),
T2bWC (upper right), and
T2cc (lower)
simplified models as a function of the top squark mass
and the difference between the top squark and LSP masses.
The solid black curves represent the observed exclusion contour
with respect to approximate NNLO+NNLL signal cross sections
and the change in this contour due to variation of these
cross sections within their theoretical uncertainties ($\sigma_{\text{theory}}$)~\cite{Beenakker:2016lwe,
bib-nlo-nll-01,bib-nlo-nll-02,bib-nlo-nll-03,bib-nlo-nll-04,
Beenakker:2011sf,Beenakker:2013mva,Beenakker:2014sma,
Beenakker:1997ut,Beenakker:2010nq,Beenakker:2016gmf}.
The dashed red curves indicate the mean expected exclusion contour
and the region containing 68\% ($\pm 1\,\sigma_{\text{experiment}}$) of the distribution of expected
exclusion limits under the background-only hypothesis.
}
\label{fig:limits_T2_lowdm}
\end{figure*}

Figure~\ref{fig:limits_T2_lowdm} shows the 95\% \CL upper limits on the production cross section in the plane of \mstop versus \dm for the T2ttC, T2bWC, and T2cc models.
Signal events with \dm below \mW in the range of 10--80\GeV are considered.
Top squark masses up to 640\GeV are excluded at the 95\% \CL in the context of the T2ttC model, 740\GeV for the T2bWC model, and 630\GeV for the T2cc model.

\begin{figure}[t!]
\centering
\includegraphics[width=0.48\textwidth]{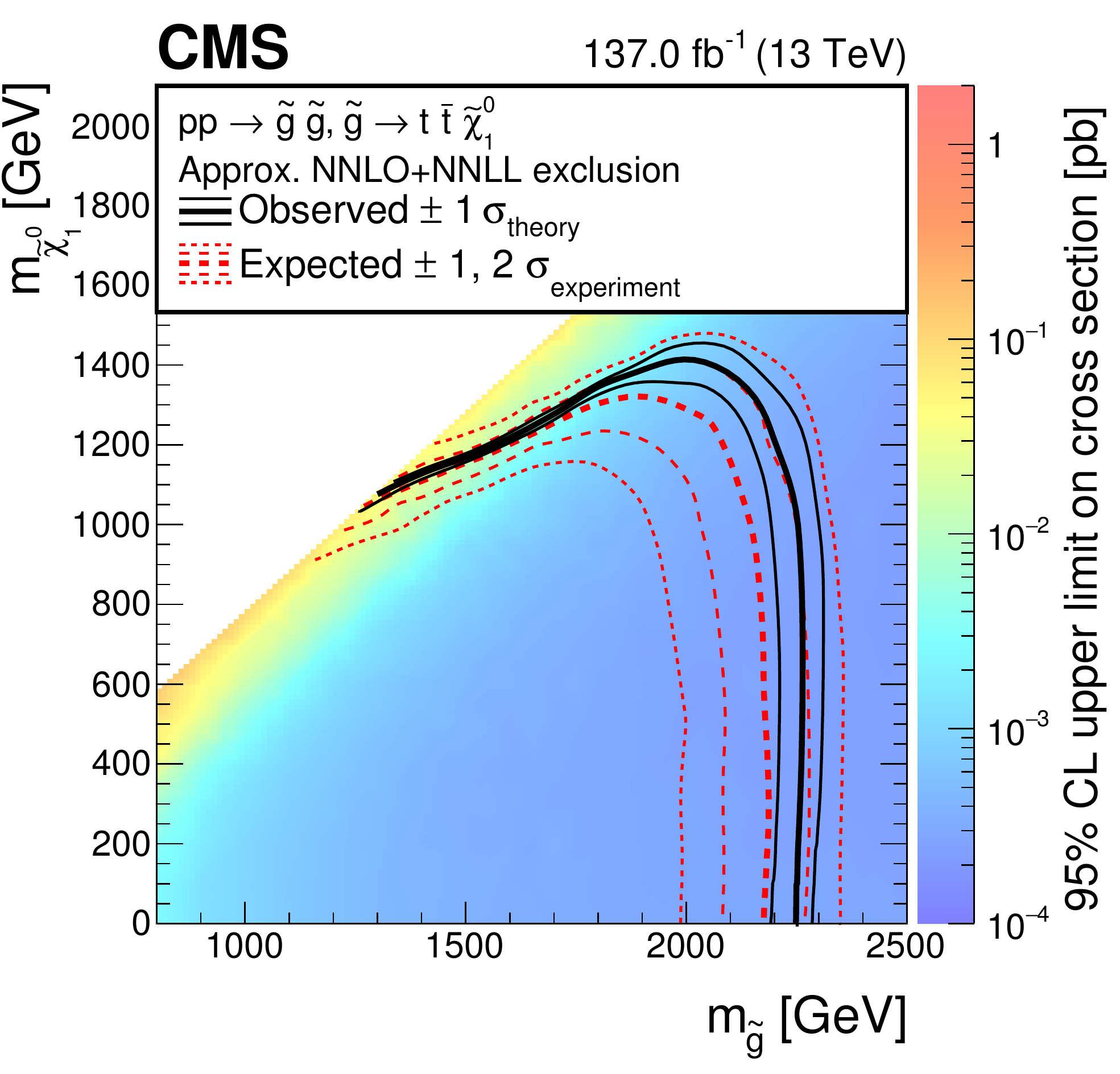}
\includegraphics[width=0.48\textwidth]{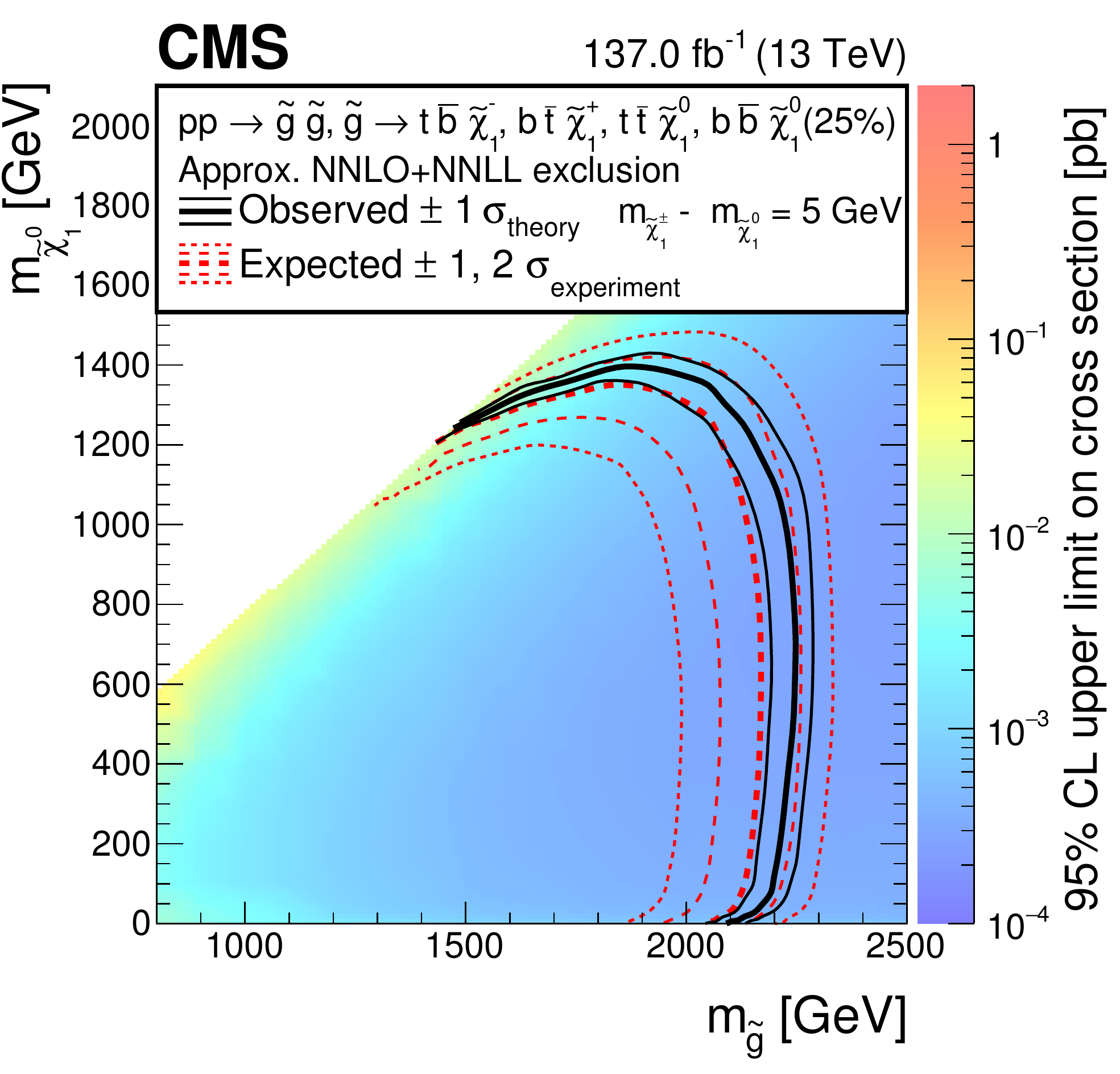}
\caption{
The 95\% \CL upper limit on the production cross section of the
T1tttt (\cmsLeft) and T1ttbb (\cmsRight)
simplified models as a function of the gluino and LSP masses.
The solid black curves represent the observed exclusion contour
with respect to approximate NNLO+NNLL signal cross sections
and the change in this contour due to variation of these
cross sections within their theoretical uncertainties ($\sigma_{\text{theory}}$)~\cite{Beenakker:2016lwe,
bib-nlo-nll-01,bib-nlo-nll-02,bib-nlo-nll-03,bib-nlo-nll-04,
Beenakker:2011sf,Beenakker:2013mva,Beenakker:2014sma,
Beenakker:1997ut,Beenakker:2010nq,Beenakker:2016gmf}.
The dashed red curves indicate the mean expected exclusion contour
and the region containing 68 and 95\% ($\pm 1$ and $2\,\sigma_{\text{experiment}}$) of the distribution of expected
exclusion limits under the background-only hypothesis.
}
\label{fig:limits_T1}
\end{figure}

\begin{figure}[t!]
\centering
\includegraphics[width=0.48\textwidth]{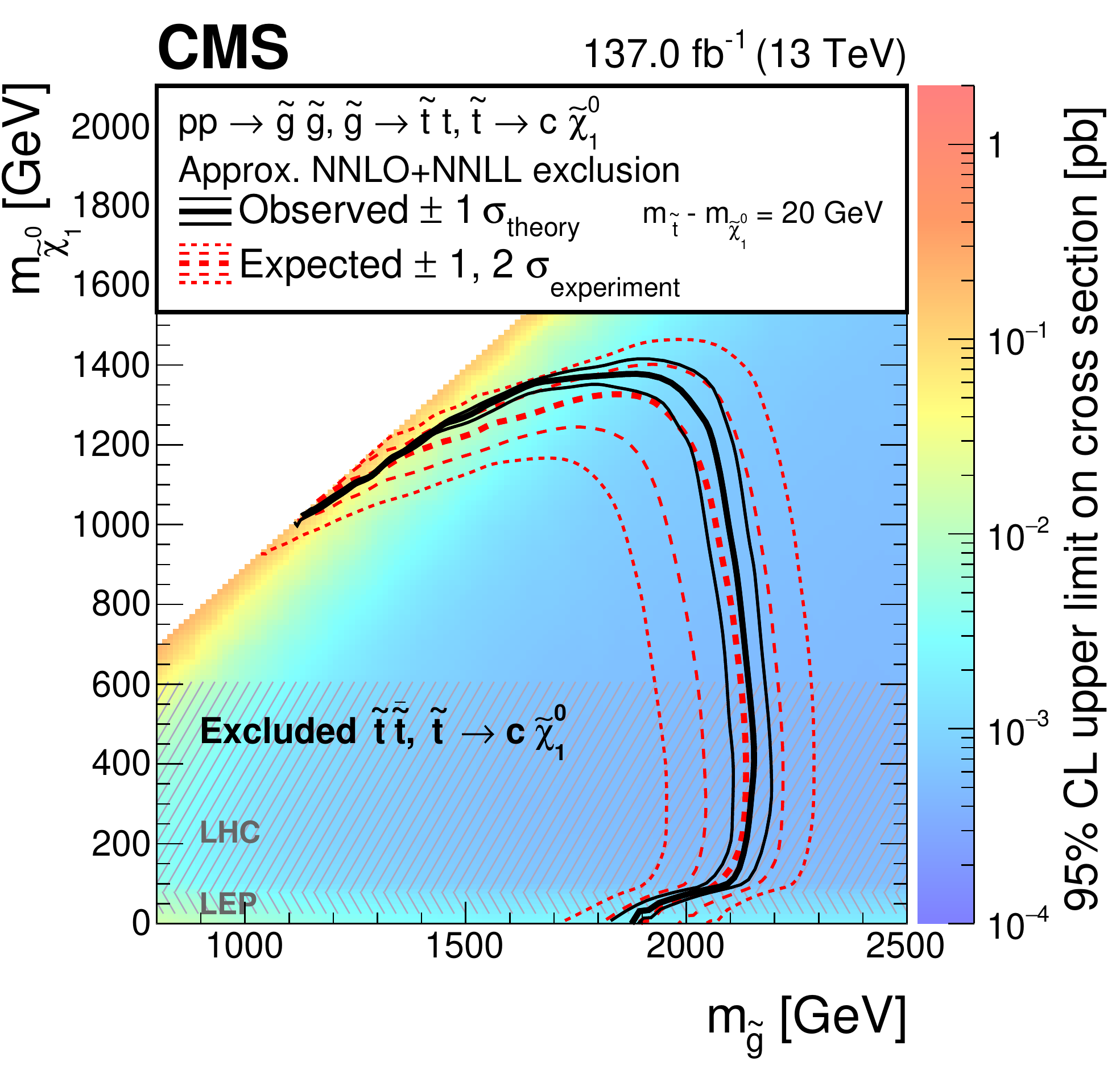}
\caption{The 95\% \CL upper limit on the production cross section of the
T5ttcc simplified model as a function of the gluino and LSP masses.  The
solid black curves represent the observed exclusion contour with respect to
approximate NNLO+NNLL signal cross sections and the change in this contour due
to variation of these cross sections within their theoretical uncertainties ($\sigma_{\text{theory}}$)~\cite{Beenakker:2016lwe,
bib-nlo-nll-01,bib-nlo-nll-02,bib-nlo-nll-03,bib-nlo-nll-04,
Beenakker:2011sf,Beenakker:2013mva,Beenakker:2014sma,
Beenakker:1997ut,Beenakker:2010nq,Beenakker:2016gmf}.  The dashed red curves
indicate the mean expected exclusion contour and the region containing 68\%
and 95\% ($\pm 1$ and $2\,\sigma_{\text{experiment}}$) of the distribution of expected exclusion limits under the
background-only hypothesis.
The expected and observed upper limits do not take into account contributions from direct top squark pair production; however,
its effect is small for $\mlsp > 600\GeV$, which corresponds to the phase space beyond the exclusions based on direct top squark pair production.
The excluded regions based on direct top squark pair production from this search and earlier searches by
the ATLAS~\cite{Aaboud:2018zjf} and CMS~\cite{Sirunyan:2016jpr,SUS-16-049,Sirunyan:2017kiw} experiments,
as well as by the LEP experiments~\cite{ALEPH,DELPHI,L3,OPAL} are indicated by the hatched areas.}
\label{fig:limits_T5}
\end{figure}

Exclusion limits for the models of gluino pair production, T1tttt, T1ttbb, and T5ttcc, are shown in the \mgluino-\mlsp plane in Figs.~\ref{fig:limits_T1} and \ref{fig:limits_T5}.
Gluino masses up to 2260\GeV and LSP masses up to 1410\GeV are excluded for the T1tttt model, up to 2250 and 1400\GeV for the T1ttbb model,
and up to 2150 and 1380\GeV for the T5ttcc model.
In the case of the T5ttcc model there is a reduction in sensitivity as \mlsp approaches zero.
This is due to the kinematic properties of the top squark decay $\stopq \to \PQc\lsp$.
The LSP in this situation carries only a small fraction of the top squark momentum, and this results in reduced \ptmiss and reduced signal acceptance.
With the SUSY particle spectrum assumed in the T5ttcc model, direct top squark production should also occur as in the T2cc model.
For $\mlsp < 600\GeV$, the T2cc model is excluded by this search and by earlier searches by the ATLAS~\cite{Aaboud:2018zjf} and CMS~\cite{Sirunyan:2016jpr,SUS-16-049,Sirunyan:2017kiw} experiments as well as by the LEP experiments~\cite{ALEPH,DELPHI,L3,OPAL}.
For $\mlsp > 600\GeV$, where the T2cc model is not excluded, adding the T2cc direct top squark production contributions to the gluino pair production contributions already present in T5ttcc does not have a significant effect on the sensitivity.
For simplicity, Fig.~\ref{fig:limits_T5} shows the exclusion based on the T5ttcc model without contributions from direct top squark production.

\section{Summary}
\label{sec:summary}

Results are presented from a search for direct and gluino-mediated top squark production in proton-proton collisions at a center-of-mass energy of 13\TeV.
The analysis includes deep neural network based tagging algorithms for top quarks and \PW bosons both at low and high transverse momentum.
The search is based on events with at least two jets and large imbalance in transverse momentum \ptmiss.
The data set corresponds to an integrated luminosity of 137\fbinv collected with the CMS detector at the LHC in 2016--2018.
A set of 183 search bins is defined based on several kinematic variables and the number of reconstructed top quarks, bottom quarks, and \PW bosons.
No statistically significant excess of events is observed with respect to the expectation from the standard model.

Upper limits at the 95\% confidence level are established on the cross section for several simplified models of
direct and gluino-mediated top squark pair production as a function of the masses of the supersymmetric particles.
Using the predicted cross sections, which are calculated with approximate next-to-next-to-leading order plus next-to-next-to-leading logarithmic accuracy,
lower limits at the 95\% confidence level are established on the top squark, lightest supersymmetric particle (LSP), and gluino masses.
In the case of the direct top squark production models,
top squark masses are excluded below a limit ranging from 1150 to 1310\GeV in the region of parameter space where the mass difference between the top squark and the LSP is larger than the \PW boson mass, depending on the top squark decay scenario.
In the region of parameter space where the mass difference between the top squark and the LSP is smaller than the mass of the \PW boson, top squark masses are excluded below a limit ranging from 630 to 740\GeV,
depending on the top squark decay scenario.
In the case of the gluino-mediated top squark production models,
gluino masses are excluded below a limit ranging from 2150 to 2260\GeV, depending on the signal model.
These results significantly extend the mass exclusions of the previous top squark searches in the fully-hadronic final state from CMS~\cite{SUS-16-049,SUS-16-050} by about 100--300\GeV,
benefiting not only from the larger data set, but also from improved analysis methods.
For models of direct top squark production, the results obtained in this analysis are the most stringent constraints to date, regardless of the final state.

\begin{acknowledgments}
    We congratulate our colleagues in the CERN accelerator departments for the excellent performance of the LHC and thank the technical and administrative staffs at CERN and at other CMS institutes for their contributions to the success of the CMS effort. In addition, we gratefully acknowledge the computing centers and personnel of the Worldwide LHC Computing Grid and other centers for delivering so effectively the computing infrastructure essential to our analyses. Finally, we acknowledge the enduring support for the construction and operation of the LHC, the CMS detector, and the supporting computing infrastructure provided by the following funding agencies: BMBWF and FWF (Austria); FNRS and FWO (Belgium); CNPq, CAPES, FAPERJ, FAPERGS, and FAPESP (Brazil); MES (Bulgaria); CERN; CAS, MoST, and NSFC (China); COLCIENCIAS (Colombia); MSES and CSF (Croatia); RIF (Cyprus); SENESCYT (Ecuador); MoER, ERC PUT and ERDF (Estonia); Academy of Finland, MEC, and HIP (Finland); CEA and CNRS/IN2P3 (France); BMBF, DFG, and HGF (Germany); GSRT (Greece); NKFIA (Hungary); DAE and DST (India); IPM (Iran); SFI (Ireland); INFN (Italy); MSIP and NRF (Republic of Korea); MES (Latvia); LAS (Lithuania); MOE and UM (Malaysia); BUAP, CINVESTAV, CONACYT, LNS, SEP, and UASLP-FAI (Mexico); MOS (Montenegro); MBIE (New Zealand); PAEC (Pakistan); MSHE and NSC (Poland); FCT (Portugal); JINR (Dubna); MON, RosAtom, RAS, RFBR, and NRC KI (Russia); MESTD (Serbia); SEIDI, CPAN, PCTI, and FEDER (Spain); MOSTR (Sri Lanka); Swiss Funding Agencies (Switzerland); MST (Taipei); ThEPCenter, IPST, STAR, and NSTDA (Thailand); TUBITAK and TAEK (Turkey); NASU (Ukraine); STFC (United Kingdom); DOE and NSF (USA).
    
    \hyphenation{Rachada-pisek} Individuals have received support from the Marie-Curie program and the European Research Council and Horizon 2020 Grant, contract Nos.\ 675440, 724704, 752730, and 765710 (European Union); the Leventis Foundation; the Alfred P.\ Sloan Foundation; the Alexander von Humboldt Foundation; the Belgian Federal Science Policy Office; the Fonds pour la Formation \`a la Recherche dans l'Industrie et dans l'Agriculture (FRIA-Belgium); the Agentschap voor Innovatie door Wetenschap en Technologie (IWT-Belgium); the F.R.S.-FNRS and FWO (Belgium) under the ``Excellence of Science -- EOS" -- be.h project n.\ 30820817; the Beijing Municipal Science \& Technology Commission, No. Z191100007219010; the Ministry of Education, Youth and Sports (MEYS) of the Czech Republic; the Deutsche Forschungsgemeinschaft (DFG), under Germany's Excellence Strategy -- EXC 2121 ``Quantum Universe" -- 390833306, and under project number 400140256 - GRK2497; the Lend\"ulet (``Momentum") Program and the J\'anos Bolyai Research Scholarship of the Hungarian Academy of Sciences, the New National Excellence Program \'UNKP, the NKFIA research grants 123842, 123959, 124845, 124850, 125105, 128713, 128786, and 129058 (Hungary); the Council of Science and Industrial Research, India; the Ministry of Science and Higher Education and the National Science Center, contracts Opus 2014/15/B/ST2/03998 and 2015/19/B/ST2/02861 (Poland); the National Priorities Research Program by Qatar National Research Fund; the Ministry of Science and Higher Education, project no. 0723-2020-0041 (Russia); the Programa Estatal de Fomento de la Investigaci{\'o}n Cient{\'i}fica y T{\'e}cnica de Excelencia Mar\'{\i}a de Maeztu, grant MDM-2015-0509 and the Programa Severo Ochoa del Principado de Asturias; the Thalis and Aristeia programs cofinanced by EU-ESF and the Greek NSRF; the Rachadapisek Sompot Fund for Postdoctoral Fellowship, Chulalongkorn University and the Chulalongkorn Academic into Its 2nd Century Project Advancement Project (Thailand); the Kavli Foundation; the Nvidia Corporation; the SuperMicro Corporation; the Welch Foundation, contract C-1845; and the Weston Havens Foundation (USA).\end{acknowledgments}

\bibliography{auto_generated}

\clearpage
\appendix
\numberwithin{table}{section}
\section{Background predictions for the full set of search bins}
\label{appdx}

In this appendix we present, in Tables~\ref{tab:pred-0}--\ref{tab:pred-6}, numerical values of the background predictions for the 183 search bins shown in Figs.~\ref{fig:pred-lm&pred-hm-nb1} and \ref{fig:pred-hm-nb2&pred-hm-nb3}.

\begin{table*}[!htb]
\centering
\topcaption[Prediction for bins 0--27]{Observed number of events and SM background predictions in search bins 0--27.}
\label{tab:pred-0}
\cmsTable{
\renewcommand*{\arraystretch}{1.25}
\begin{scotch}{cccccccr}
Search bin & \ptmiss [{\GeVns}]  &  Lost lepton  &  \zparennunujets  & Rare & QCD multijet &  Total SM  &  \ND{}  \\
\hline

\multicolumn{8}{c}{Low \dm, $\Nb = 0$, $\Nsv = 0$, $\ptISR > 500\GeV$, $2 \leq \Nj \leq 5$} \\[\cmsTabSkip]
0 & 450--550 & $2240\,^{+160}_{-150}$  & $5220\,^{+590}_{-520}$  & $103\,^{+11}_{-10}$  & $280\,^{+110}_{-80}$  & $7840\,^{+620}_{-550}$  & 7538 \\
1 & 550--650 & $1128\,^{+75}_{-70}$  & $3830\,^{+440}_{-390}$  & $81 \pm 10$  & $77\,^{+34}_{-23}$  & $5120\,^{+450}_{-400}$  & 4920 \\
2 & 650--750 & $446\,^{+32}_{-30}$  & $1790\,^{+230}_{-200}$  & $41.1 \pm 5.5$  & $29\,^{+12}_{-9}$  & $2300\,^{+230}_{-210}$  & 2151 \\
3 & ${>}750$ & $301 \pm 23$  & $1600\,^{+200}_{-180}$  & $38.0 \pm 5.0$  & $12.9\,^{+8.0}_{-5.2}$  & $1950\,^{+200}_{-180}$  & 1780 \\[\cmsTabSkip]

\multicolumn{8}{c}{Low \dm, $\Nb = 0$, $\Nsv = 0$, $\ptISR > 500\GeV$, $\Nj \geq 6$} \\[\cmsTabSkip]
4 & 450--550 & $115\,^{+12}_{-11}$  & $113\,^{+16}_{-15}$  & $5.2\,^{+1.6}_{-1.8}$  & $37\,^{+13}_{-10}$  & $270\,^{+25}_{-22}$  & 277 \\
5 & 550--650 & $45.7\,^{+5.8}_{-5.5}$  & $74\,^{+11}_{-10}$  & $4.5\,^{+1.8}_{-1.6}$  & $18.5\,^{+7.8}_{-6.0}$  & $143\,^{+15}_{-13}$  & 146 \\
6 & 650--750 & $19.5 \pm 3.0$  & $49 \pm 31$  & $3.5 \pm 1.6$  & $3.9\,^{+1.8}_{-1.5}$  & $76 \pm 32$  & 63 \\
7 & ${>}750$ & $20.6\,^{+3.3}_{-3.1}$  & $42.9\,^{+7.3}_{-6.8}$  & $1.35\,^{+0.38}_{-0.44}$  & $4.5\,^{+2.9}_{-2.1}$  & $69.4\,^{+8.5}_{-7.9}$  & 85 \\[\cmsTabSkip]

\multicolumn{8}{c}{Low \dm, $\Nb = 0$, $\Nsv \geq 1$, $\ptISR > 500\GeV$, $2 \leq \Nj \leq 5$} \\[\cmsTabSkip]
8 & 450--550 & $80.1 \pm 8.9$  & $115\,^{+17}_{-16}$  & $3.5\,^{+1.3}_{-1.1}$  & $5.9\,^{+2.6}_{-2.2}$  & $205\,^{+20}_{-18}$  & 161 \\
9 & 550--650 & $27.7 \pm 4.4$  & $83\,^{+13}_{-12}$  & $1.33\,^{+0.49}_{-0.45}$  & $1.4\,^{+1.1}_{-1.0}$  & $113\,^{+14}_{-13}$  & 126 \\
10 & 650--750 & $14.9 \pm 3.1$  & $41.6\,^{+7.6}_{-7.1}$  & $2.5 \pm 1.2$  & $1.3 \pm 1.1$  & $60.3 \pm 8.2$  & 67 \\
11 & ${>}750$ & $9.7 \pm 2.5$  & $29.4\,^{+5.7}_{-5.3}$  & $0.41 \pm 0.10$  & $0.45\,^{+0.35}_{-0.27}$  & $40.0 \pm 6.1$  & 39 \\[\cmsTabSkip]

\multicolumn{8}{c}{Low \dm, $\Nb = 0$, $\Nsv \geq 1$, $\ptISR > 500\GeV$, $\Nj \geq 6$} \\[\cmsTabSkip]
12 & 450--550 & $4.2 \pm 1.3$  & $2.5 \pm 1.2$  & $0.06 \pm 0.03$  & $1.08 \pm 0.58$  & $7.8\,^{+1.8}_{-1.9}$  & 12 \\
13 & 550--650 & $1.77 \pm 0.84$  & $1.41 \pm 0.81$  & $0.05 \pm 0.03$  & $0.53 \pm 0.33$  & $3.8 \pm 1.2$  & 4 \\
14 & 650--750 & $0.84 \pm 0.63$  & $1.7 \pm 1.1$  & $0.04 \pm 0.02$  & $0.05 \pm 0.03$  & $2.6 \pm 1.3$  & 2 \\
15 & ${>}750$ & $1.75 \pm 0.85$  & $1.9 \pm 1.3$  & $0.06 \pm 0.04$  & $0.14\,^{+0.10}_{-0.08}$  & $3.8\,^{+1.6}_{-1.5}$  & 3 \\[\cmsTabSkip]

\multicolumn{8}{c}{Low \dm, $\Nb = 1$, $\Nsv = 0$, $\mTb < 175\GeV$, $300 < \ptISR < 500\GeV$, $20 < \ptb < 40\GeV$} \\[\cmsTabSkip]
16 & 300--400 & $1302\,^{+92}_{-86}$  & $1110\,^{+130}_{-110}$  & $14.6 \pm 2.3$  & $118\,^{+43}_{-30}$  & $2540\,^{+180}_{-150}$  & 2383 \\
17 & 400--500 & $226 \pm 22$  & $246\,^{+32}_{-29}$  & $2.7\,^{+1.1}_{-0.5}$  & $27\,^{+16}_{-14}$  & $501\,^{+43}_{-39}$  & 456 \\
18 & 500--600 & $23.4 \pm 5.1$  & $32.4 \pm 6.2$  & $0.96\,^{+0.58}_{-0.66}$  & $6.3\,^{+4.8}_{-4.4}$  & $63.0 \pm 9.7$  & 68 \\
19 & ${>}600$ & $3.5 \pm 1.5$  & $5.9 \pm 2.0$  & $0.13\,^{+0.10}_{-0.03}$  & $0.14 \pm 0.15$  & $9.7 \pm 2.5$  & 14 \\[\cmsTabSkip]

\multicolumn{8}{c}{Low \dm, $\Nb = 1$, $\Nsv = 0$, $\mTb < 175\GeV$, $300 < \ptISR < 500\GeV$, $40 < \ptb < 70\GeV$} \\[\cmsTabSkip]
20 & 300--400 & $789\,^{+65}_{-59}$  & $427 \pm 51$  & $9.0\,^{+1.7}_{-1.6}$  & $70\,^{+28}_{-26}$  & $1295\,^{+91}_{-84}$  & 1250 \\
21 & 400--500 & $113 \pm 15$  & $80\,^{+12}_{-11}$  & $4.6\,^{+1.9}_{-2.2}$  & $3.7\,^{+2.7}_{-2.4}$  & $201 \pm 21$  & 222 \\
22 & 500--600 & $8.0 \pm 2.7$  & $10.2 \pm 3.7$  & $0.12 \pm 0.05$  & $0.31 \pm 0.28$  & $18.6 \pm 4.7$  & 29 \\
23 & ${>}600$ & $3.0 \pm 1.4$  & $0.76 \pm 0.60$  & $0.01 \pm 0.02$  & $0.05 \pm 0.04$  & $3.8 \pm 1.6$  & 5 \\[\cmsTabSkip]

\multicolumn{8}{c}{Low \dm, $\Nb = 1$, $\Nsv = 0$, $\mTb < 175\GeV$, $\ptISR > 500\GeV$, $20 < \ptb < 40\GeV$} \\[\cmsTabSkip]
24 & 450--550 & $82.6 \pm 9.9$  & $91 \pm 13$  & $1.64 \pm 0.98$  & $8.9\,^{+4.0}_{-3.3}$  & $185 \pm 17$  & 164 \\
25 & 550--650 & $30.5 \pm 5.5$  & $46.7 \pm 8.1$  & $1.58 \pm 0.97$  & $3.2\,^{+1.6}_{-1.4}$  & $82 \pm 10$  & 72 \\
26 & 650--750 & $7.2 \pm 2.2$  & $22.7 \pm 5.3$  & $0.20 \pm 0.06$  & $0.28 \pm 0.52$  & $30.4 \pm 5.9$  & 33 \\
27 & ${>}750$ & $8.8 \pm 2.4$  & $17.7\,^{+5.5}_{-5.2}$  & $0.23\,^{+0.15}_{-0.11}$  & $0.12 \pm 0.21$  & $26.8\,^{+6.1}_{-5.8}$  & 29 \\
\end{scotch}
}
\end{table*}

\begin{table*}[!h]
\centering
\topcaption[Prediction for bins 28--52]{Observed number of events and SM background predictions in search bins 28--52.}
\label{tab:pred-1}
\cmsTable{
\renewcommand*{\arraystretch}{1.25}
\begin{scotch}{cccccccr}
Search bin & \ptmiss [{\GeVns}]  &  Lost lepton  &  \zparennunujets  & Rare & QCD multijet &  Total SM  &  \ND{}  \\
\hline

\multicolumn{8}{c}{Low \dm, $\Nb = 1$, $\Nsv = 0$, $\mTb < 175\GeV$, $\ptISR > 500\GeV$, $40 < \ptb < 70\GeV$} \\[\cmsTabSkip]
28 & 450--550 & $72 \pm 10$  & $49.0 \pm 8.3$  & $1.28\,^{+0.56}_{-0.52}$  & $2.4\,^{+1.3}_{-1.1}$  & $125 \pm 13$  & 81 \\
29 & 550--650 & $17.2 \pm 4.0$  & $16.9 \pm 4.0$  & $0.27\,^{+0.07}_{-0.06}$  & $0.69\,^{+0.51}_{-0.46}$  & $35.0 \pm 5.7$  & 34 \\
30 & 650--750 & $7.3 \pm 2.5$  & $11.6 \pm 3.8$  & $0.56\,^{+0.69}_{-0.42}$  & $0.08 \pm 0.21$  & $19.5 \pm 4.5$  & 18 \\
31 & ${>}750$ & $3.1\,^{+1.5}_{-1.4}$  & $9.0 \pm 3.3$  & $0.12 \pm 0.04$  & $0.05 \pm 0.13$  & $12.2 \pm 3.7$  & 12 \\[\cmsTabSkip]

\multicolumn{8}{c}{Low \dm, $\Nb = 1$, $\Nsv \geq 1$, $\mTb < 175\GeV$, $\ptISR > 300\GeV$, $20 < \ptb < 40\GeV$} \\[\cmsTabSkip]
32 & 300--400 & $73 \pm 11$  & $45 \pm 13$  & $0.74 \pm 0.14$  & $7.2 \pm 4.3$  & $127 \pm 19$  & 128 \\
33 & 400--500 & $14.2\,^{+3.9}_{-3.7}$  & $13.4 \pm 3.8$  & $0.22\,^{+0.15}_{-0.09}$  & $1.5 \pm 1.2$  & $29.3\,^{+5.8}_{-5.4}$  & 42 \\
34 & ${>}500$ & $10.0 \pm 3.1$  & $7.5 \pm 2.6$  & $0.09 \pm 0.05$  & $0.33 \pm 0.35$  & $17.9 \pm 4.2$  & 16 \\[\cmsTabSkip]

\multicolumn{8}{c}{Low \dm, $\Nb \geq 2$, $\mTb < 175\GeV$, $300 < \ptISR < 500\GeV$, $40 < \ptb < 80\GeV$} \\[\cmsTabSkip]
35 & 300--400 & $154 \pm 17$  & $88\,^{+17}_{-16}$  & $2.43\,^{+0.81}_{-0.65}$  & $8.9\,^{+6.3}_{-5.9}$  & $253\,^{+26}_{-24}$  & 244 \\
36 & 400--500 & $26.5 \pm 5.8$  & $21.2 \pm 8.4$  & $0.69\,^{+0.11}_{-0.10}$  & $1.4\,^{+1.7}_{-1.3}$  & $50 \pm 11$  & 47 \\
37 & ${>}500$ & $5.6 \pm 2.6$  & $4.7 \pm 2.6$  & $0.10 \pm 0.04$  & $0.18\,^{+0.18}_{-0.17}$  & $10.6 \pm 3.8$  & 9 \\[\cmsTabSkip]

\multicolumn{8}{c}{Low \dm, $\Nb \geq 2$, $\mTb < 175\GeV$, $300 < \ptISR < 500\GeV$, $80 < \ptb < 140\GeV$} \\[\cmsTabSkip]
38 & 300--400 & $360 \pm 31$  & $93 \pm 21$  & $5.07\,^{+0.46}_{-0.42}$  & $35\,^{+20}_{-17}$  & $493\,^{+46}_{-40}$  & 443 \\
39 & 400--500 & $77 \pm 11$  & $19.0 \pm 4.7$  & $1.34\,^{+0.16}_{-0.18}$  & $9.4 \pm 6.9$  & $107 \pm 14$  & 82 \\
40 & ${>}500$ & $8.5 \pm 2.5$  & $4.5\,^{+2.0}_{-1.9}$  & $0.70 \pm 0.44$  & $0.83 \pm 0.80$  & $14.5 \pm 3.3$  & 8 \\[\cmsTabSkip]

\multicolumn{8}{c}{Low \dm, $\Nb \geq 2$, $\mTb < 175\GeV$, $300 < \ptISR < 500\GeV$, $\ptb > 140\GeV$, $\Nj \geq 7$} \\[\cmsTabSkip]
41 & 300--400 & $59.7 \pm 7.4$  & $0.90 \pm 0.82$  & $0.31\,^{+0.08}_{-0.09}$  & $4.2 \pm 4.0$  & $65.1 \pm 8.4$  & 54 \\
42 & 400--500 & $13.5 \pm 3.1$  & $0.80 \pm 0.57$  & $0.09 \pm 0.05$  & $0.30 \pm 0.34$  & $14.7 \pm 3.2$  & 15 \\
43 & ${>}500$ & $4.6 \pm 1.9$  & $5.4 \pm 5.9$  & $0.05 \pm 0.03$  & $0.06 \pm 0.06$  & $10.0 \pm 6.2$  & 2 \\[\cmsTabSkip]

\multicolumn{8}{c}{Low \dm, $\Nb \geq 2$, $\mTb < 175\GeV$, $\ptISR > 500\GeV$, $40 < \ptb < 80\GeV$} \\[\cmsTabSkip]
44 & 450--550 & $7.9 \pm 2.3$  & $4.3 \pm 2.5$  & $0.16\,^{+0.07}_{-0.06}$  & $0.31 \pm 0.29$  & $12.7 \pm 3.5$  & 22 \\
45 & 550--650 & $3.7\,^{+1.6}_{-1.7}$  & $3.5 \pm 1.9$  & $0.14 \pm 0.04$  & $0.22 \pm 0.22$  & $7.6 \pm 2.5$  & 9 \\
46 & ${>}650$ & $0.98 \pm 0.71$  & $2.7\,^{+1.9}_{-1.8}$  & $0.10 \pm 0.04$  & $0.02 \pm 0.02$  & $3.8 \pm 2.0$  & 4 \\[\cmsTabSkip]

\multicolumn{8}{c}{Low \dm, $\Nb \geq 2$, $\mTb < 175\GeV$, $\ptISR > 500\GeV$, $80 < \ptb < 140\GeV$} \\[\cmsTabSkip]
47 & 450--550 & $28.4\,^{+5.1}_{-4.8}$  & $6.1 \pm 2.2$  & $0.52 \pm 0.09$  & $0.35\,^{+0.32}_{-0.26}$  & $35.4\,^{+5.7}_{-5.3}$  & 41 \\
48 & 550--650 & $9.5 \pm 2.8$  & $5.5 \pm 2.5$  & $0.22\,^{+0.06}_{-0.07}$  & $0.12\,^{+0.11}_{-0.10}$  & $15.4\,^{+3.8}_{-3.6}$  & 14 \\
49 & ${>}650$ & $4.6 \pm 1.9$  & $4.1 \pm 1.9$  & $0.25\,^{+0.06}_{-0.07}$  & $0.09\,^{+0.08}_{-0.07}$  & $9.0 \pm 2.7$  & 8 \\[\cmsTabSkip]

\multicolumn{8}{c}{Low \dm, $\Nb \geq 2$, $\mTb < 175\GeV$, $\ptISR > 500\GeV$, $\ptb > 140\GeV$, $\Nj \geq 7$} \\[\cmsTabSkip]
50 & 450--550 & $16.6 \pm 3.3$  & $1.4 \pm 1.1$  & $0.06 \pm 0.04$  & $0.96\,^{+0.91}_{-0.85}$  & $19.0 \pm 3.6$  & 20 \\
51 & 550--650 & $6.1 \pm 1.9$  & $0.25\,^{+0.38}_{-0.32}$  & $0.05 \pm 0.02$  & $0.14 \pm 0.25$  & $6.5\,^{+2.0}_{-1.9}$  & 6 \\
52 & ${>}650$ & $2.1 \pm 1.3$  & $2.0 \pm 2.9$  & $0.04 \pm 0.03$  & $0.06 \pm 0.10$  & $4.2 \pm 3.2$  & 4 \\
\end{scotch}
}
\end{table*}

\begin{table*}[!h]
\centering
\topcaption[Prediction for bins 53--80]{Observed number of events and SM background predictions in search bins 53--80.}
\label{tab:pred-2}
\cmsTable{
\renewcommand*{\arraystretch}{1.25}
\begin{scotch}{cccccccr}
Search bin & \ptmiss [{\GeVns}]  &  Lost lepton  &  \zparennunujets  & Rare & QCD multijet &  Total SM  &  \ND{}  \\
\hline

\multicolumn{8}{c}{High \dm, $\Nb = 1$, $\mTb < 175\GeV$, $\Nj \geq 7$, $\Nres \geq 1$} \\[\cmsTabSkip]
53 & 250--300 & $199\,^{+17}_{-16}$  & $9.3 \pm 3.0$  & $3.83\,^{+0.53}_{-0.61}$  & $19\,^{+11}_{-10}$  & $231 \pm 21$  & 227 \\
54 & 300--400 & $105 \pm 11$  & $9.0 \pm 3.0$  & $3.37 \pm 0.62$  & $4.8\,^{+2.3}_{-2.1}$  & $122 \pm 12$  & 130 \\
55 & 400--500 & $25.4 \pm 5.0$  & $0.68\,^{+0.46}_{-0.41}$  & $0.68\,^{+0.16}_{-0.15}$  & $2.7 \pm 2.2$  & $29.5 \pm 5.5$  & 26 \\
56 & ${>}500$ & $7.2 \pm 2.6$  & $2.0 \pm 1.3$  & $0.30\,^{+0.08}_{-0.09}$  & $0.15 \pm 0.22$  & $9.7 \pm 2.9$  & 9 \\[\cmsTabSkip]

\multicolumn{8}{c}{High \dm, $\Nb \geq 2$, $\mTb < 175\GeV$, $\Nj \geq 7$, $\Nres \geq 1$} \\[\cmsTabSkip]
57 & 250--300 & $639 \pm 42$  & $7.3\,^{+1.9}_{-2.0}$  & $10.1 \pm 1.6$  & $11.6\,^{+9.0}_{-7.1}$  & $668 \pm 44$  & 669 \\
58 & 300--400 & $344 \pm 25$  & $5.2\,^{+1.6}_{-1.5}$  & $9.1\,^{+1.5}_{-1.3}$  & $4.9\,^{+5.3}_{-3.6}$  & $363 \pm 26$  & 345 \\
59 & 400--500 & $58.6 \pm 7.8$  & $2.7 \pm 1.4$  & $2.21\,^{+0.32}_{-0.36}$  & $6.5\,^{+7.6}_{-6.1}$  & $70\,^{+11}_{-10}$  & 54 \\
60 & ${>}500$ & $16.6 \pm 3.5$  & $1.01 \pm 0.54$  & $0.79\,^{+0.18}_{-0.15}$  & $0.89\,^{+0.85}_{-0.74}$  & $19.3 \pm 3.7$  & 21 \\[\cmsTabSkip]

\multicolumn{8}{c}{High \dm, $\Nb = 1$, $\mTb > 175\GeV$, $\Nt = 0$, $\Nres = 0$, $\Nw = 0$, $\HT > 1000\GeV$} \\[\cmsTabSkip]
61 & 250--350 & $214\,^{+21}_{-19}$  & $189\,^{+35}_{-33}$  & $4.9 \pm 1.0$  & $118\,^{+28}_{-24}$  & $526\,^{+50}_{-47}$  & 639 \\
62 & 350--450 & $88.0\,^{+9.8}_{-9.0}$  & $98\,^{+19}_{-18}$  & $3.12\,^{+0.61}_{-0.58}$  & $16.8\,^{+4.8}_{-4.1}$  & $206 \pm 22$  & 233 \\
63 & 450--550 & $39.5 \pm 5.2$  & $71\,^{+15}_{-14}$  & $1.62\,^{+0.35}_{-0.30}$  & $5.7\,^{+2.0}_{-1.7}$  & $118\,^{+16}_{-15}$  & 124 \\
64 & ${>}550$ & $40.1\,^{+5.2}_{-4.9}$  & $128\,^{+29}_{-27}$  & $5.3\,^{+1.1}_{-1.2}$  & $3.5\,^{+1.4}_{-1.1}$  & $177\,^{+30}_{-28}$  & 179 \\[\cmsTabSkip]

\multicolumn{8}{c}{High \dm, $\Nb \geq 2$, $\mTb > 175\GeV$, $\Nt = 0$, $\Nres = 0$, $\Nw = 0$, $\HT > 1000\GeV$} \\[\cmsTabSkip]
65 & 250--350 & $68.1 \pm 7.8$  & $30.4\,^{+5.7}_{-5.4}$  & $2.11 \pm 0.40$  & $35\,^{+11}_{-10}$  & $135 \pm 15$  & 139 \\
66 & 350--450 & $19.3 \pm 3.1$  & $21.4 \pm 4.2$  & $1.04\,^{+0.19}_{-0.16}$  & $2.48\,^{+0.97}_{-0.80}$  & $44.2\,^{+5.6}_{-5.3}$  & 64 \\
67 & 450--550 & $8.9 \pm 2.2$  & $12.5\,^{+3.2}_{-3.0}$  & $0.91 \pm 0.16$  & $0.89\,^{+0.40}_{-0.34}$  & $23.2\,^{+4.0}_{-3.7}$  & 23 \\
68 & ${>}550$ & $10.8 \pm 2.3$  & $21.8\,^{+5.2}_{-4.9}$  & $1.37 \pm 0.21$  & $0.90\,^{+0.77}_{-0.48}$  & $34.8\,^{+6.0}_{-5.5}$  & 45 \\[\cmsTabSkip]

\multicolumn{8}{c}{High \dm, $\Nb = 1$, $\mTb > 175\GeV$, $\Nt \geq 1$, $\Nres = 0$, $\Nw = 0$, $300 < \HT < 1000\GeV$} \\[\cmsTabSkip]
69 & 250--550 & $376 \pm 65$  & $35.3\,^{+7.6}_{-6.9}$  & $12.2 \pm 1.8$  & $4.7\,^{+2.2}_{-1.9}$  & $428 \pm 68$  & 340 \\
70 & 550--650 & $7.6 \pm 1.8$  & $5.1\,^{+1.4}_{-1.3}$  & $1.99 \pm 0.32$  & $0.13 \pm 0.13$  & $14.9 \pm 2.5$  & 17 \\
71 & ${>}650$ & $2.57 \pm 0.86$  & $3.6\,^{+1.1}_{-1.0}$  & $1.28\,^{+0.25}_{-0.23}$  & $0.09 \pm 0.12$  & $7.5\,^{+1.5}_{-1.4}$  & 6 \\[\cmsTabSkip]

\multicolumn{8}{c}{High \dm, $\Nb = 1$, $\mTb > 175\GeV$, $\Nt \geq 1$, $\Nres = 0$, $\Nw = 0$, $1000 < \HT < 1500\GeV$} \\[\cmsTabSkip]
72 & 250--550 & $82\,^{+13}_{-14}$  & $12.0\,^{+2.5}_{-2.3}$  & $4.66 \pm 0.70$  & $1.8\,^{+1.4}_{-1.3}$  & $101\,^{+14}_{-15}$  & 94 \\
73 & 550--650 & $2.84 \pm 0.84$  & $1.79\,^{+0.58}_{-0.55}$  & $0.53 \pm 0.12$  & ${<} 0.01$  & $5.2\,^{+1.1}_{-1.0}$  & 2 \\
74 & ${>}650$ & $3.13\,^{+0.99}_{-0.94}$  & $2.74\,^{+0.81}_{-0.76}$  & $0.94 \pm 0.17$  & $0.07\,^{+0.06}_{-0.05}$  & $6.9\,^{+1.4}_{-1.3}$  & 4 \\[\cmsTabSkip]

\multicolumn{8}{c}{High \dm, $\Nb = 1$, $\mTb > 175\GeV$, $\Nt \geq 1$, $\Nres = 0$, $\Nw = 0$, $\HT > 1500\GeV$} \\[\cmsTabSkip]
75 & 250--550 & $23.5 \pm 4.5$  & $3.84\,^{+0.91}_{-0.86}$  & $0.97\,^{+0.20}_{-0.19}$  & $3.9 \pm 1.1$  & $32.2 \pm 5.0$  & 28 \\
76 & 550--650 & $0.87 \pm 0.36$  & $0.28\,^{+0.17}_{-0.16}$  & $0.18\,^{+0.06}_{-0.05}$  & $0.05\,^{+0.06}_{-0.05}$  & $1.38 \pm 0.42$  & 4 \\
77 & ${>}650$ & $1.20 \pm 0.41$  & $0.49\,^{+0.22}_{-0.20}$  & $0.30 \pm 0.08$  & ${<} 0.01$  & $1.99 \pm 0.48$  & 3 \\[\cmsTabSkip]

\multicolumn{8}{c}{High \dm, $\Nb = 1$, $\mTb > 175\GeV$, $\Nt = 0$, $\Nres = 0$, $\Nw \geq 1$, $300 < \HT < 1300\GeV$} \\[\cmsTabSkip]
78 & 250--350 & $342 \pm 35$  & $47.6\,^{+9.6}_{-9.1}$  & $11.8\,^{+1.7}_{-1.6}$  & $4.8 \pm 2.5$  & $406 \pm 39$  & 351 \\
79 & 350--450 & $62.4 \pm 7.1$  & $24.1\,^{+5.2}_{-4.8}$  & $8.4 \pm 1.7$  & $3.5\,^{+2.9}_{-2.7}$  & $98\,^{+11}_{-10}$  & 90 \\
80 & ${>}450$ & $17.1\,^{+2.7}_{-2.5}$  & $13.0\,^{+2.8}_{-2.6}$  & $2.92 \pm 0.46$  & $3.3\,^{+2.3}_{-2.0}$  & $36.4\,^{+5.2}_{-4.8}$  & 29 \\
\end{scotch}
}
\end{table*}

\begin{table*}[!h]
\centering
\topcaption[Prediction for bins 81--107]{Observed number of events and SM background predictions in search bins 81--107.}
\label{tab:pred-3}
\cmsTable{
\renewcommand*{\arraystretch}{1.25}
\begin{scotch}{cccccccr}
Search bin & \ptmiss [{\GeVns}]  &  Lost lepton  &  \zparennunujets  & Rare & QCD multijet &  Total SM  &  \ND{}  \\
\hline

\multicolumn{8}{c}{High \dm, $\Nb = 1$, $\mTb > 175\GeV$, $\Nt = 0$, $\Nres = 0$, $\Nw \geq 1$, $\HT > 1300\GeV$} \\[\cmsTabSkip]
81 & 250--350 & $6.71 \pm 0.98$  & $2.10\,^{+0.54}_{-0.51}$  & $0.37 \pm 0.10$  & $1.77\,^{+0.69}_{-0.64}$  & $11.0\,^{+1.5}_{-1.4}$  & 13 \\
82 & 350--450 & $2.16\,^{+0.46}_{-0.41}$  & $1.04\,^{+0.32}_{-0.30}$  & $0.22\,^{+0.07}_{-0.06}$  & $0.75 \pm 0.52$  & $4.16\,^{+0.84}_{-0.79}$  & 4 \\
83 & ${>}450$ & $2.18 \pm 0.47$  & $1.53 \pm 0.41$  & $0.36 \pm 0.09$  & $0.49\,^{+0.40}_{-0.38}$  & $4.56 \pm 0.81$  & 4 \\[\cmsTabSkip]

\multicolumn{8}{c}{High \dm, $\Nb = 1$, $\mTb > 175\GeV$, $\Nt = 0$, $\Nres \geq 1$, $\Nw = 0$, $300 < \HT < 1000\GeV$} \\[\cmsTabSkip]
84 & 250--350 & $2260\,^{+160}_{-170}$  & $262\,^{+51}_{-47}$  & $68.5\,^{+8.7}_{-9.2}$  & $82\,^{+30}_{-25}$  & $2670\,^{+180}_{-190}$  & 2506 \\
85 & 350--450 & $343\,^{+30}_{-33}$  & $100\,^{+20}_{-18}$  & $26.3 \pm 3.8$  & $20.8\,^{+9.9}_{-8.1}$  & $490 \pm 42$  & 483 \\
86 & 450--550 & $50.5\,^{+6.8}_{-6.4}$  & $35.4\,^{+7.7}_{-7.1}$  & $8.0\,^{+1.4}_{-1.2}$  & $5.7\,^{+3.1}_{-2.5}$  & $100\,^{+12}_{-11}$  & 92 \\
87 & 550--650 & $9.2 \pm 1.6$  & $12.2\,^{+3.1}_{-2.8}$  & $2.22\,^{+0.34}_{-0.38}$  & $0.81\,^{+0.84}_{-0.75}$  & $24.4 \pm 3.8$  & 25 \\
88 & ${>}650$ & $2.34 \pm 0.66$  & $5.1\,^{+1.4}_{-1.3}$  & $0.95\,^{+0.18}_{-0.16}$  & $0.44 \pm 0.51$  & $8.8\,^{+1.7}_{-1.6}$  & 10 \\[\cmsTabSkip]

\multicolumn{8}{c}{High \dm, $\Nb = 1$, $\mTb > 175\GeV$, $\Nt = 0$, $\Nres \geq 1$, $\Nw = 0$, $1000 < \HT < 1500\GeV$} \\[\cmsTabSkip]
89 & 250--350 & $54.6 \pm 6.0$  & $8.4\,^{+2.0}_{-1.8}$  & $1.28\,^{+0.28}_{-0.24}$  & $2.7\,^{+1.7}_{-1.3}$  & $67.0 \pm 7.3$  & 69 \\
90 & 350--450 & $20.4 \pm 3.1$  & $4.9\,^{+1.2}_{-1.1}$  & $1.09\,^{+0.20}_{-0.23}$  & $1.77 \pm 0.85$  & $28.2 \pm 4.0$  & 34 \\
91 & 450--550 & $7.2 \pm 1.3$  & $3.50\,^{+0.97}_{-0.89}$  & $0.81 \pm 0.29$  & $0.33\,^{+0.20}_{-0.17}$  & $11.8 \pm 1.8$  & 9 \\
92 & 550--650 & $2.83 \pm 0.68$  & $2.89\,^{+0.88}_{-0.81}$  & $0.23 \pm 0.07$  & $0.15\,^{+0.09}_{-0.08}$  & $6.1\,^{+1.2}_{-1.1}$  & 7 \\
93 & ${>}650$ & $2.85 \pm 0.60$  & $4.1\,^{+1.2}_{-1.1}$  & $0.63\,^{+0.12}_{-0.14}$  & $0.66\,^{+0.39}_{-0.33}$  & $8.2\,^{+1.6}_{-1.5}$  & 3 \\[\cmsTabSkip]

\multicolumn{8}{c}{High \dm, $\Nb = 1$, $\mTb > 175\GeV$, $\Nt = 0$, $\Nres \geq 1$, $\Nw = 0$, $\HT > 1500\GeV$} \\[\cmsTabSkip]
94 & 250--350 & $6.8\,^{+1.1}_{-1.2}$  & $1.33\,^{+0.46}_{-0.41}$  & $0.12 \pm 0.06$  & $2.2 \pm 1.3$  & $10.5 \pm 2.1$  & 8 \\
95 & 350--450 & $2.77\,^{+0.62}_{-0.58}$  & $0.82\,^{+0.31}_{-0.29}$  & $0.08 \pm 0.04$  & $0.40\,^{+0.42}_{-0.24}$  & $4.07\,^{+0.97}_{-0.79}$  & 1 \\
96 & 450--550 & $0.96 \pm 0.32$  & $0.64 \pm 0.27$  & $0.03 \pm 0.03$  & $0.07\,^{+0.05}_{-0.04}$  & $1.70 \pm 0.45$  & 1 \\
97 & 550--650 & $0.37 \pm 0.14$  & $0.31\,^{+0.23}_{-0.14}$  & $0.05 \pm 0.03$  & $0.05\,^{+0.04}_{-0.03}$  & $0.78\,^{+0.30}_{-0.21}$  & 0 \\
98 & ${>}650$ & $1.12 \pm 0.39$  & $0.78\,^{+0.29}_{-0.27}$  & $0.14 \pm 0.05$  & $0.05\,^{+0.04}_{-0.03}$  & $2.09 \pm 0.52$  & 4 \\[\cmsTabSkip]

\multicolumn{8}{c}{High \dm, $\Nb = 1$, $\mTb > 175\GeV$, $\Nt \geq 1$, $\Nres = 0$, $\Nw \geq 1$} \\[\cmsTabSkip]
99 & 250--550 & $4.8 \pm 1.0$  & $0.36 \pm 0.15$  & $1.15 \pm 0.21$  & $0.06 \pm 0.06$  & $6.3 \pm 1.1$  & 2 \\
100 & ${>}550$ & $0.24 \pm 0.15$  & ${<} 0.03$ & $0.42\,^{+0.10}_{-0.09}$  & $0.05\,^{+0.05}_{-0.04}$  & $0.71\,^{+0.22}_{-0.20}$  & 1 \\[\cmsTabSkip]

\multicolumn{8}{c}{High \dm, $\Nb = 1$, $\mTb > 175\GeV$, $\Nt \geq 1$, $\Nres \geq 1$, $\Nw = 0$} \\[\cmsTabSkip]
101 & 250--550 & $7.3 \pm 1.3$  & $0.70 \pm 0.24$  & $2.56 \pm 0.42$  & $0.37 \pm 0.25$  & $10.9\,^{+1.7}_{-1.6}$  & 15 \\
102 & ${>}550$ & $0.51 \pm 0.19$  & $0.32\,^{+0.17}_{-0.14}$  & $0.84\,^{+0.18}_{-0.19}$  & $0.01 \pm 0.01$  & $1.68 \pm 0.34$  & 1 \\[\cmsTabSkip]

\multicolumn{8}{c}{High \dm, $\Nb = 1$, $\mTb > 175\GeV$, $\Nt = 0$, $\Nres \geq 1$, $\Nw \geq 1$} \\[\cmsTabSkip]
103 & 250--550 & $25.5 \pm 3.6$  & $2.12\,^{+0.63}_{-0.59}$  & $4.51 \pm 0.78$  & $0.02 \pm 0.02$  & $32.2 \pm 4.2$  & 34 \\
104 & ${>}550$ & $0.32 \pm 0.13$  & $0.32\,^{+0.15}_{-0.14}$  & $0.33 \pm 0.08$  & $0.08\,^{+0.07}_{-0.06}$  & $1.05\,^{+0.23}_{-0.28}$  & 1 \\[\cmsTabSkip]

\multicolumn{8}{c}{High \dm, $\Nb = 2$, $\mTb > 175\GeV$, $\Nt = 1$, $\Nres = 0$, $\Nw = 0$, $300 < \HT < 1000\GeV$} \\[\cmsTabSkip]
105 & 250--550 & $80\,^{+15}_{-14}$  & $9.9\,^{+1.9}_{-1.7}$  & $7.2 \pm 1.1$  & $0.20\,^{+0.17}_{-0.13}$  & $97\,^{+16}_{-15}$  & 79 \\
106 & 550--650 & $1.69 \pm 0.60$  & $1.84 \pm 0.88$  & $1.45 \pm 0.24$  & $0.14 \pm 0.21$  & $5.1\,^{+1.2}_{-1.1}$  & 3 \\
107 & ${>}650$ & $1.21 \pm 0.57$  & $1.28 \pm 0.46$  & $0.95\,^{+0.18}_{-0.19}$  & ${<} 0.01$  & $3.45 \pm 0.78$  & 2 \\
\end{scotch}
}
\end{table*}

\begin{table*}[!h]
\centering
\topcaption[Prediction for bins 108--136]{Observed number of events and SM background predictions in search bins 108--136.}
\label{tab:pred-4}
\cmsTable{
\renewcommand*{\arraystretch}{1.25}
\begin{scotch}{cccccccr}
Search bin & \ptmiss [{\GeVns}]  &  Lost lepton  &  \zparennunujets  & Rare & QCD multijet &  Total SM  &  \ND{}  \\
\hline

\multicolumn{8}{c}{High \dm, $\Nb = 2$, $\mTb > 175\GeV$, $\Nt = 1$, $\Nres = 0$, $\Nw = 0$, $1000 < \HT < 1500\GeV$} \\[\cmsTabSkip]
108 & 250--550 & $23.5 \pm 4.0$  & $3.57\,^{+0.87}_{-0.71}$  & $2.67 \pm 0.46$  & $0.50 \pm 0.45$  & $30.2 \pm 4.3$  & 36 \\
109 & 550--650 & $0.73 \pm 0.36$  & $0.24\,^{+0.15}_{-0.13}$  & $0.33 \pm 0.08$  & ${<} 0.01$  & $1.30 \pm 0.41$  & 3 \\
110 & ${>}650$ & $1.18\,^{+0.52}_{-0.49}$  & $0.75 \pm 0.28$  & $0.53 \pm 0.12$  & ${<} 0.01$  & $2.46\,^{+0.64}_{-0.60}$  & 4 \\[\cmsTabSkip]

\multicolumn{8}{c}{High \dm, $\Nb = 2$, $\mTb > 175\GeV$, $\Nt = 1$, $\Nres = 0$, $\Nw = 0$, $\HT > 1500\GeV$} \\[\cmsTabSkip]
111 & 250--550 & $8.4 \pm 1.8$  & $0.67\,^{+0.23}_{-0.25}$  & $0.60 \pm 0.13$  & $0.95\,^{+0.57}_{-0.52}$  & $10.7\,^{+1.9}_{-2.0}$  & 9 \\
112 & 550--650 & $0.52 \pm 0.35$  & $0.23 \pm 0.20$  & $0.09 \pm 0.04$  & $0.02 \pm 0.03$  & $0.86 \pm 0.41$  & 1 \\
113 & ${>}650$ & $0.43 \pm 0.25$  & $0.37 \pm 0.21$  & $0.14\,^{+0.04}_{-0.05}$  & $0.02 \pm 0.02$  & $0.96 \pm 0.34$  & 0 \\[\cmsTabSkip]

\multicolumn{8}{c}{High \dm, $\Nb = 2$, $\mTb > 175\GeV$, $\Nt = 0$, $\Nres = 0$, $\Nw = 1$, $300 < \HT < 1300\GeV$} \\[\cmsTabSkip]
114 & 250--350 & $67.0 \pm 8.0$  & $7.2\,^{+1.6}_{-1.5}$  & $3.61 \pm 0.55$  & $0.62 \pm 0.46$  & $78.4 \pm 8.7$  & 44 \\
115 & 350--450 & $11.4\,^{+2.5}_{-2.0}$  & $3.7\,^{+1.1}_{-1.3}$  & $2.05 \pm 0.37$  & $0.28\,^{+0.24}_{-0.22}$  & $17.5\,^{+3.1}_{-2.8}$  & 19 \\
116 & ${>}450$ & $3.27 \pm 0.72$  & $1.91\,^{+0.47}_{-0.44}$  & $1.43\,^{+0.28}_{-0.26}$  & $0.23 \pm 0.24$  & $6.8\,^{+1.1}_{-1.0}$  & 10 \\[\cmsTabSkip]

\multicolumn{8}{c}{High \dm, $\Nb = 2$, $\mTb > 175\GeV$, $\Nt = 0$, $\Nres = 0$, $\Nw = 1$, $\HT > 1300\GeV$} \\[\cmsTabSkip]
117 & 250--350 & $2.44\,^{+0.55}_{-0.63}$  & $0.08 \pm 0.05$  & $0.08 \pm 0.04$  & $0.26 \pm 0.21$  & $2.86\,^{+0.62}_{-0.69}$  & 0 \\
118 & 350--450 & $0.98\,^{+0.48}_{-0.42}$  & $0.24\,^{+0.14}_{-0.13}$  & $0.05 \pm 0.03$  & ${<} 0.01$  & $1.27\,^{+0.51}_{-0.45}$  & 0 \\
119 & ${>}450$ & $0.94 \pm 0.35$  & $0.09\,^{+0.07}_{-0.06}$  & $0.09 \pm 0.04$  & ${<} 0.01$  & $1.13\,^{+0.38}_{-0.36}$  & 2 \\[\cmsTabSkip]

\multicolumn{8}{c}{High \dm, $\Nb = 2$, $\mTb > 175\GeV$, $\Nt = 0$, $\Nres = 1$, $\Nw = 0$, $300 < \HT < 1000\GeV$} \\[\cmsTabSkip]
120 & 250--350 & $374\,^{+29}_{-32}$  & $69\,^{+12}_{-11}$  & $38.9 \pm 5.5$  & $9.0\,^{+4.9}_{-4.2}$  & $492\,^{+37}_{-40}$  & 454 \\
121 & 350--450 & $64.6 \pm 6.8$  & $24.6\,^{+4.6}_{-4.3}$  & $17.9 \pm 2.6$  & $5.8\,^{+3.9}_{-3.6}$  & $113 \pm 11$  & 114 \\
122 & 450--550 & $11.8 \pm 2.0$  & $8.0\,^{+1.9}_{-1.6}$  & $6.2\,^{+1.0}_{-1.1}$  & $3.2\,^{+2.2}_{-2.0}$  & $29.3\,^{+4.5}_{-3.6}$  & 35 \\
123 & 550--650 & $2.21 \pm 0.78$  & $3.7 \pm 1.0$  & $1.50 \pm 0.28$  & $0.9 \pm 1.2$  & $8.3 \pm 1.8$  & 6 \\
124 & ${>}650$ & $1.50 \pm 0.75$  & $1.38 \pm 0.47$  & $0.74 \pm 0.14$  & $0.31 \pm 0.45$  & $3.9 \pm 1.0$  & 4 \\[\cmsTabSkip]

\multicolumn{8}{c}{High \dm, $\Nb = 2$, $\mTb > 175\GeV$, $\Nt = 0$, $\Nres = 1$, $\Nw = 0$, $1000 < \HT < 1500\GeV$} \\[\cmsTabSkip]
125 & 250--350 & $15.9\,^{+2.4}_{-2.7}$  & $2.13\,^{+0.62}_{-0.58}$  & $0.79\,^{+0.15}_{-0.18}$  & $3.1 \pm 2.0$  & $21.9\,^{+3.8}_{-4.0}$  & 27 \\
126 & 350--450 & $3.56 \pm 0.85$  & $1.52\,^{+0.44}_{-0.41}$  & $0.38\,^{+0.11}_{-0.12}$  & $2.3\,^{+2.6}_{-2.1}$  & $7.8\,^{+3.1}_{-2.4}$  & 5 \\
127 & 450--550 & $1.76 \pm 0.55$  & $1.10\,^{+0.40}_{-0.38}$  & $0.50 \pm 0.11$  & $0.09 \pm 0.06$  & $3.45\,^{+0.76}_{-0.71}$  & 4 \\
128 & 550--650 & $0.84 \pm 0.37$  & $0.58\,^{+0.32}_{-0.28}$  & $0.28\,^{+0.09}_{-0.08}$  & $0.07 \pm 0.06$  & $1.77 \pm 0.51$  & 2 \\
129 & ${>}650$ & $1.14 \pm 0.43$  & $0.64 \pm 0.23$  & $0.90 \pm 0.46$  & ${<} 0.01$  & $2.68 \pm 0.69$  & 1 \\[\cmsTabSkip]

\multicolumn{8}{c}{High \dm, $\Nb = 2$, $\mTb > 175\GeV$, $\Nt = 0$, $\Nres = 1$, $\Nw = 0$, $\HT > 1500\GeV$} \\[\cmsTabSkip]
130 & 250--350 & $2.67 \pm 0.61$  & $0.45\,^{+0.22}_{-0.20}$  & $0.05 \pm 0.04$  & $0.28\,^{+0.18}_{-0.16}$  & $3.44 \pm 0.71$  & 4 \\
131 & 350--450 & $1.26 \pm 0.40$  & $0.26 \pm 0.14$  & $0.01\,^{+0.04}_{-0.03}$  & $0.06 \pm 0.06$  & $1.59 \pm 0.45$  & 2 \\
132 & 450--550 & $0.16\,^{+0.13}_{-0.12}$  & $0.22\,^{+0.15}_{-0.14}$  & $0.04 \pm 0.03$  & $0.03 \pm 0.02$  & $0.46\,^{+0.22}_{-0.20}$  & 1 \\
133 & 550--650 & $0.17 \pm 0.11$  & $0.20 \pm 0.14$  & $0.03 \pm 0.02$  & ${<} 0.01$  & $0.40 \pm 0.18$  & 0 \\
134 & ${>}650$ & $0.31\,^{+0.19}_{-0.17}$  & $0.37\,^{+0.20}_{-0.19}$  & $0.08 \pm 0.04$  & ${<} 0.01$  & $0.76 \pm 0.28$  & 0 \\[\cmsTabSkip]

\multicolumn{8}{c}{High \dm, $\Nb = 2$, $\mTb > 175\GeV$, $\Nt = 1$, $\Nres = 0$, $\Nw = 1$} \\[\cmsTabSkip]
135 & 250--550 & $0.81 \pm 0.23$  & $0.04 \pm 0.04$  & $0.70 \pm 0.13$  & ${<} 0.01$  & $1.54 \pm 0.29$  & 3 \\
136 & ${>}550$ & $0.10 \pm 0.05$  & $0.05 \pm 0.04$  & $0.21 \pm 0.05$  & ${<} 0.01$  & $0.36 \pm 0.09$  & 0 \\
\end{scotch}
}
\end{table*}

\begin{table*}[!h]
\centering
\topcaption[Prediction for bins 137--161]{Observed number of events and SM background predictions in search bins 137--161.}
\label{tab:pred-5}
\cmsTable{
\renewcommand*{\arraystretch}{1.25}
\begin{scotch}{cccccccr}
Search bin & \ptmiss [{\GeVns}]  &  Lost lepton  &  \zparennunujets  & Rare & QCD multijet &  Total SM  &  \ND{}  \\
\hline

\multicolumn{8}{c}{High \dm, $\Nb = 2$, $\mTb > 175\GeV$, $\Nt = 1$, $\Nres = 1$, $\Nw = 0$, $300 < \HT < 1300\GeV$} \\[\cmsTabSkip]
137 & 250--350 & $4.5\,^{+1.1}_{-1.2}$  & $0.07\,^{+0.06}_{-0.05}$  & $1.40\,^{+0.25}_{-0.23}$  & ${<} 0.01$  & $5.9\,^{+1.2}_{-1.3}$  & 5 \\
138 & 350--450 & $1.10\,^{+0.50}_{-0.43}$  & $0.14\,^{+0.10}_{-0.09}$  & $1.28\,^{+0.24}_{-0.22}$  & ${<} 0.01$  & $2.52\,^{+0.59}_{-0.52}$  & 5 \\
139 & ${>}450$ & $0.62\,^{+0.27}_{-0.24}$  & $0.17 \pm 0.10$  & $2.09 \pm 0.39$  & $1.2 \pm 1.4$  & $4.1 \pm 1.5$  & 3 \\[\cmsTabSkip]

\multicolumn{8}{c}{High \dm, $\Nb = 2$, $\mTb > 175\GeV$, $\Nt = 1$, $\Nres = 1$, $\Nw = 0$, $\HT > 1300\GeV$} \\[\cmsTabSkip]
140 & 250--350 & $0.75 \pm 0.19$  & ${<} 0.01$  & $0.16\,^{+0.06}_{-0.05}$  & ${<} 0.01$  & $0.90 \pm 0.20$  & 2 \\
141 & 350--450 & $0.31 \pm 0.12$  & $0.02 \pm 0.02$  & $0.05 \pm 0.04$  & ${<} 0.01$  & $0.38 \pm 0.13$  & 0 \\
142 & ${>}450$ & $0.21\,^{+0.11}_{-0.10}$  & $0.10 \pm 0.08$  & $0.33 \pm 0.08$  & ${<} 0.01$  & $0.64\,^{+0.17}_{-0.16}$  & 0 \\[\cmsTabSkip]

\multicolumn{8}{c}{High \dm, $\Nb = 2$, $\mTb > 175\GeV$, $\Nt = 0$, $\Nres = 1$, $\Nw = 1$} \\[\cmsTabSkip]
143 & 250--550 & $7.3\,^{+1.4}_{-1.3}$  & $0.40 \pm 0.16$  & $3.18\,^{+0.62}_{-0.58}$  & ${<} 0.01$  & $10.9 \pm 1.7$  & 6 \\
144 & ${>}550$ & $0.09 \pm 0.03$  & $0.05 \pm 0.05$  & $0.24\,^{+0.07}_{-0.06}$  & ${<} 0.01$  & $0.37 \pm 0.09$  & 0 \\[\cmsTabSkip]

\multicolumn{8}{c}{High \dm, $\Nb = 2$, $\mTb > 175\GeV$, $\Nt = 2$, $\Nres = 0$, $\Nw = 0$} \\[\cmsTabSkip]
145 & 250--450 & $0.92\,^{+0.37}_{-0.33}$  & $0.04 \pm 0.04$  & $0.78 \pm 0.16$  & ${<} 0.01$  & $1.74\,^{+0.44}_{-0.41}$  & 2 \\
146 & ${>}450$ & $0.20\,^{+0.13}_{-0.17}$  & ${<} 0.01$  & $0.36 \pm 0.09$  & ${<} 0.01$  & $0.56\,^{+0.17}_{-0.21}$  & 0 \\[\cmsTabSkip]

\multicolumn{8}{c}{High \dm, $\Nb = 2$, $\mTb > 175\GeV$, $\Nt = 0$, $\Nres = 0$, $\Nw = 2$} \\[\cmsTabSkip]
147 & ${>}250$ & $0.46 \pm 0.23$  & $0.04 \pm 0.04$  & $0.24 \pm 0.06$  & ${<} 0.01$  & $0.74 \pm 0.26$  & 0 \\[\cmsTabSkip]

\multicolumn{8}{c}{High \dm, $\Nb = 2$, $\mTb > 175\GeV$, $\Nt = 0$, $\Nres = 2$, $\Nw = 0$, $300 < \HT < 1300\GeV$} \\[\cmsTabSkip]
148 & 250--450 & $15.1\,^{+2.2}_{-2.9}$  & $0.82 \pm 0.35$  & $10.6 \pm 1.9$  & ${<} 0.01$  & $26.5\,^{+3.5}_{-4.3}$  & 19 \\
149 & ${>}450$ & $0.89 \pm 0.29$  & $0.16\,^{+0.09}_{-0.08}$  & $1.81\,^{+0.44}_{-0.35}$  & $0.58 \pm 0.59$  & $3.45\,^{+0.85}_{-0.79}$  & 3 \\[\cmsTabSkip]

\multicolumn{8}{c}{High \dm, $\Nb = 2$, $\mTb > 175\GeV$, $\Nt = 0$, $\Nres = 2$, $\Nw = 0$, $\HT > 1300\GeV$} \\[\cmsTabSkip]
150 & 250--450 & $0.43\,^{+0.19}_{-0.18}$  & ${<} 0.01$  & $0.03 \pm 0.03$  & ${<} 0.01$  & $0.46\,^{+0.20}_{-0.18}$  & 0 \\
151 & ${>}450$ & $0.19 \pm 0.15$  & $0.02 \pm 0.02$  & $0.04\,^{+0.03}_{-0.02}$  & ${<} 0.01$  & $0.24 \pm 0.15$  & 0 \\[\cmsTabSkip]

\multicolumn{8}{c}{High \dm, $\Nb = 2$, $\mTb > 175\GeV$, $(\Nt+\Nres+\Nw) \geq 3$} \\[\cmsTabSkip]
152 & ${>}250$ & $0.38\,^{+0.20}_{-0.28}$  & ${<} 0.01$  & $0.06\,^{+0.04}_{-0.03}$  & ${<} 0.01$  & $0.44\,^{+0.21}_{-0.29}$  & 1 \\[\cmsTabSkip]

\multicolumn{8}{c}{High \dm, $\Nb \geq 3$, $\mTb > 175\GeV$, $\Nt = 1$, $\Nres = 0$, $\Nw = 0$, $300 < \HT < 1000\GeV$} \\[\cmsTabSkip]
153 & 250--350 & $10.5\,^{+2.2}_{-2.0}$  & $0.20\,^{+0.11}_{-0.14}$  & $0.41 \pm 0.08$  & $0.02 \pm 0.02$  & $11.1 \pm 2.2$  & 8 \\
154 & 350--550 & $8.1 \pm 1.9$  & $0.41\,^{+0.15}_{-0.16}$  & $0.82 \pm 0.15$  & ${<} 0.01$  & $9.3 \pm 1.9$  & 6 \\
155 & ${>}550$ & $1.10 \pm 0.60$  & $0.27 \pm 0.15$  & $0.45\,^{+0.12}_{-0.10}$  & ${<} 0.01$  & $1.82 \pm 0.65$  & 4 \\[\cmsTabSkip]

\multicolumn{8}{c}{High \dm, $\Nb \geq 3$, $\mTb > 175\GeV$, $\Nt = 1$, $\Nres = 0$, $\Nw = 0$, $1000 < \HT < 1500\GeV$} \\[\cmsTabSkip]
156 & 250--350 & $5.0 \pm 1.2$  & $0.24 \pm 0.14$  & $0.32\,^{+0.08}_{-0.09}$  & $0.31 \pm 0.32$  & $5.9 \pm 1.3$  & 4 \\
157 & 350--550 & $1.64 \pm 0.61$  & $0.24\,^{+0.14}_{-0.15}$  & $0.25\,^{+0.07}_{-0.06}$  & ${<} 0.01$  & $2.13\,^{+0.67}_{-0.63}$  & 1 \\
158 & ${>}550$ & $0.12 \pm 0.12$  & $0.18 \pm 0.12$  & $0.20 \pm 0.05$  & $0.01 \pm 0.02$  & $0.52 \pm 0.18$  & 1 \\[\cmsTabSkip]

\multicolumn{8}{c}{High \dm, $\Nb \geq 3$, $\mTb > 175\GeV$, $\Nt = 1$, $\Nres = 0$, $\Nw = 0$, $\HT > 1500\GeV$} \\[\cmsTabSkip]
159 & 250--350 & $4.0\,^{+1.4}_{-1.3}$  & $0.04\,^{+0.05}_{-0.06}$  & $0.03 \pm 0.03$  & $0.10 \pm 0.08$  & $4.1 \pm 1.4$  & 9 \\
160 & 350--550 & $0.59 \pm 0.33$  & $0.19 \pm 0.24$  & $0.04 \pm 0.03$  & ${<} 0.01$  & $0.82 \pm 0.42$  & 2 \\
161 & ${>}550$ & $0.15 \pm 0.10$  & $0.07\,^{+0.10}_{-0.09}$  & $0.08 \pm 0.04$  & ${<} 0.01$  & $0.30\,^{+0.15}_{-0.14}$  & 0 \\
\end{scotch}
}
\end{table*}

\begin{table*}[!h]
\centering
\topcaption[Prediction for bins 162--182]{Observed number of events and SM background predictions in search bins 162--182.}
\label{tab:pred-6}
\cmsTable{
\renewcommand*{\arraystretch}{1.25}
\begin{scotch}{cccccccr}
Search bin & \ptmiss [{\GeVns}]  &  Lost lepton  &  \zparennunujets  & Rare & QCD multijet &  Total SM  &  \ND{}  \\
\hline

\multicolumn{8}{c}{High \dm, $\Nb \geq 3$, $\mTb > 175\GeV$, $\Nt = 0$, $\Nres = 0$, $\Nw = 1$} \\[\cmsTabSkip]
162 & 250--350 & $17.9\,^{+2.7}_{-2.5}$  & $0.64\,^{+0.27}_{-0.39}$  & $0.82 \pm 0.16$  & $0.40\,^{+0.49}_{-0.41}$  & $19.8\,^{+2.9}_{-2.7}$  & 7 \\
163 & 350--550 & $3.22\,^{+0.80}_{-0.90}$  & $0.5\,^{+1.3}_{-0.2}$  & $0.55\,^{+0.10}_{-0.11}$  & $0.16\,^{+0.18}_{-0.17}$  & $4.5\,^{+1.4}_{-1.1}$  & 2 \\
164 & ${>}550$ & $0.46 \pm 0.28$  & $0.06 \pm 0.05$  & $0.14 \pm 0.04$  & $0.12 \pm 0.13$  & $0.78 \pm 0.33$  & 0 \\[\cmsTabSkip]

\multicolumn{8}{c}{High \dm, $\Nb \geq 3$, $\mTb > 175\GeV$, $\Nt = 0$, $\Nres = 1$, $\Nw = 0$, $300 < \HT < 1000\GeV$} \\[\cmsTabSkip]
165 & 250--350 & $82.5 \pm 7.8$  & $5.0\,^{+1.5}_{-2.6}$  & $5.83 \pm 0.92$  & $1.2\,^{+1.1}_{-1.0}$  & $94.4 \pm 8.9$  & 105 \\
166 & 350--550 & $18.4\,^{+3.5}_{-3.8}$  & $4.5 \pm 1.3$  & $3.62\,^{+0.59}_{-0.63}$  & ${<} 0.01$  & $26.5\,^{+4.1}_{-4.5}$  & 20 \\
167 & ${>}550$ & $0.66 \pm 0.34$  & $0.13 \pm 0.08$  & $0.40 \pm 0.09$  & $0.01 \pm 0.01$  & $1.20 \pm 0.36$  & 1 \\[\cmsTabSkip]

\multicolumn{8}{c}{High \dm, $\Nb \geq 3$, $\mTb > 175\GeV$, $\Nt = 0$, $\Nres = 1$, $\Nw = 0$, $1000 < \HT < 1500\GeV$} \\[\cmsTabSkip]
168 & 250--350 & $6.5 \pm 1.6$  & $0.55 \pm 0.27$  & $0.15 \pm 0.06$  & $0.02 \pm 0.02$  & $7.2 \pm 1.7$  & 7 \\
169 & 350--550 & $1.61 \pm 0.56$  & $0.23\,^{+0.13}_{-0.14}$  & $0.30\,^{+0.08}_{-0.07}$  & $0.01 \pm 0.01$  & $2.15 \pm 0.61$  & 3 \\
170 & ${>}550$ & $0.22 \pm 0.18$  & $0.31 \pm 0.17$  & $0.11\,^{+0.05}_{-0.04}$  & $0.09 \pm 0.13$  & $0.73 \pm 0.29$  & 1 \\[\cmsTabSkip]

\multicolumn{8}{c}{High \dm, $\Nb \geq 3$, $\mTb > 175\GeV$, $\Nt = 0$, $\Nres = 1$, $\Nw = 0$, $\HT > 1500\GeV$} \\[\cmsTabSkip]
171 & 250--350 & $1.46 \pm 0.50$  & $0.03 \pm 0.04$  & ${<} 0.01$  & $0.03\,^{+0.03}_{-0.02}$  & $1.53 \pm 0.51$  & 4 \\
172 & 350--550 & $0.45 \pm 0.29$  & $0.20\,^{+0.27}_{-0.23}$  & $0.03 \pm 0.02$  & $0.02 \pm 0.02$  & $0.70\,^{+0.39}_{-0.37}$  & 1 \\
173 & ${>}550$ & $0.47 \pm 0.39$  & $0.03 \pm 0.03$  & ${<} 0.02$ & $0.02 \pm 0.02$  & $0.53 \pm 0.40$  & 0 \\[\cmsTabSkip]

\multicolumn{8}{c}{High \dm, $\Nb \geq 3$, $\mTb > 175\GeV$, $\Nt = 1$, $\Nres = 0$, $\Nw = 1$} \\[\cmsTabSkip]
174 & ${>}250$ & $0.45\,^{+0.19}_{-0.21}$  & $0.03\,^{+0.03}_{-0.04}$  & $0.18 \pm 0.05$  & ${<} 0.01$  & $0.66 \pm 0.21$  & 0 \\[\cmsTabSkip]

\multicolumn{8}{c}{High \dm, $\Nb \geq 3$, $\mTb > 175\GeV$, $\Nt = 1$, $\Nres = 1$, $\Nw = 0$} \\[\cmsTabSkip]
175 & 250--350 & $2.37 \pm 0.71$  & $0.04 \pm 0.04$  & $0.30\,^{+0.08}_{-0.07}$  & ${<} 0.03$ & $2.72 \pm 0.73$  & 2 \\
176 & ${>}350$ & $1.48 \pm 0.49$  & $0.18 \pm 0.09$  & $0.56 \pm 0.12$  & ${<} 0.01$  & $2.23 \pm 0.55$  & 0 \\[\cmsTabSkip]

\multicolumn{8}{c}{High \dm, $\Nb \geq 3$, $\mTb > 175\GeV$, $\Nt = 0$, $\Nres = 1$, $\Nw = 1$} \\[\cmsTabSkip]
177 & ${>}250$ & $0.84\,^{+0.63}_{-0.52}$  & $0.04 \pm 0.05$  & $0.45\,^{+0.11}_{-0.10}$  & $0.06 \pm 0.07$  & $1.39\,^{+0.66}_{-0.56}$  & 1 \\[\cmsTabSkip]

\multicolumn{8}{c}{High \dm, $\Nb \geq 3$, $\mTb > 175\GeV$, $\Nt = 2$, $\Nres = 0$, $\Nw = 0$} \\[\cmsTabSkip]
178 & ${>}250$ & $0.56 \pm 0.23$  & $0.06 \pm 0.06$  & $0.27 \pm 0.07$  & ${<} 0.01$  & $0.90 \pm 0.27$  & 1 \\[\cmsTabSkip]

\multicolumn{8}{c}{High \dm, $\Nb \geq 3$, $\mTb > 175\GeV$, $\Nt = 0$, $\Nres = 0$, $\Nw = 2$} \\[\cmsTabSkip]
179 & ${>}250$ & $0.04 \pm 0.02$  & ${<} 0.01$  & $0.04 \pm 0.02$  & ${<} 0.01$  & $0.08 \pm 0.03$  & 0 \\[\cmsTabSkip]

\multicolumn{8}{c}{High \dm, $\Nb \geq 3$, $\mTb > 175\GeV$, $\Nt = 0$, $\Nres = 2$, $\Nw = 0$} \\[\cmsTabSkip]
180 & 250--350 & $2.9\,^{+1.1}_{-0.9}$  & $0.02\,^{+0.02}_{-0.03}$  & $0.44\,^{+0.13}_{-0.11}$  & ${<} 0.01$  & $3.4\,^{+1.2}_{-1.0}$  & 6 \\
181 & ${>}350$ & $0.88\,^{+0.36}_{-0.33}$  & $0.03\,^{+0.03}_{-0.02}$  & $0.42 \pm 0.12$  & ${<} 0.01$  & $1.33\,^{+0.42}_{-0.37}$  & 0 \\[\cmsTabSkip]

\multicolumn{8}{c}{High \dm, $\Nb \geq 3$, $\mTb > 175\GeV$, $(\Nt+\Nres+\Nw) \geq 3$} \\[\cmsTabSkip]
182 & ${>}250$ & $0.07 \pm 0.02$  & ${<} 0.01$  & $0.04 \pm 0.02$  & ${<} 0.01$  & $0.11 \pm 0.03$  & 0 \\
\end{scotch}
}
\end{table*}
\cleardoublepage \section{The CMS Collaboration \label{app:collab}}\begin{sloppypar}\hyphenpenalty=5000\widowpenalty=500\clubpenalty=5000\vskip\cmsinstskip
\textbf{Yerevan Physics Institute, Yerevan, Armenia}\\*[0pt]
A.M.~Sirunyan$^{\textrm{\dag}}$, A.~Tumasyan
\vskip\cmsinstskip
\textbf{Institut f\"{u}r Hochenergiephysik, Wien, Austria}\\*[0pt]
W.~Adam, J.W.~Andrejkovic, T.~Bergauer, S.~Chatterjee, M.~Dragicevic, A.~Escalante~Del~Valle, R.~Fr\"{u}hwirth\cmsAuthorMark{1}, M.~Jeitler\cmsAuthorMark{1}, N.~Krammer, L.~Lechner, D.~Liko, I.~Mikulec, F.M.~Pitters, J.~Schieck\cmsAuthorMark{1}, R.~Sch\"{o}fbeck, M.~Spanring, S.~Templ, W.~Waltenberger, C.-E.~Wulz\cmsAuthorMark{1}
% ...
\vskip\cmsinstskip
\dag: Deceased\\
1:  Also at Vienna University of Technology, Vienna, Austria\\
% ...
\end{sloppypar}
%%% END EDITABLE REGION %%%
% skeleton_end
\end{document}